\documentclass[fleqn,usenatbib]{mnras}

\usepackage{newtxtext,newtxmath}

\usepackage[T1]{fontenc}

\DeclareRobustCommand{\VAN}[3]{#2}
\let\VANthebibliography\thebibliography
\def\thebibliography{\DeclareRobustCommand{\VAN}[3]{##3}\VANthebibliography}

\usepackage{ae,aecompl}

\usepackage{graphicx}	
\usepackage{amsmath}	
\usepackage{aas_macros}
\usepackage{cancel}
\usepackage{soul,xcolor}
\usepackage{pdflscape}
\usepackage{deluxetable}

\newcommand{\Msun}{M$_\odot$}

\setstcolor{red}

\renewcommand\ion[2]{#1$\;${\small\rmfamily \uppercase\expandafter{\romannumeral #2 \relax}}}

\title[Abundances and IMF in massive ETGs]{Recovery of the low- and high-mass end slopes of the IMF in massive early-type galaxies using detailed elemental abundances}

\author[M. den Brok et al.]{Mark den Brok,$^{1}$ Davor Krajnovi{\'c},$^{1}$\thanks{E-mail: dkrajnovic@aip.de}  Eric Emsellem,$^{2,3}$ Wilfried Mercier,$^{4,5}$ 
\newauthor Matthias Steinmetz$^{1}$ and Peter M. Weilbacher$^{1}$\\
$^1$Leibniz-Institut f\"ur Astrophysik Potsdam (AIP), An der Sternwarte 16, 14482, Potsdam, Germany\\
$^2$European Southern Observatory, Karl-Schwarzschild-Str. 2, D-85748 Garching, Germany\\
$^3$Univ Lyon, Univ Lyon1, ENS de Lyon, CNRS, Centre de Recherche Astrophysique de Lyon UMR5574, 69230 Saint-Genis- Laval, France\\
$^4$Institut de Recherche en Astrophysique et Plan\'etologie (IRAP), Universit\'e de Toulouse, CNRS, UPS, CNES, 31400 Toulouse, France\\
$^5$Aix Marseille Univ, CNRS, CNES, LAM, Marseille, France\\
}

\date{Accepted 2024 March 22. Received 2024 February 20; in original form 2023 July 7}

\pubyear{2023}

\begin{document}
\label{firstpage}
\pagerange{\pageref{firstpage}--\pageref{lastpage}}
\maketitle

\begin{abstract}
Star formation in the early Universe has left its imprint on the chemistry of observable stars in galaxies. We derive elemental abundances and the slope of the low-mass end of the initial mass function (IMF) for a sample of 25 very massive galaxies, separated into brightest cluster galaxies (BCGs) and their massive satellites. The elemental abundances of BGCs and their satellites are similar, but for some elements, satellite galaxies show a correlation with the global velocity dispersion. Using a subset of derived elemental abundances, we model the star formation histories of these galaxies with chemical evolution models, and predict the high-mass end slope of the IMF and star formation timescales. The high-mass end IMF slope of the satellite galaxies correlates with the global velocity dispersion. The low- and the high-mass end IMF slopes are weakly correlated in a general sense that top heavy IMFs are paired with bottom heavy IMFs. Our results do not necessarily imply that the IMF was simultaneously bottom and top heavy. Instead, our findings can be considered consistent with a temporal variation in the IMF, where, for massive galaxies, the high-mass end IMF slope is representative of the very early age and the low-mass end slope of the later star formation. The small but noticeable differences between the BCGs and the satellites in terms of their elemental abundances and IMF slopes, together with their stellar kinematical properties, suggest somewhat different formation pathways, where BCGs experience more major, gas-free mergers.
\end{abstract}




\section{Introduction}
\label{sec:intro}

The elemental abundances in the atmospheres of stars are indicative for the chemical composition of the gas from which these stars formed. Different chemical elements can be related to different types of stellar end-products, which generally correspond to different time scales. Understanding the origin of these elements therefore helps interpreting the evolution of galaxies \citep[e.g.][]{Tin79}. Type II, or core collapse (CC) supernovae (SNe) produce primarily $\alpha$ elements (e.g. Na, Mg, Si, Ca). Type Ia supernovae mainly contribute iron peak elements (e.g. Fe, Cr, Co, Ni, Cu, Mn). The third important channel for the nucleosynthesis of light elements is the formation of these elements in asymptotic giant branch (AGB) stars. Depending on the initial mass of these stars, they can be important contributors of C or N and many of the {\it s}-process elements {e.g. Sr, Y, Nb, Ba} \citep[e.g.][]{NomKobTom13}.

The time scales at which the ISM is enriched by these processes vary: CC SNe, such as type Ib, type Ic and type II SNe, happen almost instantly ($\sim$few Myr) after star formation starts. Type Ia SNe happen on much longer time scales \citep{MaoBad10}, probably with a delay time distribution that has a slope declining somewhat steeper than $t^{-1}$. Observationally the delay time distribution has however been difficult to constrain. Theoretically, delay time distributions can be predicted from binary population synthesis calculations. Depending on whether the progenitor of the SN Ia consists of a single degenerate (SD) or a double degenerate (DD) object, different time scales are predicted. For example, \citet{MenvanDeG10} quote 50 Myr as earliest supernova for the DD mechanism, whereas their fig.~4 shows a $\sim 150-200$ Myr delay for the onset of supernovae through the SD channel.

AGB enrichment sets in after the heaviest progenitors evolve away from the main sequence ($\sim 8$\Msun, corresponding to $\sim 40$ Myr). The high equivalent widths of CO and CN indices have been used to argue extended star formation histories \citep[e.g.][]{SanGorCar03,MarCarSan09}, as higher equivalent widths can result from enrichment by AGB stars.

The observed amount of $\alpha$ elements has been widely used as a measure of star formation duration. This is usually done using Mg as a proxy and normalising it by the solar abundance ([Mg/H]), with respect to the amount of iron peak elements ([Fe/H]). The $\alpha$ abundance is then written as the solar abundance-normalised logarithm [Mg/Fe]. It is now well established that there exists a relation between $\alpha$ abundance and the mass (or velocity dispersion) of early-type galaxies \citep[e.g.][]{WorFabGon92,DavSadPel93,CarDan94,MarGreRen03,ThoMarBen03}. Over the past 15 years it has however become clear that abundances and ages are not the only stellar population parameters that vary with galaxy mass; more and more evidence is pointing in the direction of a non-universal initial mass function \citep[IMF; e.g.][for a review]{2018PASA...35...39H}.

The IMF in the Milky Way was inferred for the first time by \citet{Sal55} using bright stars ($M_V < 3.5$). The \citet{Sal55} IMF is described by a power law with a negative slope of $\alpha=2.35$\footnote{ Throughout the paper we will use the convention where the IMF is given by the formula $\psi(m) = \frac{dN}{dm}\sim m^{-\alpha}$, as in \citet{2010ARA&A..48..339B}.}. Later work, in particular the papers by \citet{MilSca79} and \citet{Kro01}, changed the idea of a single power law describing both the high-mass end and the low-mass end of the IMF in the Milky Way. The high-mass end of the IMF has long been suspected to be top heavy (i.e. having a power law with index shallower than Salpeter's $\alpha=2.35$) at early times, based on several lines of evidence \citep[see e.g.][]{Lar98,CalMen09,DeMVinMat19}. Observations of nearby star forming galaxies \citep{GunHopSha11} as well as observations of galaxies at high redshift \citep{NanGlaKac17,NanBriGla20} confirm that the high-mass end depends on star formation rate.

The analysis of old stellar populations suggests that in the centres of early type galaxies, the slope of the IMF  with respect to mass is  steeper than the IMF of the Milky Way, i.e., they show an excess of low-mass stars (bottom heavy), with the most bottom heavy IMFs occurring in the more massive galaxies. The evidence for this is based independently on mass-to-light ratios obtained from stellar population modelling and Jeans models \citep[e.g.][]{CapMcDAla12,CapMcDAla13}, from lensing studies \citep[e.g.][]{TreAugKoo10,SpiKooTra11,OldAug18}, from stellar population analysis of dwarf sensitive features in spectra \citep[e.g.][]{vanCon10,SpiTraKoo12,LaBFerVaz13,GuGreNew21,FelLonFre21}, but also from alternative methods that count the amount of dwarf stars \citep[e.g.][]{vanCon21}. It is, however, not fully understood what is the main driver of the variable low-mass end of the IMF. The IMF slope is known to correlate with velocity dispersion and [$\alpha$/Fe] abundance \citep{CapMcDAla13,2014ApJ...792L..37M}. It is also known that IMF appears to be most bottom-heavy in the centre of galaxies and bottom-lighter at larger radii \citep{MarLaBVaz15, vanConVil17, SarSpiLaB18, ParThoMar18,BarSpiArn20,FelLonFre21}, and it is therefore not clear how to reconcile this with the rather flat [Mg/Fe] gradients observed in early-type galaxies \citep{MehThoSag03,SanForStr07,KunEmsBac10,GreJanMa15,2020A&A...635A.129K,ParThoMar21,FelLonFre21}.
We note that although there exists evidence from multiple methods for a variable IMF, each of these methods carries a considerable amount of uncertainty with it. Both lensing and stellar dynamics studies have to deal with the problem that they have to separate any non-stellar mass contribution from a stellar mass contribution. In addition, non-homology in elliptical galaxies may cause variation in mass-to-light ratio with increasing galaxy mass that is not always  captured by the dynamical models at the required accuracy. For example, axisymmetric models can severely underestimate the mass-to-light ratio of triaxial galaxies \citep[e.g.][]{ThoJesNaa07,denKraEms21} and also the assumed shape of the velocity ellipsoid may bias the stellar mass-to-light ratio \citep{ThaKraWei22}. \citet{ClaSchFra15} argued that the IMF trends seen in the ATLAS$^{\rm 3D}$ sample may in fact arise from a universal IMF combined with underestimated modelling errors. The simultaneous analysis of mass-to-light ratio excesses from SSP modelling and from kinematics by \citet{Smi14} showed no evidence for any correlation between the two methods; a similar result was found for the lensing/SSP analysis by \citet{NewSmiCon17}. The lensing analysis of 23 strong lenses by \citet{SonJaeCha19} shows that on scales of 5-10 kpc an IMF heavier than in the Milky Way is ruled out. On somewhat smaller scale ($<5$ kpc), \citet{ColSmiLuc18}, using strong lensing, find no evidence for a large mass excess in two lensed galaxies;  \citet{ColSmiLuc18a} show that this is even the case for a lensed galaxy for which the mass is constrained within $R < 1.5$kpc.

From a theoretical perspective, there are several reasons to expect an environmental dependence for the IMF. \citet{PadNor02}, \citet{HenCha08} and \citet{Hop12} derive an IMF based on the turbulent Jeans mass of molecular clouds. \citet{NamFedKru21} show that flatter turbulent spectra produce top-heavier IMFs. \citet{WeiFerVaz13} proposed a temporal variation of the IMF, which changes during the formation of early-type galaxies from a short duration top-heavy IMF to a long duration bottom heavy IMF, motivated by the increased pressure in the interstellar medium from the high amount of high-mass stars formed because of the top-heavy IMF. \citet{LacBauFre16} show that semi-analytic galaxy formation models do indeed fit better to data when they allow for a somewhat top-heavy IMF during starburst phases. \citet{BarCraSch18} implement a pressure dependent IMF in the EAGLE simulations, which leads to weak trends of IMF slope with global metallicity \citep{BarSchCra19} and  negative radial IMF gradients \citep{BarSchCra19a}.

Contrary to the direct measurement of the low-mass slope of the IMF in early-type galaxies (ETGs), which can be determined with advanced stellar population models directly from the observed spectra, the determination of the slope of the high-mass end of the IMF can almost never be done directly, as high-mass stars have lifetimes that are  much shorter than the typical age of the stellar populations in elliptical galaxies. Through their latest evolutionary stages, the high-mass stars have, however, left an imprint on the chemistry of the ISM, which is still observable in the low-mass stars formed out of this gas. By analyzing the abundances of the low-mass stars and using chemical evolution models, it is, therefore, still possible to get an inference on the high-mass IMF slope, even after the high-mass stars have died. Moreover, this inference can be directly compared to the low-mass slope of the IMF.

In this paper we derive abundances and stellar population parameters for a sample of 25 {\it very} massive ETGs (stellar mass $M \gtrapprox 10^{12}$\Msun), which we observed with the Multi Unit Spectroscopic Explorer \citep[MUSE,][]{2010SPIE.7735E..08B} at the Very Large Telescope, and derive slopes for the high {\it and} low-mass end of the IMF.  The sample selection and data reduction are presented in Section~\ref{sec:data}. The methods used for the analysis of the data and the chemical modelling are presented in Section~\ref{sec:analysis}. The resulting measurement of the elemental abundances, predictions of the chemical evolution modelling, and the derivation of the low- and high-mass end IMF parameters are presented in Section~\ref{sec:results}. We discuss the uncertainties on the measurements and the chemical evolution models, as well as the implication of the results in Section~\ref{sec:discussion}, presenting conclusions in Section~\ref{s:sum}.

\section{Sample and data}
\label{sec:data}

\begin{figure}
 \includegraphics[angle=0, width=0.8\columnwidth]{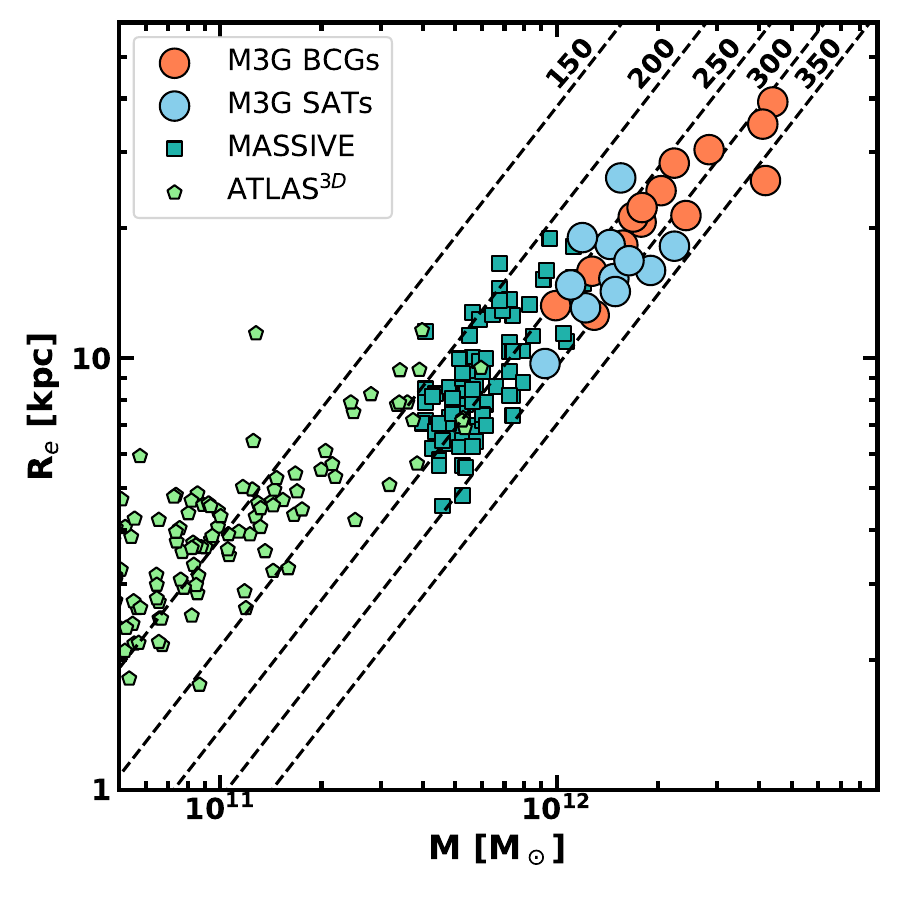}
 \caption{Mass -- size relation for M3G galaxies (circles), divided into BCGs (orange) and satellites (light blue). Squares are galaxies from the MASSIVE survey \citep[values taken from][]{2014ApJ...795..158M}, and pentagons show massive part of the ATLAS$^{\rm 3D}$ Survey of nearby early-type galaxies \citep[values taken from][]{CapMcDAla13}. Sizes of galaxies from all surveys are based on the 2MASS observations. The masses of the MASSIVE galaxies are based on the 2MASS K-band absolute magnitude and calibration eq.~(2) from \citet{2013ApJ...778L...2C}. ATLAS$^{\rm 3D}$ masses come from dynamical models. M3G masses are based on the virial estimate and taken from \citet{KraEmsden18}. The dashed lines are the lines of the constant velocity dispersion, based on the \citet{CapBacBur06} parametrisation of the virial mass estimator.}
\label{fig:ms}
\end{figure}

The data used in this work come from the MUSE observations of massive galaxies, collected in the M3G sample \citep{KraEmsden18}. Both the sample selection as well as the data reduction of the M3G galaxies has been presented in \citet{KraEmsden18} and will be elaborated in more details in Krajnovi\'c et al (in prep.). We briefly repeat that the sample consists of 25 very massive early-type galaxies found in some of the densest environments. Galaxies are located in central region of rich clusters, with 14 galaxies spread over the three main clusters of the Shapley Super Cluster (SSC), including the three brightest cluster galaxies (BCGs). The remaining 11 galaxies were chosen to be BCGs from a sample of galaxy clusters with richness above 40 \citep{1989ApJS...70....1A}. 
The sample, therefore consists of 14 BCGs (3 in SSC and 11 in various clusters) and 11 non-BCGs, which we will refer to as satellites (all in SSC). Among satellites there are also second brightest galaxies in the three SSC clusters covered by the survey. Fig.~\ref{fig:ms} presents mass -- size relation for the M3G galaxies, divided into BCGs and satellites, together with galaxies from the MASSIVE Survey \citep{2014ApJ...795..158M} and the more massive galaxies from the ATLAS$^{\rm 3D}$ Survey \citep{2011MNRAS.413..813C} sample. Noteworthy for this work is that, while there is some overlap, M3G BCGs are typically more massive and larger than the M3G satellites. 

Galaxies were observed with MUSE, with exposure times between 2 to 6 hr. All data were reduced using v1.2 to v1.6 of the MUSE pipeline \citep{WeiPalStr20}, except PGC\,046832, for which the reduction has been presented in \citet{denKraEms21} and for which the data reduction included the use of the Zurich Atmospheric Purge \citep[ZAP,][]{SotLilBac16}. Basic reduction included bias subtraction, flat-fielding, wavelength calibration, flux calibration, sky subtraction (using offset fields), alignment and drizzling on a common reference frame. The major difference in this work with respect to the steps presented in \citet{KraEmsden18} is that we switched of the telluric correction in the pipeline, but instead determined telluric absorption with the \textsc{molecfit} tool \citep[v1.5.9;][]{SmeSanNol15}, as in \citet{PagKraden21}. We do this by selecting foreground field stars, depending on availability, with a as featureless spectrum as possible, in each (non telluric subtracted) reduced and combined cube. For a few galaxies that have no available field stars, we use instead the centre of the galaxy as the input spectrum for \textsc{molecfit} to calculate a telluric absorption spectrum. We apply the telluric spectra produced by \textsc{molecfit} for correcting the final data cubes. This approach helped in reducing the residual telluric features, but we still track their location and mask them when necessary. 

The M3G sample can be put in perspective by comparison with the ATLAS$^{\rm 3D}$ and the MASSIVE samples (Fig.~\ref{fig:ms}). The former is a sample of 260 nearby ($<$40Mpc) early-type galaxies, less massive than a few $\times10^{11} M_{\odot}$, of which only one is a BCG (M49 in Virgo, but there is also M87 as the second brightest galaxy), while the sample galaxies are in typically less dense environments compared to the clusters that host M3G galaxies. The latter is a sample of the most massive galaxies within 108 Mpc. The M3G galaxies are larger and more massive than galaxies in the ATLAS$^{\rm 3D}$ sample, while there is an overlap with galaxy properties of the MASSIVE sample. M3G galaxies are also more distant, with a mean distance of about 190 Mpc. Therefore, M3G can be considered as an extension of these nearby galaxy samples, both in mass and size.

\section{Analysis}
\label{sec:analysis}

\subsection{Abundances within an effective radius}
\label{ss:cent_data}

We bin spectra for each galaxy within an elliptical aperture of a semi-major axis radius equal to the half-light (effective) radius of the galaxy. The shape of the elliptical aperture (position angle, ellipticity and size) is based on values from Table 1 of \citet{KraEmsden18}. The same aperture was used by \citet{KraEmsden18} to extract the global stellar velocity dispersion, $\sigma_e$, which we use throughout this work as a characteristic property of each galaxy. The combined central spectra have average S/N per pixel above 200 in all but one case where it is just under 100. The mean S/N is  235, while in three cases it is higher than 300. Although the data have a high S/N, the use of separate sky observations, and the data reduction methods for sky removal, nevertheless lead to imperfect sky emission-line subtraction. We solve this by using an outer-bin spectrum obtained by combining the spectra outside the central effective radius (elliptical) aperture. We first perform a stellar population fit for each galaxy on this outer-bin, and use it to estimate the residual sky-lines. These sky-lines are then removed from the central effective-radius bin before the final stellar population fit. Figure~\ref{fig:skysub} shows a typical example of the residual sky spectrum and the effect of its removal from the galaxy spectra. See Appendix~\ref{a:skysub} for more details. 

Two commonly used approaches for extracting information about the properties of stellar populations from galaxy spectra are the index fitting \citep[e.g.,][]{1994ApJS...95..107W, 2000AJ....119.1645T, 2003MNRAS.339..897T, 2006MNRAS.369..497K, 2007ApJS..171..146S} and the full spectral fitting \citep[e.g.][]{2004PASP..116..138C, Cap17, 2005MNRAS.358..363C, 2006MNRAS.365...74O, 2007MNRAS.381.1252T, 2009A&A...501.1269K, 2017MNRAS.472.4297W} methods, while in the recent literature there are also hybrid methods combining spectra and photometric (broad or narrow band) magnitudes \citep{2023MNRAS.526.3273C}, or defining specific indices but fitting each pixels within the index wavelength ranges \citep{2019A&A...626A.124M}. All methods have their advantages, and in some cases limiting the range of the pixel fitting methods is beneficial \citep[e.g.][]{2015MNRAS.449.1177V, 2020MNRAS.499.2327G}, while in some cases index fitting might be a more appropriate method \citep[e.g.][]{2024MNRAS.tmp..170Z}. Nevertheless, we select the full spectral fitting as it does not depend on the specific definition of the spectral features and their pseudo-continua, and benefits from the higher S/N of our spectra. \citet{2021A&A...649A..93B} present an educational overview of recent choices and an extensive references, advocating the full spectral fitting for estimating elemental abundances and the IMF of galaxies. A comparison of our results, and those from index fitting and full pixel fitting studies is provided in Appendix~\ref{a:el} and Fig~\ref{fig:alf_comp}.

We model the central effective-radius bin of each galaxy with \textsc{alf} \citep{Convan12a,Convan12,ConVilvan18}, which fits synthesized stellar population models to galaxy spectra. The underlying models are based on MIST isochrones \citep{ChoDotCon16}, MILES stellar spectra \citep{SanPelJim06}, IRTF stellar spectra \citep{VilConJoh17} and theoretical response curves to change the abundance of individual elements \citep[see references in ][]{ConVilvan18}. We use the `full' mode of \textsc{alf}. This means that besides metallicity, 19 elemental abundances are left as free parameters. In addition to a dominant old population, two parameters (age and light fraction) govern the presence of a second (subdominant) younger population with the same elemental abundances. We explore single power law IMFs with a slope fixed to the Salpeter slope ($\alpha=2.35$), as well as with a free power-law slope, both with a lower mass cut-off at 0.08 \Msun. Although it is possible to run \textsc{alf} with a more complex shape for the IMF, we chose not to do this for the reason of simplicity, but also because recent observations prefer uni-modal power laws over more complex shapes \citep[e.g.][]{ConvanVil17,ZhoMoLi19}. Therefore, we fit for 23 free parameters (age, metallicity, light fraction of the younger component, 19 elements and the IMF slope). \textsc{alf} uses an implementation of the MCMC sampler \textsc{emcee} \citep{ForHogLan13}. We use \texttt{nwalkers=1024} and \texttt{nburn=20000}, for the number of walkers used by the MCMC sampler and the duration of the burn-in phase.

\begin{figure}
  \includegraphics[angle=0, width=\columnwidth]{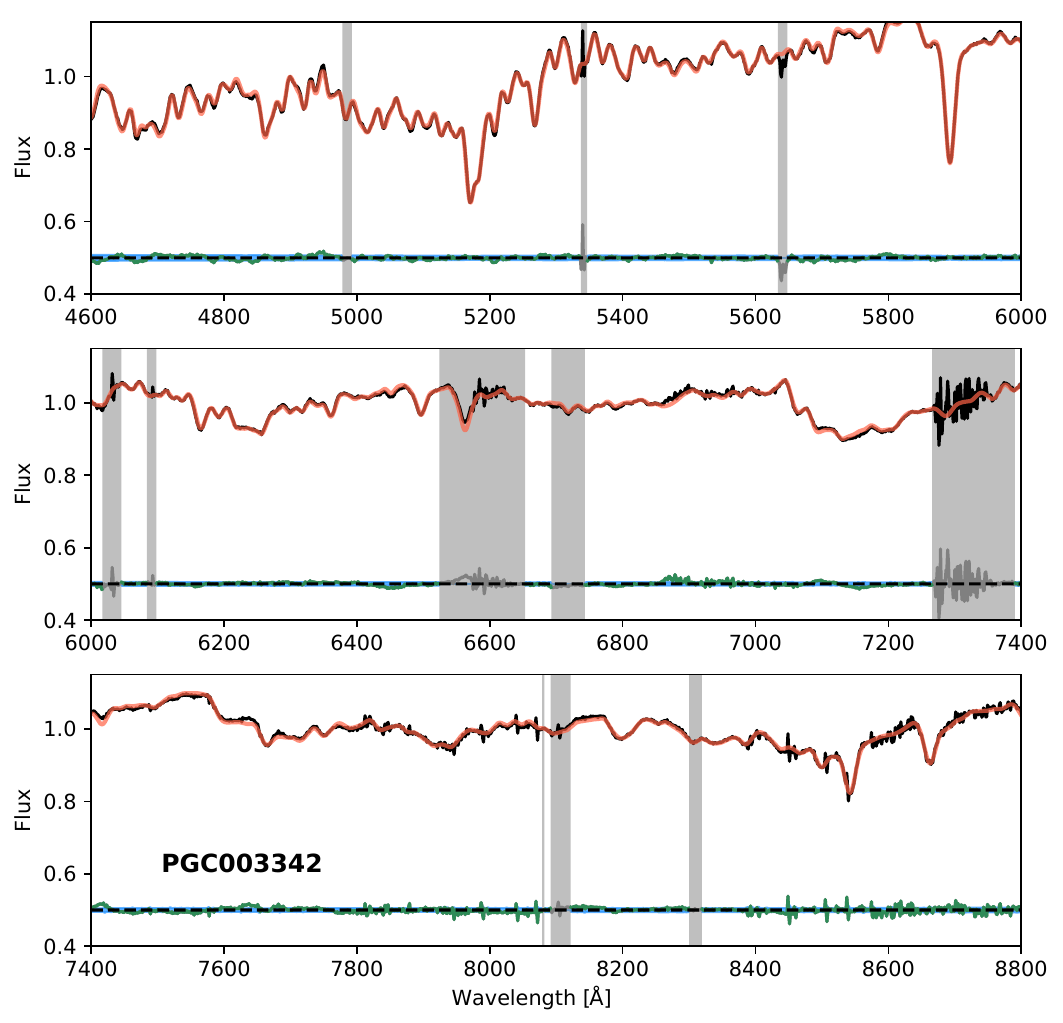}
 \caption{Example of the best-fit \textsc{alf} model to the MUSE data of PGC003342. Grey patches show spectral regions that were masked because of telluric absorption, sky emission or line emission. The data-model residual is shown in green (grey) in unmasked (masked) regions, and shifted to 0.5 the average flux value. The blue region denotes the 1-$\sigma$ uncertainty on the data. Similar spectra for other galaxies are presented in Fig.~\ref{fig:alf_fits}.}
 \label{fig:example_spec}
\end{figure}

For all galaxies we fit the spectral range between $4600-8800$ \AA. Despite the care we take in subtracting the sky lines and correcting the telluric absorption, several spectra still show not fully or over-subtracted sky emission, or over or under-corrected telluric absorption. We therefore mask features that are poorly subtracted (e.g. the [\ion{O}{1}] emission at 6300\AA) and additionally mask any remaining strong features by hand. We also mask H$\alpha$, the surrounding  [\ion{N}{2}] lines and the [\ion{S}{2}] $\sim150$\AA\ redward of  H$\alpha$, all with a width of 800 km/s on each side of the systemic velocity of the galaxy. A few of the M3G galaxies show line emission from ionized gas \citep{PagKraden21}. For these galaxies, we also masked the H$\beta$, NaD, [\ion{O}{1}] 6300\AA\ and [\ion{O}{3}] 4960,5008\AA\ lines, with the same window width.

We show an example fit in Fig. \ref{fig:example_spec}, and for all other galaxies in Fig.~\ref{fig:alf_fits}. Clearly the model spectra reproduces the observed data very well. Some features are however visible: i.e. positive residuals are visible around $\sim$7630\AA\ and $\sim$7680\AA. We see this in almost all fits, and it may be due to the absorption feature around 7665\AA\ being underestimated. The same holds for the feature at $\sim$7416\AA. 

Distribution of elemental abundances as a function of global velocity dispersion $\sigma_e$ (measured within the same central elliptical aperture) can be seen in Fig.~\ref{fig:alf_elem_sig}. The abundance values for each galaxy are presented in Table~\ref{t:alf}.

\begin{figure*}
\includegraphics[angle=0, width=\textwidth]{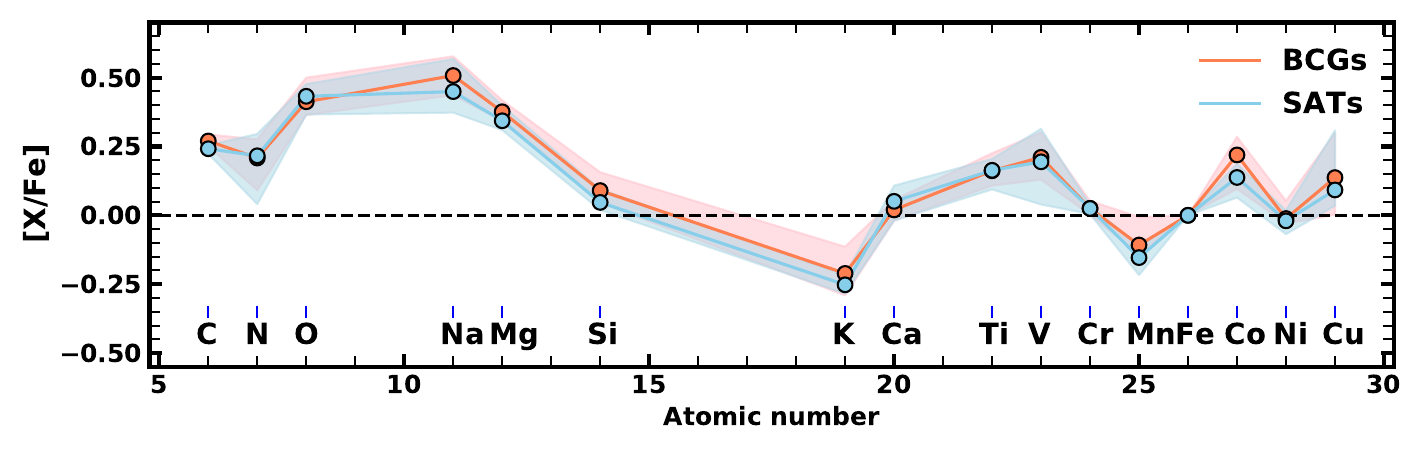}
 \caption{Abundances of elements in M3G sample galaxies, with respect to [Fe/H] within the effective radius for BCGs (orange) and satellite galaxies (light blue), normalised to the solar abundances. Shaded regions show the limits of the distributions estimated as the 16th and 84th percentiles for each element.}
 \label{fig:abundances_vs_sat}
\end{figure*}

\subsection{Chemo-dynamical modelling}
\label{ss:chempy}

As the second step in this work, we use the derived element abundances to constrain the chemical evolution of each galaxy using the code \textsc{Chempy}\footnote{Available from \url{https://github.com/jan-rybizki/Chempy}.} \citep{RybJusRix17}. In our default setup, which is very similar to the one in \citet{RybJusRix17}, stars are formed according to a $\Gamma(k, \tau)$ distribution \citep[eq.(2) in][]{RybJusRix17}, with a shape parameter $k$ and time scale $\tau$. In this set up \textsc{chempy} is a single-zone open box chemical model. Gas with primordial abundances is added to a corona with the same rate as star formation. Corona is defined as a reservoir of a well mixed gas surrounding the zone that we are modelling chemically. It is slowly enriched by the stellar feedback and diluted by the inflow of primordial gas, and, therefore, it can be visualised as a circum-galactic medium surrounding our galaxies. Feedback from stars is distributed over the central star forming zone and the corona according to an outflow feedback fraction parameter $x_{\mathrm{out}}$. The gas mass in the corona at the beginning of the simulation starts out with a mass $f_{\mathrm{cor}}$ times the total mass in star formation. Star formation in the central zone happens with a star formation efficiency $\mathrm{SFE}$, which defines the mass of interstellar medium (ISM) gas required to form $\mathrm{SFE}\times$mass of stars. Supernovae Ia are assumed to have no mass dependence. The delay time distribution has an initial delay of $\tau_{\rm Ia}$ Gyr, for which we assume a Gaussian prior width  $\mu_{\rm \tau_{Ia}} = 0.5$ Gyr and  $\sigma_{\rm \tau_{Ia}} = 0.3$ Gyr. The delay time distribution has a shape $t^{-1.12}$ based on \citet{MaoShaGal10} and a normalisation of $N_{\rm Ia}$ SN Ia explosions per solar mass formed integrated over a time range of 15 Gyr. The model does not allow for accretion beyond the corona and does not account for merging. 

\textsc{Chempy} divides a simulation up in discrete time steps. Since the stellar populations in our galaxies are mostly old, we limit the total simulation time length to 3.5 Gyr. The chosen simulation time length is tuned to reach the peak of the star formation epoch of today massive disks and elliptical galaxies at z $\sim2$ \citep{2014ARA&A..52..415M}. As visible from the top left panel of Fig.~\ref{fig:alf_elem_sig}, this covers the span of stellar ages for most of our galaxies, except for a few youngest (see discussion at the end of Section~\ref{ss:abund}). We also run a chemical model of 7 Gyr duration to test for possible effects of extended star formation (see Appendix~\ref{a:chm}). We use 200 time steps, leading to a resolution of 17.5 Myr. At each time step, stars are formed according to the assumed star formation rate, with a power-law IMF with a slope $\alpha$ and (M$_{\mathrm{min}}$, M$_{\mathrm{max}}$) = (0.08, 100) M$_{\odot}$. At the end of its lifetime, every formed star enriches the ISM according to its mass, metallicity and explosion mechanism. Stellar lifetimes are taken from \citet{ArgSamGer00}. In our default setup, stars above 8 \Msun\ explode as core-collapse supernovae (CC-SNe). This ignores the fact that stars with masses between 8 and 10 \Msun\ are expected to explode as electron capture supernovae (EC-SN). These EC-SN will mainly contribute to iron peak elements. Yields for the CC-SN explosions, as well as for hypernova, are taken from \citet{NomKobTom13}, original presented by \citet{2011MNRAS.414.3231K}.  Above 25 \Msun, a fraction $x_{HN}$  of CC-SNe explodes as hypernovae. This fraction is a free parameter of the model. Stars with masses between 1 and 8\Msun\ evolve as AGB stars. We use the yields of \citet{KarLug16} for AGB enrichment. Yields for SN Ia are from \citet{SeiCiaRop13}.

As we will see in Section~\ref{ss:chem}, it is difficult to reproduce all elemental abundances for our galaxies, regardless of the input parameters for \textsc{chempy} (we have investigated several different set up, but present below only the most relevant). We, therefore, limit ourselves to reproducing (fitting) only selected abundances, [O/H], [Fe/H], [Mg/H] and [Na/H], in our chemo-dynamical run with \textsc{chempy}.

\begin{table*}
   \caption{Median elemental abundances for M3G galaxies.}
   \label{t:elem}
\begin{tabular}{c l  rrr  rrr  rrr}
   \hline
    \noalign{\smallskip}
Z & [X/Fe] & ALL &  &  & BCGs &  &  & SATs &  &  \\
  &             &        & lower & upper & & lower & upper & & lower & upper \\
    \noalign{\smallskip} 
    \hline \hline
    \noalign{\smallskip}
6 & [C/Fe] & 0.254 & 0.23 & 0.29 & 0.27 & 0.25 & 0.29 & 0.242 & 0.22 & 0.25 \\
7 & [N/Fe] & 0.216 & 0.07 & 0.29 & 0.208 & 0.09 & 0.28 & 0.216 & 0.04 & 0.3 \\
8 & [O/Fe] & 0.43 & 0.36 & 0.48 & 0.412 & 0.36 & 0.5 & 0.432 & 0.37 & 0.48 \\
11 & [Na/Fe] & 0.489 & 0.39 & 0.58 & 0.507 & 0.44 & 0.58 & 0.449 & 0.37 & 0.57 \\
12 & [Mg/Fe] & 0.366 & 0.34 & 0.42 & 0.376 & 0.34 & 0.42 & 0.343 & 0.31 & 0.4 \\
14 & [Si/Fe] & 0.078 & 0.03 & 0.13 & 0.09 & 0.05 & 0.16 & 0.047 & 0.02 & 0.09 \\
19 & [K/Fe] & -0.234 & -0.29 & -0.12 & -0.211 & -0.29 & -0.11 & -0.252 & -0.28 & -0.22 \\
20 & [Ca/Fe] & 0.043 & -0.02 & 0.11 & 0.02 & -0.02 & 0.06 & 0.051 & -0.02 & 0.11 \\
22 & [Ti/Fe] & 0.163 & 0.11 & 0.21 & 0.162 & 0.11 & 0.23 & 0.163 & 0.09 & 0.2 \\
23 & [V/Fe] & 0.2 & 0.05 & 0.31 & 0.211 & 0.13 & 0.3 & 0.194 & 0.04 & 0.31 \\
24 & [Cr/Fe] & 0.025 & 0.0 & 0.04 & 0.025 & 0.0 & 0.05 & 0.025 & 0.0 & 0.04 \\
25 & [Mn/Fe] & -0.129 & -0.17 & -0.08 & -0.108 & -0.14 & -0.01 & -0.153 & -0.22 & -0.11 \\
26 & [Fe/Fe] & 0.0 & 0.0 & 0.0 & 0.0 & 0.0 & 0.0 & 0.0 & 0.0 & 0.0 \\
27 & [Co/Fe] & 0.184 & 0.07 & 0.28 & 0.219 & 0.1 & 0.28 & 0.138 & 0.06 & 0.19 \\
28 & [Ni/Fe] & -0.013 & -0.05 & 0.04 & -0.012 & -0.03 & 0.05 & -0.02 & -0.07 & 0.02 \\
29 & [Cu/Fe] & 0.109 & 0.02 & 0.31 & 0.137 & 0.0 & 0.3 & 0.092 & 0.03 & 0.31 \\
38 & [Sr/Fe] & 0.156 & -0.07 & 0.3 & 0.155 & -0.09 & 0.24 & 0.161 & -0.01 & 0.31 \\
56 & [Ba/Fe] & -0.352 & -0.5 & -0.19 & -0.385 & -0.54 & -0.18 & -0.351 & -0.45 & -0.27 \\
63 & [Eu/Fe] & 0.201 & 0.02 & 0.33 & 0.216 & 0.07 & 0.32 & 0.157 & -0.0 & 0.34 \\
     \noalign{\smallskip}
    \hline
\end{tabular}
\\
{Notes: Columns with ``lower" and ``upper" headers show the 16th and 84th percentile values of the element abundance distributions. }

\end{table*}

\section{Results}
\label{sec:results}

\subsection{Stellar abundances}
\label{ss:abund}

We use \textsc{alf} results to estimate the average abundances of elements in M3G sample galaxies. For each elemental abundance (Fig.~\ref{fig:alf_elem_sig}), we median combine the values and estimate the 16th and 84th percentile limits of the distributions. We do this for all galaxies in the sample as well as separately for BCGs and satellites, and present the values in Table~\ref{t:elem}. We show in Fig.~\ref{fig:abundances_vs_sat} the abundances of the BCGs and satellite galaxies with respect to the iron abundance, normalised to the abundance values of the Sun \citep{AspGreSau09}. There are no significant difference between satellite galaxies and BCGs, which perhaps can be understood as a consequence of the small difference in mass between these two sets. The most notable (but still statistically insignificant) difference is observed for [Co/Fe], while for [Si/Fe], [K/Fe] and [Mn/Fe] there are BCGs with large elemental abundance values, which drive the upper limits of the distributions. In some cases, it is the satellite galaxies that show large spreads of elemental abundances, such as for [N/Fe], [Na/Fe] and [V/Fe]. Nevertheless, the lack of significant differences points to a broadly similar formation scenarios for stars in BCGs and massive satellites at these mass scales, but we will examine in more detail the differences for individual cases below. 

\begin{figure}
  \includegraphics[angle=0, width=0.45\columnwidth]{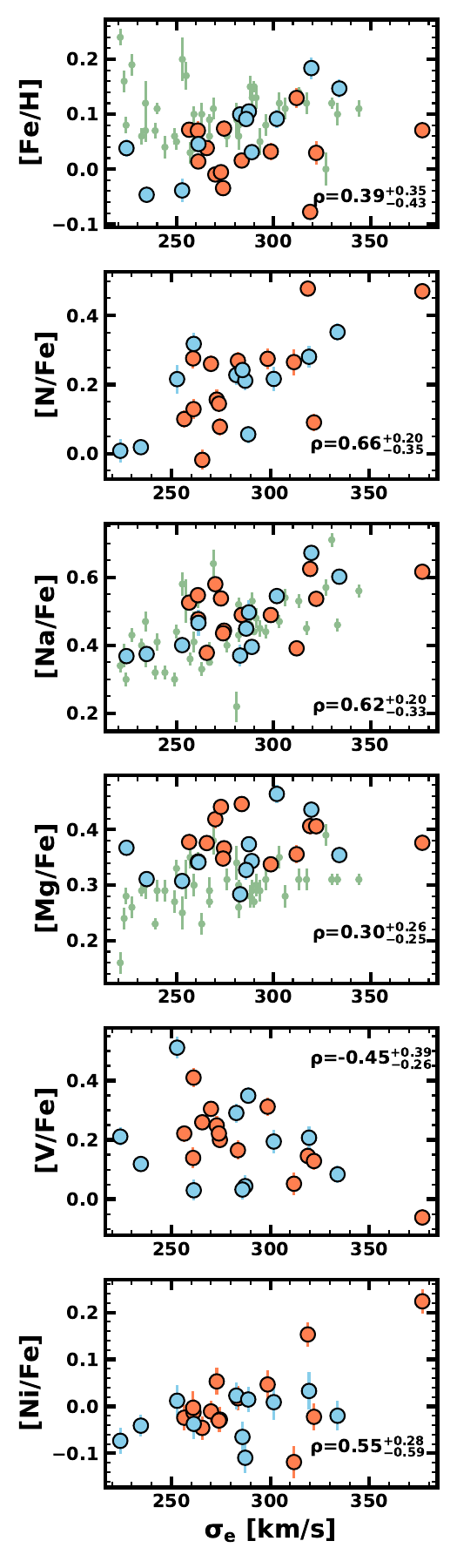}
  \includegraphics[angle=0, width=0.45\columnwidth]{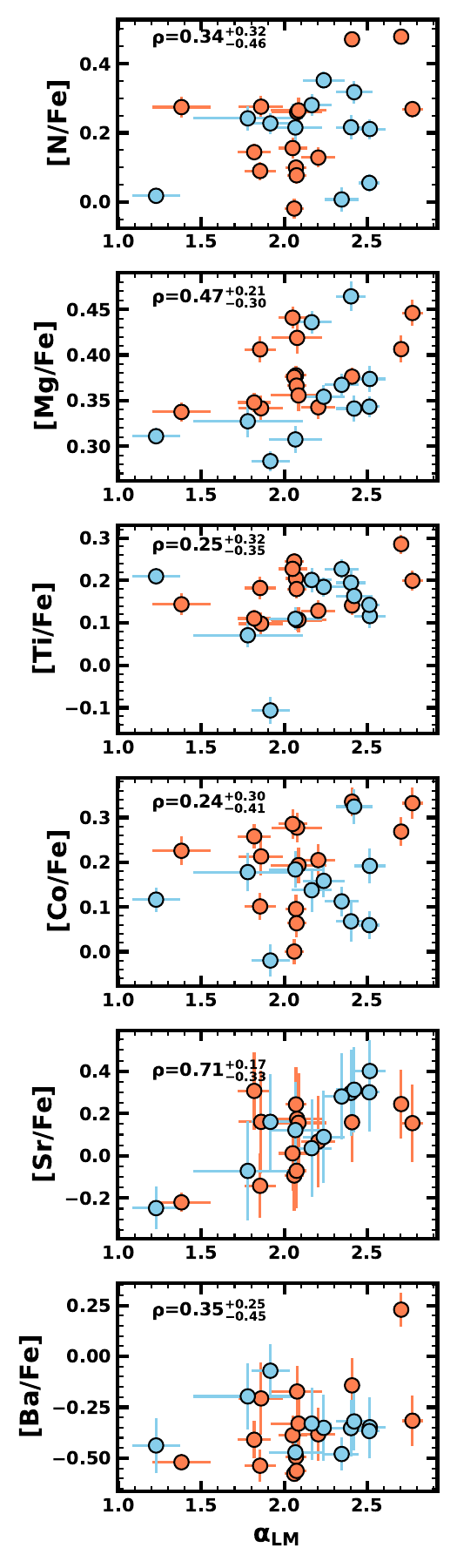}
 \caption{Selected elemental abundances as a function of global velocity dispersion (left) and low-mass end IMF slope (right). BCGs are denoted with orange points, satellite galaxies with light blue points. We plot here only elements for which a significant correlations, either for BCGs or satellite galaxies. The Pearson correlation coefficient $\rho$ (for the full sample) is indicated on each panel, with uncertainties indicating the $2\sigma$ level. The correlations for all other elements are in Figs.~\ref{fig:alf_elem_sig} and~\ref{fig:alf_xfe_imf}. For comparison we include data from the MASSIVE sample \citep{2022ApJ...932..103G} with small green symbols. }
 \label{fig:X_corr}
\end{figure}

A more detailed look at Fig. \ref{fig:abundances_vs_sat} reveals high average values for some elemental abundances, especially [O/Fe] and [Na/Fe] (for more information about the trends for individual galaxies see also Figs.~\ref{fig:X_corr} and \ref{fig:alf_elem_sig}). In general, the $\alpha$-elements have relatively high values, as expected, as the contribution of CC-SNe are much more pronounced for galaxies with short formation time scales. We note that the $\alpha$-element abundances differ from the ones measured by \citet{ConGravan14}, for their galaxies with largest velocity dispersions (their bins with $\sigma = 250$ and $300$km/s). Specifically, [Mg/Fe] is higher by up to $\sim0.1$ dex, while [Si/Fe] is significantly lower than in \citet{ConGravan14} ($\sim0.1$ dex), but in the latter case there are galaxies in the M3G sample which reach values observed in the SDSS sample of galaxies by \citet{ConGravan14}. \citet{2023arXiv230303412B} updated the \citet{ConGravan14} abundances for SDSS galaxies using the latest version of \textsc{alf} (we used the same version, but in a somewhat different set up, especially with respect to IMF parametrisation). The differences between our and their abundances are smaller, but with the same trend as for the \citet{ConGravan14} values. Given the higher value of [Mg/Fe], the values of [Si/Fe] between 0--0.16 found in this work seem somewhat low. It is unclear if this is related to observational issues, for example, because of the limited wavelength range of the data (i.e., the strongest response of Si is observed between 4100-4300\AA) for this element, or if the Si abundance is indeed low. The Ca abundance is supposed to be close in value to the iron abundance, and this is indeed the case for our data. 

We observe two peaks around the iron peak elements. The Co peak was also observed by \citet{ConGravan14}, who also noted a large spread in values as a function of velocity dispersion, with the highest $\sigma$ bin galaxies having larger values than what we observe. As mentioned above, in our case this element shows the largest difference between the BCG and satellite galaxies. Contrary to \citet{ConGravan14}, we also observe a peak in the vanadium abundance, with the values much larger than those reported by \citet{ConGravan14}, specifically for galaxies with high velocity dispersions. This discrepancy seems to decrease with an updated values for [V/Fe] by \citet{2023arXiv230303412B}, which are now positive for the most massive galaxies in the SDSS sample, and are within the distribution of values of our (satellite) galaxies. Another difference are the lower values for [Mn/Fe] compared to the \citet{ConGravan14}, which in our case are mostly negative, except in a few BCGs, while the average of the \citet{ConGravan14} galaxies are positive at all velocity dispersions. 

Abundance measurements with respect to iron as a function of the global velocity dispersion of each galaxy, again normalised by the solar values, can be used as a proxy to investigate trends with galaxy mass. We show them for M3G sample galaxies in Fig.\ref{fig:X_corr} (left column), but only for elements selected because they show correlations between the parameters, while all elements are shown in Fig.~\ref{fig:alf_elem_sig}. We estimate the Pearson correlation coefficient between the velocity dispersion, for which we ignore the small observational uncertainties, and the elemental abundance (taking into account the uncertainties returned by \textsc{alf}). Furthermore, we estimate a 95\% confidence interval on the correlation coefficient by repeatedly drawing [X/Fe] values from the \textsc{alf} MCMC chains and bootstrapping different galaxies. The correlation coefficient uncertainty is determined by taking the lower 2.5\% and upper 97.5\% limit values of all bootstraps, therefore providing a $2\sigma$ level of significance.  

Most elements show little evolution with velocity dispersion. Exceptions to this are [Fe/H], [N/Fe], [Na/Fe] and [Ni/Fe], which we find to correlate, and [V/Fe] which instead anti-correlates with velocity dispersion. For reasons that we do not understand, some of the elements slightly less massive than Fe (Ti, V, Cr, Mn) show negative trends with velocity dispersion. The same was reported for V and Cr by \citet{ConGravan14}. On the other hand, somewhat more massive elements (Co, Ni, Cu) show instead positive trends with velocity dispersion. These trends are however dominated by large spread and strong outliers. We find that the correlations between the abundance and the velocity dispersion are the most significant for Na and N in our data, but we note that we have a small range of the velocity dispersion. Nevertheless, the strong correlation trend in [Na/Fe] is also seen in the MASSIVE sample, but MASSIVE galaxies seem to have somewhat smaller [O/Fe] and [Mg/Fe] element abundances \citep{GuGreNew21}. A more detailed comparison is presented in Appendix~\ref{a:el}.

In some cases, there is a notable difference between the BCGs and satellites in terms of how strongly their elemental abundances correlate with the velocity dispersion. The largest differences are seen for [Fe/H], [Na/Fe], [Mg/Fe] and [V/Fe]. In the case of [Fe/H] abundances, BCGs do not show any evidence for a correlation ($\rho_{\rm BCG}=0.09_{-0.48}^{+0.35}$), but the satellite galaxies have a high Pearson correlation coefficient: $\rho_{\rm SAT}=0.82_{-0.11}^{+0.10}$. The situation is similar for the [Mg/Fe] abundances, where the observed weak correlation between [Mg/Fe] and $\sigma_e$ for all galaxies is due to the dilution of a relatively significant correlation for the satellite galaxies ($\rho=0.44_{-0.24}^{+0.24}$) with non-correlating BCGs ($\rho=0.08_{-0.24}^{+0.28}$). For the [Na/Fe] abundance the situation is somewhat different as both BCGs and satellite galaxies show a correlation, but the correlation is not statistically significant for BCGs ($\rho=0.41_{-0.49}^{+0.26}$), but it is for satellites ($\rho=-0.81_{-0.13}^{+0.10}$). The situation is reversed for [V/Fe], where the satellites do not show a correlation ($\rho=-0.18_{-0.26}^{+0.33}$), while BCGs have a significant one ($\rho=-0.75_{-0.14}^{+0.35}$). These results point to a tentative difference in the chemical evolution of BCGs and satellite galaxies in our sample, as Na and Mg are $\alpha$-elements produced in CC-SN explosions, while V is produced both in CC-SN and SN Ia explosions. We summarise most significant correlations in Table~\ref{t:corr}.

\begin{figure}
  \includegraphics[angle=0, width=\columnwidth]{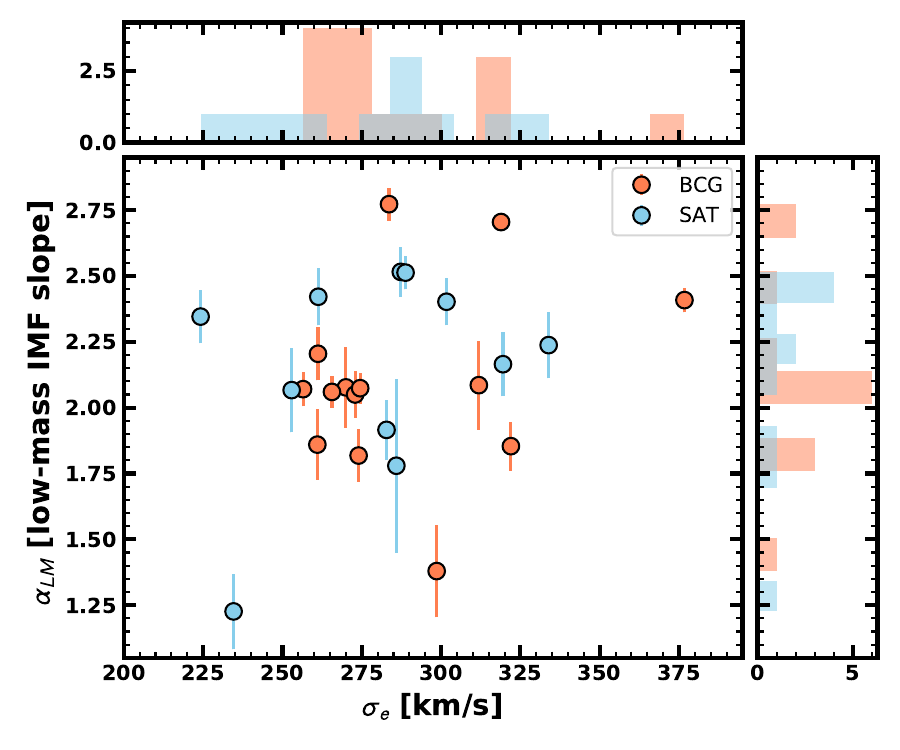}
 \caption{Low-mass end slope of the IMF, $\alpha_{\rm LM}$, versus velocity dispersion $\sigma_e$ within the effective radius aperture. BCGs are denoted with orange points, satellite galaxies with light blue points.}
 \label{fig:lowmass_alpha_sig}
\end{figure}

In our nominal \textsc{alf} extraction we left the IMF slope as a free parameter, and in Fig. \ref{fig:lowmass_alpha_sig} we show the derived IMF slopes for our galaxies. As the currently alive stars in our massive galaxies are all low mass stars, we will refer to this slope as the {\it low-mass IMF slope}, $\alpha_{\rm LM}$. Numerous studies, both based on the spectral analysis and dynamical arguments \citep[e.g.][]{Convan12a, CapMcDAla12, LaBFerVaz13, 2013MNRAS.429L..15F, 2014MNRAS.438.1483S}, showed that the low-mass IMF slope correlates with velocity dispersion. Our velocity dispersion range is somewhat small, and the data in Fig. \ref{fig:lowmass_alpha_sig} do not show a clear correlation. The Pearson correlation coefficient for the full sample is $\rho=0.27_{-0.37}^{+0.30}$, and $\rho=0.24_{-0.36}^{+0.28}$ and $\rho=0.32_{-0.40}^{+0.29}$ for BGCs and satellites, respectively. While the correlation seem to be somewhat stronger than that reported for the MASSIVE sample \citep{2022ApJ...932..103G}\footnote{Note that in that study the correlation was between the $\sigma$ and the IMF mismatch parameter $\alpha \equiv \frac{(M/L)}{(M/L)_{\rm Kroupa}}$.}, it is similarly not statistically significant, and the lack of correlation can be traced to the short $\sigma_e$ baseline. Furthermore, 8/25 (32\%) galaxies have $\alpha_{\rm LM}$ larger than the Salpeter value, while 7/25 (28\%) galaxies have $\alpha_{\rm LM}$ smaller than 2, significantly deviating from the Salpeter slope. This is in agreement with previous findings, e.g. where approximately one-third of the galaxies with $\sigma>200$ km/s in \citet{Convan12a} show evidence for an IMF with a steeper $\alpha_{\rm LM}$ (i.e. more bottom heavy) than Salpeter, while only about 1/4 of galaxies in the MASSIVE sample ($\sigma>200$ km/s) have the low-mass IMF slope below the Salpeter \citep{2022ApJ...932..103G}. We also note that the large spread of $\alpha_{\rm LM}$ values at the mean $\sigma_e$ for our sample ( $\sim290$ km/s) is also consistent with results for lensing studies \citep[e.g.][]{NewSmiCon17,ColSmiLuc18, ColSmiLuc18a}, which typically find lower (low-mass end) IMF slopes. 

There is no difference in the $\alpha_{\rm LM}$ between BCGs and satellite galaxies, with average values of $2.1\pm0.3$ and $2.1\pm0.4$, respectively, even though 5 satellite galaxies and 3 BGCs have larger than Salpeter IMF slope. The low-mass IMF slope derived by \textsc{alf}, as well as other IMF slopes used in this work are presented in Table~\ref{t:imf}.  

We show the individual correlations between elements and $\alpha_{LM}$ for selected elements in Fig.~\ref{fig:X_corr} (right column) and for all elements in Fig.~\ref{fig:alf_xfe_imf}.  As expected \citep{Convan12}, we see a correlation between [Mg/Fe] and $\alpha_{\rm LM}$, and we also find statistically significant correlations for [Sr/Fe] with a Pearson correlation coefficient of $\rho=0.71_{-0.33}^{+0.17}$. For these elements the correlations exist for the full sample and when separating into BCGs and satellite galaxies. In a few cases, the BCG galaxies (and not the satellites) have a noticeable trend between the element abundances and the low-mass end IMF slope ([N/Fe], [Ti/Fe], [Co/Fe], and [Ba/Fe]), but they are not statistically significant (i.e. in case of [Ti/Fe] $\rho = 0.45_{-0.53}^{+0.14}$, or [Ba/Fe] $\rho=0.60_{-0.70}^{+0.01}$). The most significant correlations are summarised in Table~\ref{t:corr}. 

The population modelling with \textsc{alf} provided also the luminosity-weighted stellar ages for our galaxies. Most populations are very old, reaching the limits of the models and the age of the Universe \citep[for a discussion why this is possible, see][]{McDAlaBli15}, but there are six galaxies with stellar ages lower than 10 Gyr. A particular case is PGC073000, the youngest galaxy in the sample with an age of $5.9\pm0.3$ Gyr. Stellar populations of this ages are just about old enough for the derivation of the low-mass end IMF slope based on spectral fitting \citep{Convan12a}. Nevertheless, this galaxy also has the highest value of $\alpha_{\rm LM}$ (2.77), while its overall metallicity is close to solar. Elemental abundances, in particular for elements created in CC SN, such as [Na/Fe] or [Mg/Fe], are moderate to high. The relatively low overall metallicity, low stellar age and the existence of emission-line gas in the nucleus, mass of which is the highest for M3G sample \citep{PagKraden21}, suggest this galaxy experienced a rejuvenation event, and a significant secondary star formation. The high $\alpha_{\rm LM}$ is likely not related to the actual measured elemental abundance, except to possibly Na values, as we show in Section~\ref{ss:unc_imf_low}. Importantly, this galaxy is the only one in the sample with a clear evidence for multiple star-formation phases, and, in this respect, the simple set up of our \textsc{chempy} models is likely not applicable for this galaxy. This is not necessarily because the duration of the chemical evolution model is set to 3.5 Gyr (as we show in Appendix~\ref{a:chm}), but because our model does not account for accretion beyond the corona, or any kind of mergers. While PGC073000 warrants a more detailed stellar population and chemo-dynamical modelling, we nevertheless keep the same \textsc{chempy} set up for all galaxies, and highlight PGC073000 in the text when necessary. 

\begin{table*}
   \caption{Low- and high-mass end IMF slopes for M3G galaxies.}
   \label{t:imf}
\begin{tabular}{lccccccc}
   \hline
    \noalign{\smallskip}
galaxy & $\sigma_e$ [km/s] &$\alpha_{\rm LM}$ & $\alpha_{\rm HM}$ & $\alpha_{\rm LM}$ (Na 8190) & $\alpha_{\rm LM}$ (Na full)& $\alpha_{\rm CL04}$ & BCG\\
  (1)   &       (2)                     &      (3)                    &       (4)                   &          (5)                &          (6)     & (7)   & (8) \\
    \noalign{\smallskip} 
    \hline \hline
    \noalign{\smallskip}
PGC003342 & 270 & 2.08 $\pm$  0.15 & 2.07 $\pm$  0.04 & 1.76 $\pm$  0.09 & 1.83 $\pm$  0.09 & 2.00 & 1 \\
PGC004500 & 257 & 2.07 $\pm$  0.06 & 2.07 $\pm$  0.02 & 2.08 $\pm$  0.07 & 2.07 $\pm$  0.06 & 1.98 & 1 \\
PGC007748 & 266 & 2.06 $\pm$  0.06 & 2.12 $\pm$  0.02 & 2.30 $\pm$  0.05 & 2.24 $\pm$  0.05 & 2.03 & 1 \\
PGC015524 & 273 & 2.05 $\pm$  0.09 & 2.09 $\pm$  0.03 & 2.00 $\pm$  0.10 & 1.92 $\pm$  0.11 & 1.99  & 1\\
PGC018236 & 275 & 2.07 $\pm$  0.06 & 2.10 $\pm$  0.02 & 2.27 $\pm$  0.09 & 2.31 $\pm$  0.08 & 2.02  & 1\\
PGC019085 & 261 & 2.20 $\pm$  0.10 & 2.11 $\pm$  0.04 & 1.91 $\pm$  0.17 & 1.77 $\pm$  0.11 & 2.04  & 1\\
PGC043900 & 377 & 2.41 $\pm$  0.05 & 2.04 $\pm$  0.03 & 2.33 $\pm$  0.05 & 2.42 $\pm$  0.06 & 1.95  & 1\\
PGC046785 & 334 & 2.24 $\pm$  0.12 & 2.01 $\pm$  0.04 & 1.93 $\pm$  0.12 & 1.94 $\pm$  0.11 & 2.03  & 0\\
PGC046832 & 312 & 2.09 $\pm$  0.17 & 2.08 $\pm$  0.03 & 2.33 $\pm$  0.16 & 2.52 $\pm$  0.14 & 2.12  & 1\\
PGC046860 & 283 & 1.92 $\pm$  0.12 & 2.08 $\pm$  0.03 & 1.96 $\pm$  0.11 & 2.01 $\pm$  0.10 & 2.10  & 0\\
PGC047154 & 320 & 2.17 $\pm$  0.12 & 1.97 $\pm$  0.04 & 2.07 $\pm$  0.08 & 2.03 $\pm$  0.09 & 2.05  & 0\\
PGC047177 & 287 & 2.52 $\pm$  0.10 & 2.06 $\pm$  0.04 & 2.58 $\pm$  0.08 & 2.56 $\pm$  0.10 & 1.98  & 0\\
PGC047197 & 302 & 2.40 $\pm$  0.09 & 2.03 $\pm$  0.04 & 2.41 $\pm$  0.09 & 2.43 $\pm$  0.10 & 2.01  & 0\\
PGC047202 & 319 & 2.70 $\pm$  0.02 & 2.11 $\pm$  0.05 & 2.58 $\pm$  0.06 & 2.60 $\pm$  0.05 & 2.00  & 1\\
PGC047273 & 261 & 2.42 $\pm$  0.11 & 2.08 $\pm$  0.05 & 2.06 $\pm$  0.26 & 2.34 $\pm$  0.18 & 2.03  & 0\\
PGC047355 & 253 & 2.07 $\pm$  0.16 & 2.15 $\pm$  0.05 & 1.79 $\pm$  0.13 & 1.75 $\pm$  0.14 & 2.07  & 0\\
PGC047590 & 289 & 2.51 $\pm$  0.06 & 2.13 $\pm$  0.02 & 2.59 $\pm$  0.05 & 2.62 $\pm$  0.04 & 2.02  & 0\\
PGC047752 & 261 & 1.86 $\pm$  0.13 & 2.04 $\pm$  0.04 & 1.72 $\pm$  0.13 & 1.82 $\pm$  0.13 & 1.98  & 1\\
PGC048896 & 322 & 1.85 $\pm$  0.09 & 2.09 $\pm$  0.03 & 2.09 $\pm$  0.07 & 2.08 $\pm$  0.06 & 1.87  & 1\\
PGC049940 & 299 & 1.38 $\pm$  0.17 & 2.10 $\pm$  0.03 & 1.56 $\pm$  0.15 & 1.52 $\pm$  0.14 & 1.96  & 1\\
PGC065588 & 274 & 1.82 $\pm$  0.10 & 2.17 $\pm$  0.03 & 1.63 $\pm$  0.08 & 1.63 $\pm$  0.09 & 1.93  & 1\\
PGC073000 & 284 & 2.77 $\pm$  0.06 & 2.10 $\pm$  0.03 & 2.40 $\pm$  0.08 & 2.20 $\pm$  0.09 & 2.18  & 1\\
PGC097958 & 286 & 1.78 $\pm$  0.33 & 2.07 $\pm$  0.04 & 1.59 $\pm$  0.14 & 1.59 $\pm$  0.18 & 2.00  & 0\\
PGC099188 & 224 & 2.35 $\pm$  0.10 & 2.12 $\pm$  0.02 & 2.38 $\pm$  0.15 & 2.38 $\pm$  0.14 & 2.05  & 0\\
PGC099522 & 235 & 1.23 $\pm$  0.14 & 2.16 $\pm$  0.02 & 1.58 $\pm$  0.13 & 1.58 $\pm$  0.12 & 2.10  & 0\\
     \noalign{\smallskip}
    \hline
\end{tabular}
\\
{Notes: Column (1): name of the galaxy; Column (2): global velocity dispersion estimated within an elliptical aperture of a semi-majr axis radius equal to the half-light radius of the galaxy. Values in the subsequent columns are also estimated within the same aperture. Column (3): low-mass end IMF slope from the nominal \textsc{alf} run; Column (4): high-mass end IMF slope from the nominal \textsc{chempy} run; Column (5):  low-mass end IMF slope when Na 8190 line is masked; Column (6):  low-mass end IMF slope when NA 8190 and NaD lines are masked; Column (7): high-mass end IMF slope using \citet{ChiLim04} yields. Note that for $\alpha_{\rm CL04}$ we did not run MCMC chains to calculate the uncertainties for individual galaxies, but the errors are expected to be similar to $\alpha_{\rm HM}$. Column (8): BGCs are marked with 1. }
\end{table*}

\subsection{Chemical evolution models}
\label{ss:chem}

In Fig.~\ref{fig:chempy_vs_sig} we show the results of the parameters from the chemical evolution model using the \citet{NomKobTom13} yields for core-collapse SNe. The values are listed in Table~\ref{t:chempy}. As expected, all galaxies show short star formation time scales, with profile shapes that are close to exponential, and decreasing with time. The models prefer no or little outflows to the corona. For most galaxies we find a fraction of CC-SN over all other supernovae (i.e. hypernovae plus other supernovae) close to 1, but we note that fixing this parameter to either 1.0 or 0.5 makes little difference for the results. The star formation efficiency is found to be close to one and similar for all galaxies. Most of other parameters are poorly constrained with large uncertainties, and there is little difference between BCGs and satellite galaxies. 

In the same Appendix~\ref{a:chm} and Fig.~\ref{fig:chempy_vs_sig}, we show the results of an extended duration \textsc{chempy} run, where the model set-up was kept the same as for the nominal run, except that the simulation time was extended to 7 Gyr. All model parameters were kept free, as before. There are no significant differences between the models, and we continue using the results of the 3.5 Gyr run \textsc{chempy} in the rest of this work. Furthermore we tested chemical models with respect to different outflow fraction, but there were no significant changes to the main results. We refer the discussion of the similarities between the chemical models to the appendix.

\begin{figure}
  \includegraphics[angle=0, width=\columnwidth]{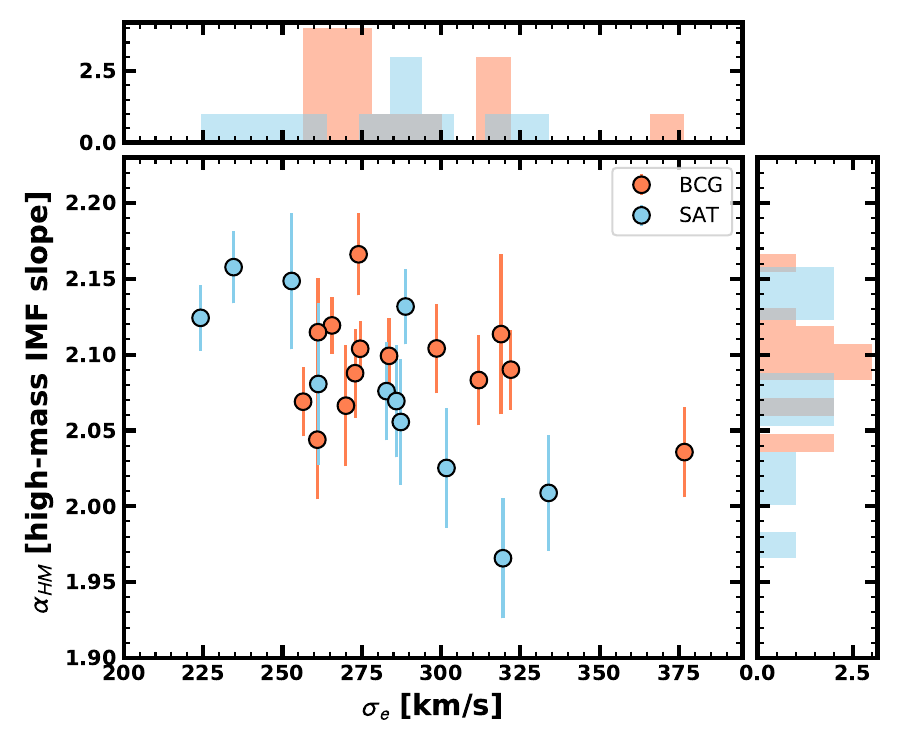}
 \caption{High-mass end slope of the IMF, $\alpha_{\rm HM}$, obtained from our chemical evolution model versus the velocity dispersion within the effective radius aperture. BCGs are denoted with orange points, satellite galaxies with light blue points. Note a tentative anti-correlation of the IMF slope with velocity dispersion for the full sample and a more significant one for the satellite galaxies.}
 \label{fig:highmass_alpha_sig}
\end{figure}

One parameter deserves some attention: the slope of the {\it high-mass IMF} ($\alpha_{\rm HM}$), which shows an anti-correlation with velocity dispersion. We show it in more details in Fig.~\ref{fig:highmass_alpha_sig}. While the Pearson correlation coefficient for the full sample is $\rho=-0.55_{-0.19}^{+0.31}$ and is statistically significant, BCGs do not show a significant correlation ($\rho=-0.30_{-0.39}^{+0.58}$). It is the satellite galaxies that drive the anti-correlation of the high-mass IMF slope with velocity dispersion for the full sample ($\rho=-0.84_{-0.09}^{+0.12}$). In the \textsc{chempy} model with an extended duration, as well with different levels of outflow fractions, the resulting $\alpha_{\rm HM}$ are consistent with the one presented here, conserving the same anti-correlations (see Appendix~\ref{a:chm}).

The origin of this anti-correlation could be related to the noticeable correlations of the [Na/Fe] and [Fe/H] with velocity dispersion (Fig.~\ref{fig:alf_elem_sig}). Specifically, as mentioned above, the satellite galaxies in our sample show a clear rise in the metallicity ([Fe/H]) with $\sigma_e$, but also show a stronger correlation for [Na/Fe] with $\sigma_e$ compared to BCGs. Furthermore, \citet{NomKobTom13} show that [Na/Fe] abundance, weighted by the initial mass function and their yields, increases with the metallicity much more than any other element (their fig. 10). Therefore, the chemical evolution models are trying to compensate for the increased metallicity and strong correlations in [Na/Fe] of the satellite galaxies, while trying to fit a slight increase in [Mg/Fe] (for both populations). 

We note that the $\chi^2$ values of the \textsc{Chempy} fits are well above one. One reason for this is the small value of the uncertainties on the abundances, which are of order $\sim0.01$ (Fig.~\ref{fig:alf_elem_sig}), but also show the general inability of the chemical model to reproduce the ``observed" element abundance obtained with \textsc{alf}. In the Section~\ref{ss:unce} we will comment more on these issues. 

\begin{figure}
  \includegraphics[angle=0, width=\columnwidth]{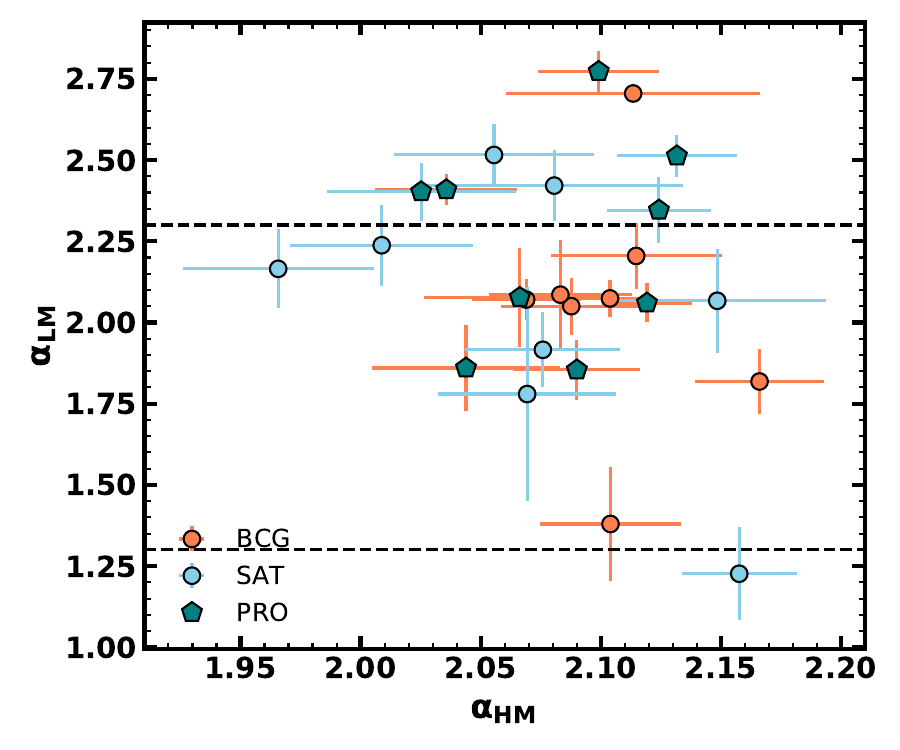}
 \caption{High-mass end slope of the IMF derived using \textsc{Chempy} with the yields of \citet{NomKobTom13} versus the low-mass end slope derived using \textsc{alf}. BCGs and satellite galaxies are shown with orange and light blue symbols, respectively. Galaxies with prolate-like rotation (with kinematic misalignments larger than 75\degr) are shown as green pentagons. The colour of the error bars of the pentagons relate them to BCGs or satellites. Horizontal dashed lines indicate the Salpeter (higher) and Kroupa (lower) values for the low-mass end slopes, for comparison purposes.
 }
 \label{fig:imf_vs_imf}
\end{figure}

\begin{figure*}
   \includegraphics[angle=0, width=\textwidth]{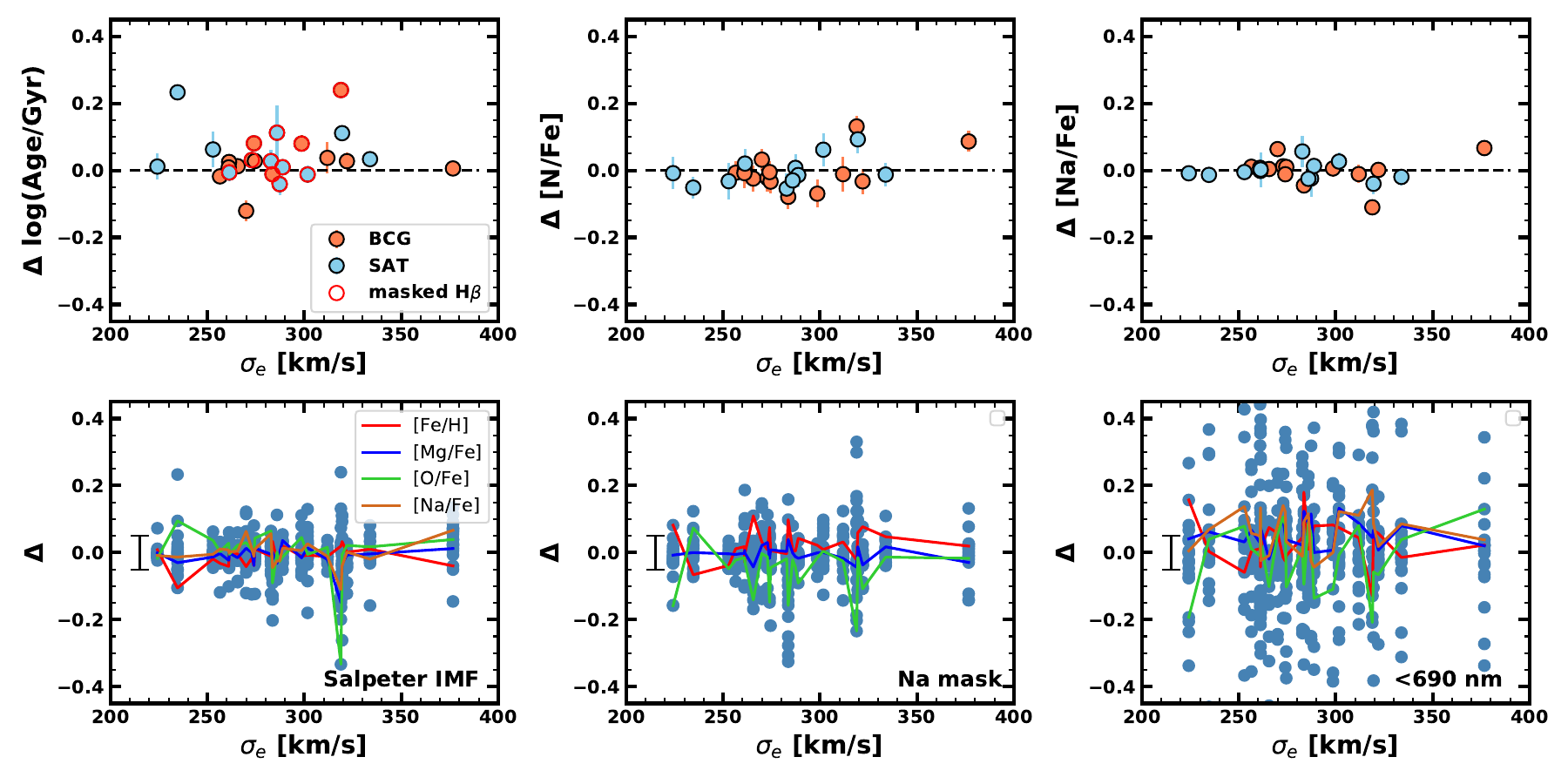}
 \caption{Top: The difference between the selected \textsc{alf} results allowing for the low-mass end IMF slope ($\alpha_{\rm LM}$) to be free or fixed to the Salpeter value ($\alpha=2.3$), respectively. From left to right: the difference in the log age, [N/Fe] and [Na/Fe] abundances (IMF free - IMF fixed). Orange symbols are for BCGs and light blue for satellite galaxies. Symbols with red edge on the left-most panel are galaxies with emission-line gas where we masked the H$\beta$ line in \textsc{alf} runs, as indicated on the legend. Note the small scatter in elemental abundance and relatively large scatter in age, where the age of the models with free IMF slope is systematically larger. Bottom: for each galaxy (indicated by their $\sigma_e$ values), we plot all free \textsc{alf} parameters, as the differences between the nominal fit (outlined in Section~\ref{ss:cent_data}) values and three \textsc{alf} setups, which are, from left to right: IMF slope fixed to the Salpeter values (as in the top three panels), the full set of Na lines masked, and limiting the extraction to the ``blue" region ($\lambda<6900$\AA). Red, blue, green and brown lines highlight the differences in values of [Fe/H], [Mg/Fe], [O/Fe] and [Na/Fe], respectively. The single vertical bar in each panel highlights an error of $\pm0.05$ dex. }
 \label{fig:age_vs_age}
\end{figure*}

An inference of the slope of the high-mass IMF for individual galaxies obtained with chemical modelling allows us a simultaneous comparison of the low and high-mass end slopes, shown in Fig.~\ref{fig:imf_vs_imf}. We note that the high-mass end of the IMF is always flatter than Salpeter regardless of the chemical evolution model set up; per definition the high-mass end is therefore always top heavy. BCGs and satellite galaxies seem to have somewhat different distributions of the low and high-mass IMF slopes. This can be seen by calculating the Pearson correlation coefficient for the two samples. For BCGs, $\rho=-0.09_{-0.29}^{+0.36}$ at the 95\% confidence level clearly shows no correlation between the IMF slopes, while for the satellite galaxies $\rho=-0.32_{-0.28}^{+0.46}$ suggest a statistically significant correlation, but not a particularly strong one. Combining the two subsamples results in a low and statistically insignificant Pearson correlation coefficient ($\rho=-0.23_{-0.32}^{+0.41}$ ) as one could judge from the figure by eye. Therefore, for the satellite galaxies there is a weak trend within their central effective radius that, in general, if a galaxy has a more bottom heavy IMF, it also {\it had} a more top heavy IMF; a galaxy that has a bottom light IMF, also {\it had} a top light IMF. Such a correlation does not seem to exist for BCGs, which for a relatively uniform high-mass IMF slope (centred on $\overline{\alpha_{\rm HM}}=2.1$) exhibit a range of low-mass IMF slopes, from Kroupa- to super-Salpeter-like ($1.4<\alpha_{\rm LM}<2.7$).

\citet{KraEmsden18} showed that almost half of M3G galaxies have rotation {\it around} their major axis, what is sometimes described as prolate-like rotation and can be quantified with kinematic misalignment angle \citep{1991ApJ...383..112F}, the angle measured between the major axis and the orientation of the velocity field\footnote{Alternatively, the angle between the minor axis and the projection of the angular mometnum vector.}. Prolate-like rotation in M3G galaxies was defined when the kinematic misalignment angle is greater than $75\degr$. Such rotation is mostly present in M3G BCGs, but some satellites also have it. In order to test if the kinematics will bring more information, we show in Fig.~\ref{fig:imf_vs_imf} galaxies with prolate-like rotation with pentagons. Galaxies with prolate-like rotation are distributed differently from both the BCGs and satellites, suggesting a positive correlation (top-heavy to bottom-heavy IMFs). However, the correlation is not significant ($\rho=0.17_{-0.35}^{+0.29}$). On the other hand, the Pearson correlation coefficient for galaxies {\it without} prolate-rotation is $\rho=-0.36_{-0.26}^{+0.36}$, somewhat stronger and more significant compared to that of the satellites, but the differences are small. 

\citet{Convan12} were the first to find a correlation between [Mg/Fe] and IMF slope, and they argued that the low-mass IMF slope could be a result of the short formation time scale of galaxies with high [Mg/Fe]. Such as correlation is also found in our sample (Fig.~\ref{fig:alf_xfe_imf}).  The \textsc{Chempy} models on M3G galaxies, however, suggest the star formation time scales are similar for galaxies with different velocity dispersions (Fig.~\ref{fig:chempy_vs_sig}, panel a), while the chemical evolution model compensates the differences in the elemental abundances by pairing different low- and high-mass end IMF slopes. The general abundance values, however, drive the general trend of pairing the bottom heavy (low-mass) with the top heavy (high-mass) IMFs.

\begin{figure}
  \includegraphics[angle=0, width=\columnwidth]{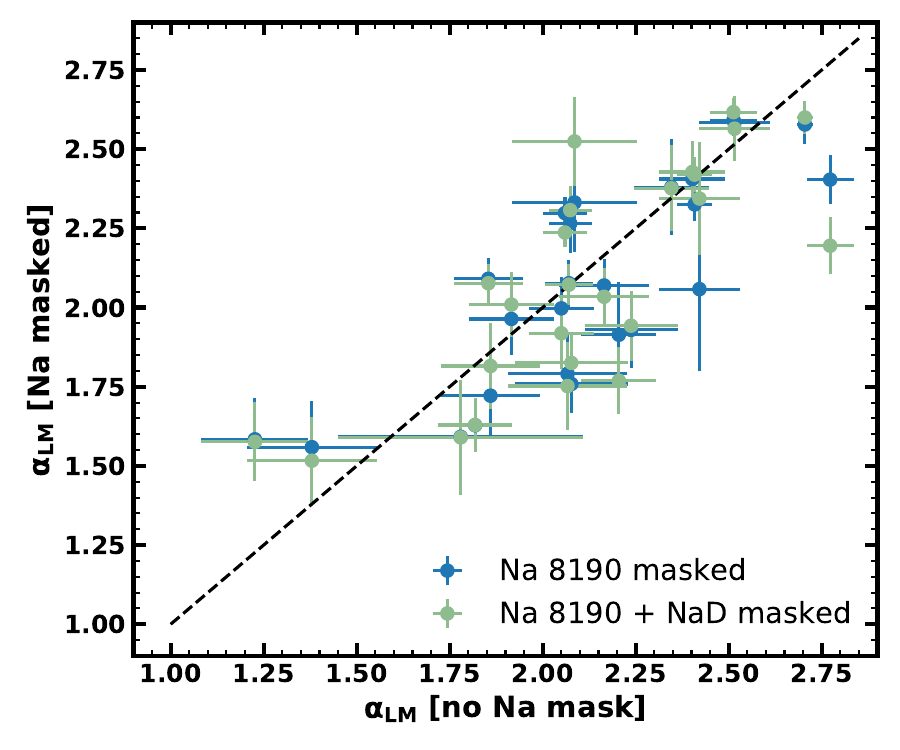}
 \caption{The relation between the low-mass end IMF slopes derived using {\tt alf} by masking Na lines ($y$-axis) and not masking ($x$-axis). We used two different masks, only applying it to the Na 8190\AA\,line, or both on 8190\AA\,line and NaD doublet. Only small differences are seen between two masks and there is an overall good agreement with the unmasked values.}
\label{fig:imf_no_Na}
\end{figure}

\section{Discussion}
\label{sec:discussion}

\subsection{Uncertainties in the values of the abundances and the low-mass end IMF slope}
\label{ss:unc_imf_low}
We found that both the average abundances of N and Na (Fig.~\ref{fig:X_corr}), as well as, to a lesser extent, the low-mass end  slope of the IMF (Fig.~\ref{fig:lowmass_alpha_sig}), change with increasing velocity dispersion. The possibilities exist therefore that an actual change in the IMF slope with velocity dispersion induces a non-existing trend in elemental abundance measured via absorption features, especially for those that are also sensitive to IMF \citep[e.g.][]{LaBFerVaz13, 2014MNRAS.438.1483S}. For example, are the trends for N or Na abundances due to shortcomings in the stellar population models, or, in the opposite way, that trends in abundances manifest itself as a change in IMF slope? 
To asses the first possibility, we have re-measured the abundances from the spectra with \textsc{alf}, but fixed the IMF slope to the Salpeter value ($\alpha =2.35$). The difference in the abundances between the free-IMF and fixed-IMF measurements are generally very small (the mean difference is typically less than 0.01 dex). We show the cases of [N/Fe] and [Na/Fe] in Fig.~\ref{fig:age_vs_age}. It is thus unlikely that the free low-mass IMF slope changes the measured abundances in such a way that they imply a strong co-variance between the low-mass end and the high-mass end slopes. 

One interesting change in the measured parameters is, however, seen for the average age of the old component. In the first panel of Fig. \ref{fig:age_vs_age}, we show that this age is almost always lower for those measurements for which the IMF slope was fixed to the Salpeter value. Fixing the dwarf-to-giant ratio in stellar models also fixes the ratio of turn-off stars to giant stars. Therefore, imposing a Salpeter slope will give a higher number of main sequence stars and deeper hydrogen absorption lines, which the stellar models perhaps compensate for by decreasing the age of the population. As the low-mass IMF slope in our sample, when left free, is often shallower than Salpeter, the result is that for many galaxies the ages of old populations decrease somewhat. Nevertheless, it is surprising that the average abundances stay however roughly constant when varying the IMF slope. Additionally, galaxies for which we masked H$\beta$, because of emission-line contamination, do not show evidence for their low-mass IMF slopes being outliers.

While relative precision is more relevant in this workl than the absolute accuracy of the derived values, we note that the statistical errors on element abundances are quite small (Table~\ref{t:alf}). They range from 0.01 dex for [Fe/H], [Mg/Fe] or [Na/Fe], to 0.1-0.3 for abundances of elements with minor spectral features in our wavelength range (i.e. [Cu/Fe], [Sr/Fe] or [Eu/Fe]). Such small errors are nevertheless not uncommon in similar studies of massive galaxies. \citet{2022ApJ...932..103G} present uncertainties of $\sim0.02$ for both [Fe/H] and [Mg/Fe] for galaxies with $\sigma>250$ km/s also using \textsc{alf}. Working with the SDSS spectra, and limiting again to galaxies with $\sigma>250$ km/s, \citet{2015A&A...582A..46W} derive uncertainties of 0.03 for both metallicity and $\alpha$-element abundances. While these two studies used full spectral fitting methods, \citet{2009MNRAS.398..133L}  perform index fitting on a sample of BCGs (long-slit spectra) and obtain average uncertainties of 0.07 and 0.04 for the metallicity and element abundances, respectively. \citet{KunEmsBac10}, working with IFS data and index fitting methods report errors of 0.02-0.04 and 0.02-0.05 for metallicity and element abundances, respectively, of early-type galaxies. The S/N of our data are likely higher than in most of the mentioned studies, while our aperture of one effective radius is also larger, both of which contribute to somewhat smaller errors in our case. We present a more detailed comparison with literature values in Appendix~\ref{a:el}.

Such small statistical errors suggest that the actual error budget could be dominated by the systematic errors. Those are hard to measure, as they depend on a number of assumptions inherent to the modelling approach. The quality of the stellar population models, fitted wavelength range, assumptions on IMF and the general number of free parameters used in the fit. Regarding \textsc{alf}, some of these aspects were tested by \citet[][their figure 17]{ConGravan14}, who conclude that one can expect a typical change in abundance patters of the order of 0.05 dex or less, while some cases can be as large as 0.1. We report similar tests in the lower panels of Fig.~\ref{fig:age_vs_age}, where we fixed the slope of the IMF (as in the top panels), masked all Na lines within our spectral range and limited the fit to $<6900$\AA. Our conclusions are consistent with \citet{ConGravan14}, where the systematic uncertainties range between 0.05 and 0.1 dex. This is highlighted by [Fe/H], [Mg/Fe], [O/Fe] and [Na/Fe], where the differences stay below or around 0.05 dex, except in the case of shorter fitting range, when they are closer to 0.1 dex. Note that these are the elemental abundances that were used to constrain the \textsc{chempy} fits. We further discuss the difference in the derived \textsc{alf} parameters in Appendix~\ref{a:el}. 

\begin{figure*}
  \includegraphics[angle=0, width=.95\textwidth]{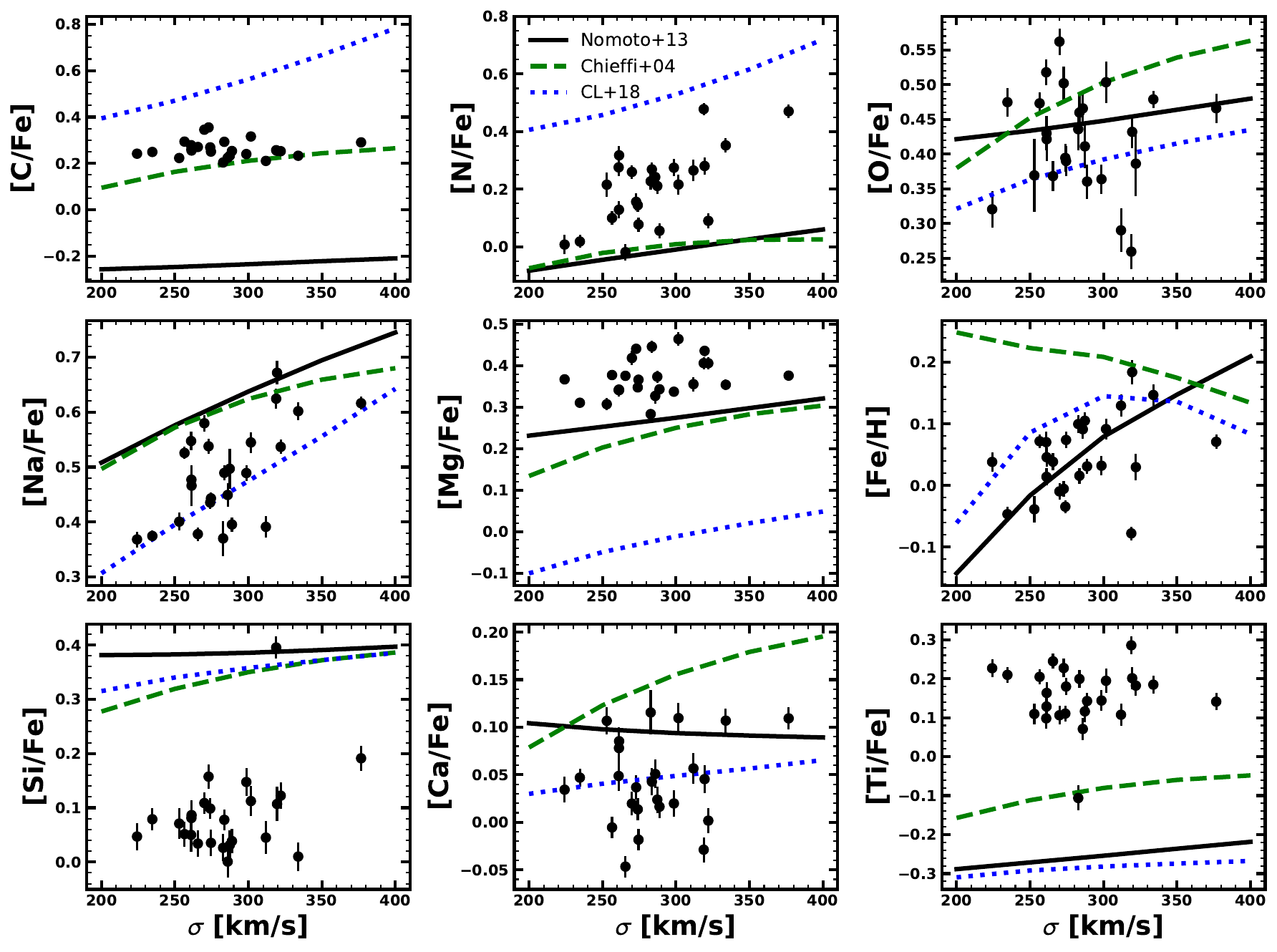}
  \caption{Elemental abundances from observations as a function of central velocity dispersion for a subset of light elements. All elements are shown with respect to the Fe abundance. Lines denote model predictions based on the average parameters from the \textsc{Chempy} fits to the observational data. Note that fits included only the elements  O, Na, Mg and Fe. Models are based on the yields of \citet{NomKobTom13}, \citet{ChiLim04} and \citet{LimChi18} (set R).}\label{fig:results_models}
\end{figure*}

\begin{figure}
  \includegraphics[angle=0, width=\columnwidth]{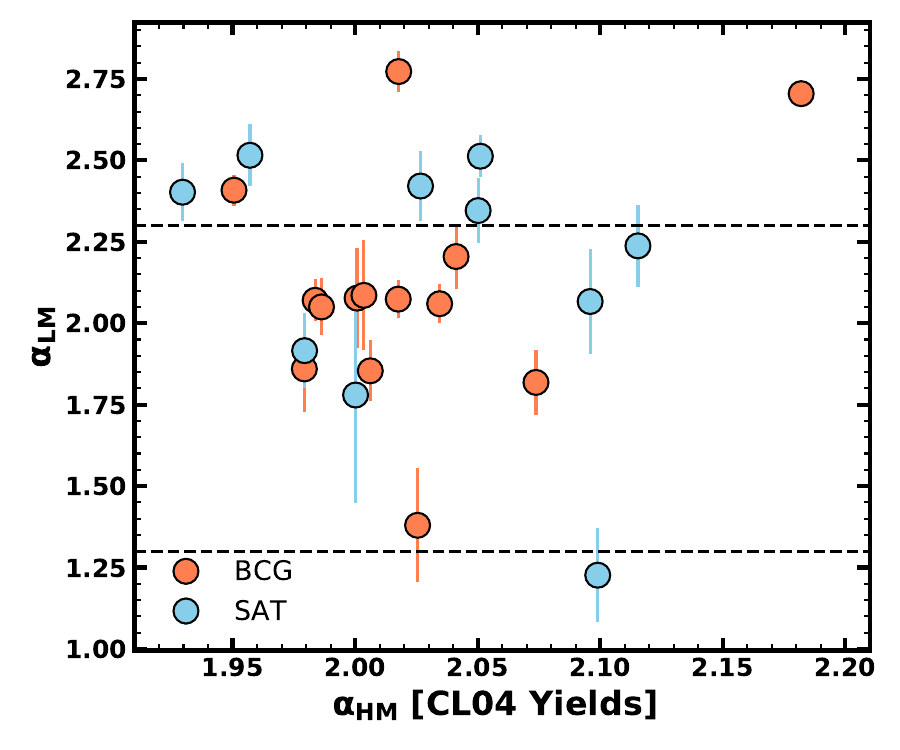}
 \caption{High mass slope of the IMF derived using the yields of \citet{ChiLim04} versus the low-mass end slope derived using \textsc{alf}. BCGs and satellite galaxies are shown with orange and light blue symbols, respectively. Note the similarity of this figure with Fig.~\ref{fig:imf_vs_imf}, once the systematic shift to lower $\alpha_{\rm HM}$ is take into account. Horizontal dashed lines indicate the Salpeter (higher) and Kroupa (lower) values for the low-mass end slopes, for comparison purposes.}
 \label{fig:imf_vs_imf_chieffi04}
\end{figure}

The second case, in which abundances cause the low-mass part of the IMF to change, is not as easily invalidated. We do however note that one of the strongest indicators for the low-mass IMF slope is the Na line at 8190\AA. The sensitivity of this line to the proportions of cool dwarfs and giants has been known since \citet{SpiTay71}, and has subsequently been used by several authors \citep[e.g.][]{CarVisPic86,FabFre80,vanCon10,SpiTraKoo12}. To show that the low-mass IMF slope is not a reflection of the Na abundance, we re-fit the low-mass part of the IMF from the stellar spectra, but with the 8190\AA\,Na line masked. We also re-run \textsc{alf} with all Na line features masked (including also the Na D doublet at 5892\AA), even though various authors \citep[e.g.][]{LaBFerVaz13, SpiTraKoo15} show that this absorption features, although sensitive to IMF slope, is more sensitive to abundance. 

The results are shown in Fig.~\ref{fig:imf_no_Na}, where only small differences are noticeable between the low-mass IMF slopes with partially or fully masked Na lines. Both Na masks result in a consistent low-mass IMF slope with respect to the one from the unmasked run. The largest discrepancy for both masks is observed for PGC073000, which has the youngest stellar populations in the sample. The low-mass end IMF slope measured without Na lines is close to the Salpeter value or lower. The reason for this change could be that the unusually high value is driven by the combination of Na abundance (which is not extreme) and young age, or due to a difficulties in fitting the line, due to emission. The panels in Fig.\ref{fig:alf_fits} pertaining to PGC073000 (bottom left), show a possible contribution of He{\small I}\,$\lambda5875$ emission-line. Similarly, the IMF slope is influenced in a few other galaxies with emission-lines \citep[i.e. PGC047273,][]{PagKraden21}, and we masked fully or partially the regions around their NaD doublets (for the nominal fit). Overall however, Fig.~\ref{fig:imf_no_Na} results suggest that the observed trends are genuine and not driven by internal inconsistencies.

\subsection{Uncertainties in the chemical evolution modelling}
\label{ss:unce}

As we noted in Section~\ref{ss:chem}, our \textsc{Chempy} fits to the elemental abundances produced by \textsc{alf} have large $\chi^2$ values. We visualise how well the chemical evolution models can reproduce observed data for our galaxies in Fig.~\ref{fig:results_models}. We first fitted the \textsc{Chempy} resulting parameters (presented in Fig.~\ref{fig:chempy_vs_sig}) versus the velocity dispersion with straight lines and used these models to predict elemental abundances as functions of the velocity dispersion. In Fig.~\ref{fig:results_models} we show the model predictions (as lines) and the \textsc{alf} abundances for the following elements: C, N, O, Na, Mg, Fe, Si, Ca and Ti. Note, however, that \textsc{chempy} models are only constrained by [O/Fe], [Na/Fe], [Mg/Fe] and [Fe/H] (see Section~\ref{ss:chempy}). Part of the reason of leaving out most elemental abundances for our chemical evolution modelling was to focus on elements for which the yields are more reliable. In general the uncertainties increase for heavier elements, which provides an explanation why we are not able to reproduce the abundances of Ti and Si. 

The chemical models reasonably reproduce the elemental abundances used to directly constrain them, although [Na/Fe] and [Mg/Fe] are typically over or under predicted, respectively. When it comes to pure prediction of elemental abundances (those not used to constrain the models), the models typically do not fit them particularly well (with a possible exception of [Ca/Fe]). Using yields from \citet{ChiLim04} or \citet{LimChi18}, as a replacement for \citet{NomKobTom13}, does not improve the predictions in general, with some possible exceptions (i.e. [C/Fe] for \citet{ChiLim04} or [Ca/Fe] for \citet{LimChi18}). An interesting case is nitrogen, which seems to be bracketed between chemical models based on the \citet{NomKobTom13} and \citet{ChiLim04} yields, and those of \citet{LimChi18}. Nitrogen can be produced in multiple ways, originating both in AGB stars and in CC-SN \citep{2020ApJ...900..179K}. It can be a by-product of the CNO cycle in stars, but could also be produced from initial hydrogen and helium present in a star. In the first case, the nitrogen is called secondary, as the production of nitrogen is dependent on the initial CNO abundance \citep{Cla83,Arn96}. This leads to a metallicity dependence on the nitrogen abundance. If on the other hand nitrogen is produced in a primary way, it would, to first order, follow the abundances of carbon and oxygen \citep{TalArn74}. One way to produce sufficient nitrogen is therefore to include stellar rotation \citep{MeyMae02a,MeyMae02}. Since the yields of \citet{NomKobTom13} do not include rotation this might be the reason why our models under-predict it. However, even without rotation (the $V=0$ km/s models of set R) the \citet{LimChi18} yields already produce more than sufficient N, and are also able to reproduce the increase seen with velocity dispersion, as can be seen from Fig. \ref{fig:results_models}.

The \citet{LimChi18} yields also over-predict the [C/Fe] abundance, possibly because a lot of carbon is ejected before entering the Wolf-Rayet phase for stars heavier than 10 \Msun. C is produced either in low mass (1-4\Msun) AGB stars or in massive ($>10$\Msun) stars. As the high-mass C yields depend very sensitively on the treatment of convective overshooting, predictions for C yields vary strongly. Additionally, C yields may be particularly sensitive to binary evolution \citep{FarLapdeM21}.   Fig.~\ref{fig:results_models} also shows that it is difficult to obtain models with [Mg/Fe] high enough to explain the observations, especially so with \citet{LimChi18} yields. Difficulties with the underproduction of Mg in the \citet{LimChi18} models were previously also noted by \citet{PraAbiLim18}, who attribute them to presently poorly modelled physical phenomena. Keeping in mind the level at which the data can be reproduced, we continue the discussion about the only parameter of the chemical evolution model of our sample of very massive galaxies that shows a clear variability with the galaxy velocity dispersion, the high-mass IMF slope $\alpha_{\rm HM}$ (Fig.~\ref{fig:highmass_alpha_sig}). 

Our preferred yield sets of \citet{NomKobTom13} result in a slight trend of $\alpha_{\rm HM}$ with $\alpha_{\rm LM}$ for the satellite galaxies (Fig.~\ref{fig:imf_vs_imf}). This trend relies largely on the abundances of Na and $\alpha$ elements. In Fig.~\ref{fig:imf_vs_imf_chieffi04}, we show the two IMF slopes when we, instead of \citet{NomKobTom13} use the yields of \citet{ChiLim04}. The high-mass IMF slopes using the \citet{ChiLim04} yields ($\alpha_{\rm HM(CL04)}$) are on average smaller (mean slope is 2.02) compared to the $\alpha_{\rm HM}$ based on the \citet{NomKobTom13} yields (mean slope 2.1). One galaxy, previously highlighted PGC073000, seems to be an outlier with $\alpha_{\rm HM(CL04)}\sim2.182$. If we neglect this galaxy, and a systematic shift to lower values for $\alpha_{\rm HM(CL04)}$, the distribution of low-mass and high-mass IMF slopes is very similar to that seen on Fig.~\ref{fig:imf_vs_imf}. The Pearson correlation coefficient in Fig.~\ref{fig:imf_vs_imf_chieffi04} for the satellite galaxies is somewhat smaller, but still statistically significant with $\rho=-0.28_{-0.25}^{+0.36}$, while the coefficient for the BCGs is still not significant ($\rho=-0.25_{-0.32}^{+0.32}$). We conclude that using different yields does not change the results from our chemical evolution modelling, and that there might exist a tentative anti-correlation between low- and high-mass end IMF slopes, specifically among massive non-BCG galaxies.  More objects are, however, needed to establish the real trends. 

We also used the yields of \citet{LimChi18} to explore the high-mass end of the IMF. Leaving [Mg/Fe] out of the analysis (as it is poorly reproduced) and only using [O/Fe], [Fe/H] and [Na/Fe], we also find a trend between the high-mass end IMF slope and velocity dispersion, which is, however, even steeper and leads to more extreme values of $\alpha \sim 1$ for the galaxies with the highest velocity dispersions. We do not consider these models any further. 

Recently, \citet{CinKar22} published yields for AGB stars at very high metallicities. Particularly at metallicities Z$>$0.04, these yields predict that AGB stars can produce very high amounts of $^{14}$N and $^{23}$Na, which could potentially explain the trends we see with velocity dispersion and might remove the need for top-heavy IMFs at the highest dispersions. To test this scenario we have incorporated these yields in our distribution of \textsc{Chempy}, and supplemented  them with the \citet{KarLug16} AGB yields at lower metallicities. In practice this does not seem to change the results of the chemical evolution models, as metallicities almost never reach values above Z = 0.04 and instead points at Na being mainly produced in massive stars, and returned via CC-SN  \citep[i.e.][]{2020ApJ...900..179K}. The N abundance in the M3G galaxies correlates weakly with the Ba abundance. Although the latter is an {\it s} process element, suggesting a possible origin in AGB stars, we note for our galaxies N abundance does not correlate with Sr abundance (another element mostly produced in AGB stars), while \citep[e.g.][]{LimChi18} show that Ba can also be produced in (rotating) massive stars.

Fig.~\ref{fig:chempy_vs_sig} presents the set of free parameters using in the chemical modelling, for both the "short" (3.5 Gyr) and the "long" (7 Gyr) runs. It is telling that while the IMF slopes of the satellite galaxies in both cases show the consistent anti-correlation with the velocity dispersions, while this is not the case for any other parameter. We stress that all \textsc{chempy} parameters were kept free during modelling, and the resulting values are mostly flat across the range of the $\sigma_e$ covered by our sample, with a possible exception of the SN Ia delay times, which in the "long" \textsc{chempy} run show a weak trend with $\sigma_e$. Furthermore, we run \textsc{chempy} models by setting the outflow fraction to zero and fixing it to x$_{out}=0.2$. This means that all enriched material is kept within the ISM, or a fixed fraction is forced to be diluted in the the corona, respectively. While some model parameters show minor changes, the high-mass end IMF slopes are essentially the same and keep the same correlations with the velocity dispersion as in our nominal run (Appendix~\ref{a:chm}). These tests offer further support that $\alpha_{\rm HM}$ is not degenerate with respect to the chemical evolution parameters.

Our high-mass end of the IMF slope is thus mainly dependent on the almost constant [O/Fe] and [Mg/Fe] abundance and increasing [Na/Fe] with velocity dispersion. The decrease of the IMF slope with velocity dispersion for satellite galaxies is then not surprising, as [Na/Fe] strongly increase with stellar mass. We stress that this assumes that we can fully rely on the assumptions of the chemical evolution code (one zone, instant mixing, accretion of gas with primordial abundances) as well as all the uncertainties associated with the yields, and our specific parametrisation of the IMF in the first place.

\subsection{Differences between BCGs and satellites}

The differences in elemental abundances between BCGs and satellite galaxies seem to be minor and at a low statistical significance (Fig. \ref{fig:abundances_vs_sat}). Nevertheless, we see three notable trends:
\begin{enumerate}
\item At fixed velocity dispersion, the mean abundances of elements in BCGs are equal or somewhat larger to those in satellite galaxies. There are some exceptions, such as the [Fe/H] and the [Ca/Fe] abundances which are larger for the satellite galaxies. On the other hand, for some elements produced in CC-SN explosions ([Si/Fe] and [K/Fe]), the values for BCGa are higher at fixed velocity dispersion. 
\item In a number of BCGs, elements primarily produced in SN Ia explosions have larger abundances than in satellite galaxies. One can see this based clearly on the values of [Mn/Fe] and [Co/Fe]. In the case of [V/Fe], while both satellite galaxies and BGCs have similar mean and high values, satellite galaxies are more likely to show lower values. 
\item Correlations between the elemental abundance and velocity dispersion are stronger for satellite galaxies, especially so for the overall metallicity ([Fe/H]) and the CC-SN produced [Na/Fe] and [Mg/Fe].
\end{enumerate}

These trends can be explained as a combination of the differences in the galaxy mass, where BCGs are somewhat more massive than non-BCGs in our sample (Fig.~\ref{fig:ms}), but also as a consequence of the specific mass assembly paths for these two galaxy types. While all galaxies have high [$\alpha$/Fe] values, indicating short starbursts (at early times), BCGs do not show dependance of the [$\alpha/$Fe], or the metallicity, on the velocity dispersion (or, therefore, galaxy mass). We note that [Fe/H] also correlates with the I-band surface brightness in the aperture of the MUSE data used for this analysis, and may thus be a result of either an averaging of the radial [Fe/H] profile due to merging or a prolonged star formation leading to a higher central surface brightness. We also note that the [Ca/H] abundance does not show any difference between BCGs and satellite galaxies, and the difference in [Ca/Fe] between BCGs and satellites we attribute to the observed difference in [Fe/H]. 

The tentative differences between BCGs and satellite galaxies are consistent with the hypothesis that BCGs have more violent formation paths, experiencing more major mergers and accreting stars from different types of galaxies. Our data probe the central half-light regions of galaxies, regions where the episodes of star formation are the most likely, but also the regions influenced the most by similar mass mergers. Our chemical evolution model showed that the star formation occurred long time ago and it was short. However, BGCs have both the $\alpha$-element abundances and Fe-peak elements larger than satellite galaxies, suggesting somewhat more extended star formation histories, or perhaps a different number of bursts of star formation. Similar conclusions were reached by other studies \citep[e.g.][]{2010MNRAS.404.1775T, 2014MNRAS.445.1977L, 2021MNRAS.502.4457G} using much larger samples of galaxies, over a larger mass range and different levels of star formation. One of the results from these studies was that central galaxies can appear somewhat younger and with lower [$\alpha$/Fe] abundances than their satellites. Nevertheless, we do not consider this inconsistent with our findings due to the very different types of studied galaxies. The above-mentioned literature studies have a much larger leverage on the environment type, spanning poor groups to large clusters, while all our satellites are in similarly rich and massive clusters of galaxies, where they are typically 2nd, 3rd or 4th brightest galaxy (Krajnovi\'c et al. in prep). In that sense, and given the low number of galaxies, it is remarkable that we detect differences between M3G galaxies. Finally, the satellites in the above-mentioned studies have significantly lower masses compared to the centrals, while in our case, even though the masses of BCGs are somewhat larger than of satellites, they are generally comparable (and $>10^{12}$ M$_\odot$).
 
Major mergers, when occurring between galaxies without gas, change $\sigma_e$, or, generally, the mass -- size relation, only mildly \citep{2006MNRAS.369.1081B,2009ApJ...697.1290B,2009ApJ...691.1424H, 2009ApJ...699L.178N}. They also flatten any existing trends in metallicity or element abundances \citep[e.g.][]{2004MNRAS.347..740K,2009A&A...499..427D}. Such an ``equalising" effect in terms of chemical properties can be expected between galaxies of the same type (i.e. BCGs) even if they have somewhat different masses (or $\sigma_e$). The differences that we observe between the satellites and BCGs suggest that, while they have similar masses and effective velocity dispersions, the satellite galaxies experienced less gas-free major mergers. This is also supported by the kinematics of M3G galaxies, as about 50\% of BCGs show prolate-like rotation around the major-axis \citep{KraEmsden18}. On the other hand, only about 30\% of satellite galaxies show the same type of rotation \citep{KraEmsden18}. Such prolate-like rotation is typically expected to originate in major, gas-free mergers \citep[e.g.][]{2017ApJ...850..144E,2018MNRAS.473.1489L}, suggesting a clear formation way for some galaxies.

The IMF differences between BGC and satellite galaxies are small, but are, nevertheless, noticeable. There is a remarkable trend of the high-mass end IMF slope with $\sigma_e$ for satellite galaxies  (Fig.~\ref{fig:highmass_alpha_sig}), which does not exist for BCGs, and can be traced back to the differences in elemental abundances. Noteworthy is also the distribution of satellite galaxies in $\alpha_{\rm LM} - \alpha_{\rm HM}$ plane (Fig.~\ref{fig:imf_vs_imf}), where, again, satellite galaxies show an anti-correlation (at 2$\sigma$ level), while BCGs do not show a statistically significant correlation. Marking the galaxies with their kinematics (with prolate-like and without prolate-like rotation) enhances this difference somewhat: galaxies without prolate-like rotation behave similarly like satellites. Prolate-like rotation is a direct evidence that the last merger was gas-free and a major merger, but galaxies without prolate-like rotation could have also experienced similar type mergers, ``diluting" the signal. Nevertheless, the trend suggest that the observed difference originate in different evolution paths taken by BCGs and satellite galaxies.

\subsection{The shape of the IMF in the most massive galaxies}

\begin{figure}
  \includegraphics[angle=0, width=\columnwidth]{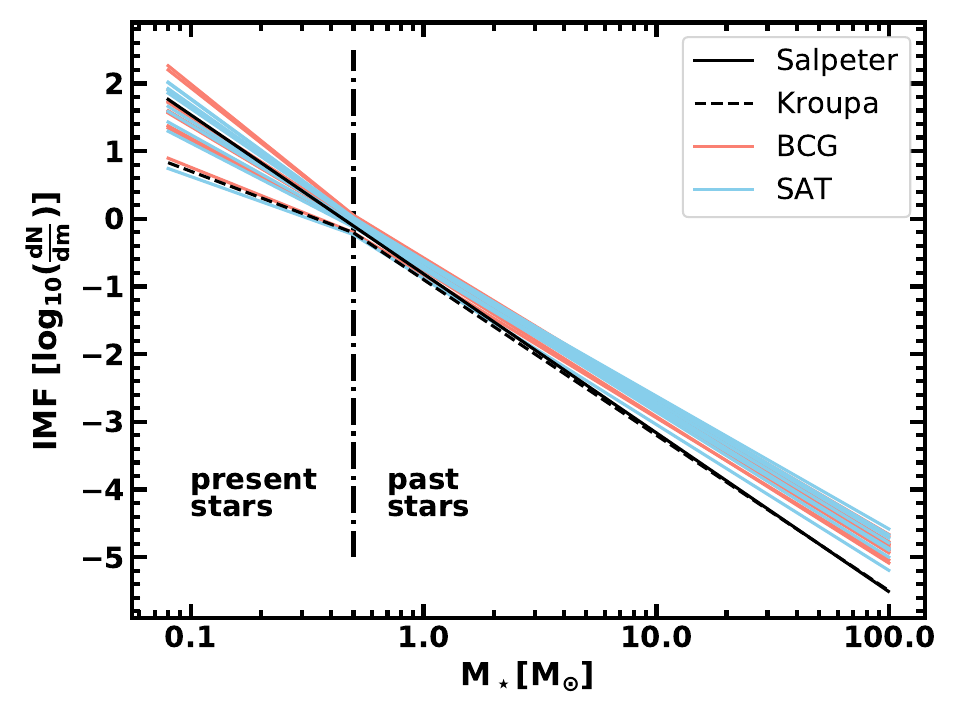}
 \caption{A composite distribution of the initial mass functions in the $\psi(m) = \frac{dN}{dm}\sim m^{-\alpha}$ form for M3G galaxies. The IMFs are constructed using the low-mass ($<0.5$ M$_\odot$) and high-mass ($>0.5$ M$_\odot$) end slopes obtained from the \textsc{alf} analysis of the MUSE spectra and the \textsc{chempy} modelling of recovered elemental abundances. For comparison, we plot also the Kroupa (dashed line) and the Salpeter (solid line) IMFs. Orange lines are BCGs and light blue lines are the satellite galaxies of the M3G sample. Vertical dashed-dotted line approximately shows the boundary between stars that still exits in M3G galaxies, and stars that ended their evolutionary paths. Note that this plot is not meant to represent a fixed IMF. Instead it shows a temporal variation of the IMF, in the sense of the IGIMF theory \citep[e.g.][]{2018A&A...620A..39J}, where, for the massive galaxies, the high-mass end is representative of the very early age, and the low-mass end of the later star formation.}
 \label{fig:m3g_imf_full}
\end{figure}

In Fig.~\ref{fig:m3g_imf_full} we illustrate the shape of the IMF recovered through \textsc{alf} and \textsc{chempy} models. We combine the low-mass and the high-mass slopes assuming the transition occurs at 0.5 M$_{\odot}$. As already visible in Fig.~\ref{fig:imf_vs_imf}, the spread in the slopes values is generally small, while it is a bit larger on the low-mass end. Both BCGs and satellite galaxies are consistent with bottom heavy low-mass IMF and a top heavy high-mass IMF, overabundant in low and high mass stars compared to the standard IMF parametrisations. This is equivalent to what \citet{Mar16} invoked as a non-canonical shape of the IMF providing both the high [Mg/Fe] values and being overabundant in low mass stars (relative to Salpeter IMF) in massive ellipticals \citep[e.g.][]{Convan12,SpiTraKoo12,CapMcDAla12,2013MNRAS.429L..15F}. 

An alternative way of looking at Fig.~\ref{fig:m3g_imf_full} is that it does not show a static picture of the IMF in any of these galaxies, but could be considered as a depiction of a time-varying IMF, where the stellar mass (the x-axis) is a proxy for time: massive stars, which do not exist anymore, signify the past, while low mass stars, still burning hydrogen, relate to the present. Here the initial IMF in progenitors of our massive galaxies had an excess of high-mass stars that were born in a short ($\sim0.3$ Gyr), but strong starburst, characterised by CC-SN and $\alpha$-element enrichment of the ISM and the subsequent stellar populations. This phase is followed by a prolonged ($\sim1$ Gyr), less violent star-burst characterised with a bottom heavy IMF \citep{1997ApJS..111..203V, 2013MNRAS.435.2274W, 2015MNRAS.448L..82F}. 

Our results are broadly consistent with the notion that the IMF is not universal or invariant, but should be considered within a specific time and a spatial volume. What we observe is a ``galaxy IMF" (gIMF) characteristic for the full galaxy, and different from the much more local ``stellar IMF" \citep[sIMF;][]{2018PASA...35...39H}. In this context, the ``galaxy wide IMF"\footnote{Different authors have different names, and we will use gIMF.} \citep[][]{2018A&A...620A..39J}, arising from an ``integrated galaxy-wide IMF" (IGIMF) model \citep{2003ApJ...598.1076K, 2005ApJ...625..754W, 2013pss5.book..115K, 2015MNRAS.446.4168R,2017A&A...607A.126Y}, is obtained as a summation of many sIMF. The gIMF found in the centres of present day massive galaxies starts with an IMF overabundant in high mass stars (top heavy), and then evolves into an IMF with an excess of low mass stars (bottom heavy), which are also enriched in metals. Crucial point is that the gIMF is not simultaneously bottom heavy and top heavy, but the observed excesses of the high and low mass stars happen sequentially. This evolution of the IMF is present in both galaxy types of our sample (BCGs and satellites), but the differences in the correlation of the IMF slope (specifically $\alpha_{\rm HM}$ vs. $\sigma_e$) suggest that their subsequent evolutionary paths are nevertheless sufficiently different, most likely in terms of the number and types of mergers. 

Given that all galaxies in our sample end up as massive, quiescent and giant ellipticals, a broadly speaking argument would be as follows. During the early formation, properties of these galaxy (e.g. gas accretion, start formation, the transition between top to bottom heavy IMF, etc) depend on their total mass and fuel availability (i.e. environment). At some point, some of these galaxies (i.e. BCGs), however, end up being in the centre of the gravitational wells and experience more mergers of all kind, with an increasing fraction of gas-free mergers as time passes. A consequence of this ``privileged" position is the accretion and mixing of stars formed in different types of galaxies and under different conditions, maintaining the chemistry, elemental abundances and metallicity of the stars, but removing the signatures of the dependence with galaxy velocity dispersion (or mass). Massive galaxies outside of this ``privileged" position are able to retain more of the ``fossil" records from the epoch of early star-formation, recognisable in the dependancies with the velocity dispersion (i.e. $\alpha_{\rm HM}$ vs. $\sigma_e$).

\section{Summary}
\label{s:sum}

We present in this work an analysis of the stellar populations in the M3G sample of 25 very massive galaxies observed with VLT/MUSE instrument. Our goal is to reconstruct the chemical evolution of these galaxies and place constraints on both the low- and high-mass IMF slopes, as an overview of the change in the star-formation. We focus on the central half-light regions of our galaxies, and spectra of very high signal-to-noise ratios. 

Using advanced stellar population tool \textsc{alf}, we measure 19 elemental abundances, age, and low-mass end slope of the IMF of the galaxies. The elemental abundances are typically high, as expected for massive galaxies, and are broadly consistent with previous observations based on SDSS galaxies, especially when compared with the measurements obtained with the same version of the tool. Next to higher average values of some  elements (e.g. [O/Fe], [Mg/Fe] and [V/Fe]), we find significant correlations between the elemental abundances and the global velocity dispersion. Notable correlations are found for [Fe/H], [N/Fe], [Na/Fe] and [Ni/Fe], while [V/Fe] shows an anti-correlation. The correlations are typically more pronounced (significant at 2$\sigma$ level) for satellite galaxies for [Fe/H] and [Mg/Fe], while in the case of [V/Fe] it is the BCGs that show a significant correlation. 

Our stellar population models are constrained with a unimodal IMF which is free to vary between galaxies. We find that in most cases the slope of this low-mass end IMF is similar to the Salpeter, and often larger, but there are at least two galaxies which have the IMF slope consistent with that of (low-mass end) Kroupa IMF. There is no clear difference in the derived slopes between BCGs and satellites. This results is consistent with previous works, both in the sense that very massive galaxies (with high $\sigma_e$) have super-Salpeter IMF slopes and that there are massive galaxies with smaller, Kroupa-like IMF slopes. 

We use the derived elemental abundances of selected elements ([Fe/H], [O/Fe], [Mg/Fe] and [Na/Fe]) to constrain the chemical evolution using the \textsc{Chempy} model. While these elements are typically well reproduced by the set of standard yields, other chemical abundance are poorly reproduced. Changing the yields does not necessarily improve the reconstruction of elemental abundances. The chemical evolution modelling results show that the very massive galaxies have similar formation time scales and star formation efficiency.

One parameter critical for the chemical evolution shows an interesting change between galaxies: the high-mass end slope of the IMF. Satellite galaxies show a clear correlation between this IMF and the velocity dispersion, while this is not true for BCGs. Having both the low- and high-mass end IMF slopes, we combine them and show that there is a general tendency that the more top heavy IMF slope (high-mass end) is paired with a more bottom heavy (low-mass end) IMF slope. This correlation is more pronounced for the satellite galaxies, but only at a 1$\sigma$ confidence level. Furthermore, the changes in [$\alpha$/Fe] and [Na/Fe] abundances can be explained by this change in slope of the high-mass end of the IMF with the global velocity dispersion. We test to what extent are the measurement of both the low- and high-mass end IMF slope susceptible to systematic errors in the models, by masking different absorption-lines, changing the wavelength range used for fitting, changing the SN yields, and modifying the the duration of the chemical evolution or their outflow feedback fraction, but the general trends in the IMF slopes remain. 

Our IMF results do not necessarily imply that the IMF was simultaneously bottom heavy and top heavy, and that we derived a static shape of the IMF. We take our results as a confirmation of the galaxy wide IMF hypothesis, where the gIMF is not universal nor static, but it changes with time. The high-mass end IMF is representative of the situation in the first few hundred Myr of galaxy evolution and is characterised by an excess of massive stars, coinciding with the first intensive star burst. As the star formation changes to a more extended and less violent phase, the gIMF transits to the low-mass end IMF that we {\it measure} in nearby (massive) galaxies. It is characterised by an excess of low mass stars, having the chemical imprint of the high mass stars. Such a temporal variation of the gIMF was previously invoked to explain the high metallicities, enhanced $\alpha$-elements and bottom heavy IMFs in massive galaxies. Our results show that a combination of top and bottom heavy IMFs, at different epochs of formation are indeed required to explain the chemistry of low mass stars in very massive galaxies. 

The small differences between the BCGs and satellite galaxies (in the elemental abundances and the high-mass IMF slope) can be understood as small variation in the formation paths of these two galaxy types. All investigated galaxies are very massive and live in dense environments, but the dependence of elemental abundance (and IMF) on the velocity dispersion, together with evidence based on stellar kinematics, suggest that BCGs experienced more gas-free similar mass mergers than the satellite galaxies.

\section*{Acknowledgements}
MdB and DK acknowledge financial support through the grant GZ: KR 4548/2-1 of the Deutsche Forschungsgemeinschaft.  PMW was supported by the BMBF through the ErUM program (project VLT-BlueMUSE, grant 05A20BAB). MdB would like to thank Charlie Conroy for sharing his SSP models. We thank Themiya Nanayakkara and Sebastian Kammann for useful comments and discussions. The MUSE observations were taken within the following observing programmes: 094.B-0592, 095.B-0127, 096.B-0062, 097.B-0776, 098.B-0240, 099.B-0148, 099.B-0242, 0102.A-0327.

\section*{Data Availability}

Raw MUSE data are available in the ESO archive. Reduced cubes are available from the authors on request.



\bibliographystyle{mnras}




\appendix

\section{An example of sky subtraction method}
\label{a:skysub}

\begin{figure*}
  \includegraphics[angle=0, width=\textwidth]{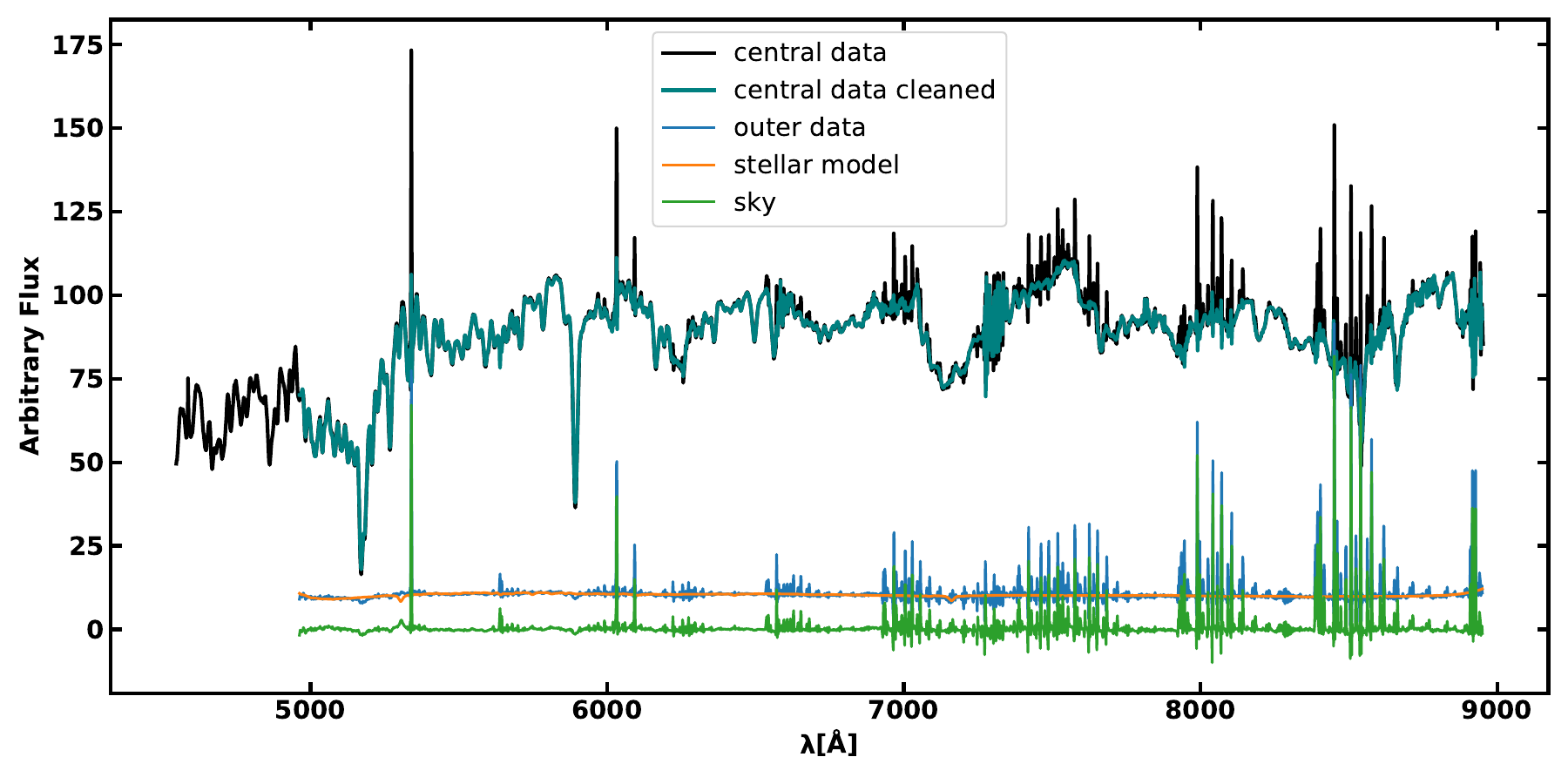}
 \caption{An example of the additional sky subtraction method applied to all galaxies in the sample. This case is for PGC003342, also shown in Fig.~\ref{fig:example_spec}. The central spectrum of PGC003342 (within 1 R$_e$) is shown with a thick black line, while the outer spectrum, expected to be dominated by the sky, is shown by thin blue line below. The orange line is an attempt to fit a stellar population model. The green line at the bottom is the residual, which we assume is fully dominated by sky spectrum. The teal thick line, over-plotted on the central spectrum, is the difference between the central spectrum and the estimated sky (green line).}
 \label{fig:skysub}
\end{figure*}

In Fig.~\ref{fig:skysub} we present an example of our method to remove the remaining sky contribution from the galaxy spectra. As briefly explained in Section~\ref{ss:cent_data}, we combine MUSE spectra outside the central effective radius aperture, and attempt to fit a stellar population model. This fit is typically not very well constrained as the outer spectra have essentially no information on the stellar continuum. The sky residuals are then obtained subtracting the stellar population fit from the outer spectrum. In principle, we could have also estimated the sky residuals by removing the median, or by a simple polynomial fit of the outer spectrum. For PGC003342 the difference in sky residuals obtained in these ways are far below 1\%, except for two features at $5304$\AA\, and $7160$\AA, possibly corresponding to some Fe and Ca features, where they just approach the 1\% level. They are carried over to the sky spectrum, but their influence to the final results are negligible. Presented sky residuals are typical for all other galaxies. 

The residual sky spectrum is subtracted from the inner spectrum and the result is shown on Fig.~\ref{fig:skysub} with the thick teal line, in comparison with the original spectrum produced by the data reduction pipeline (thick black line). The improvement is obvious, even though some spectral regions remain noisy. We mask those during the \textsc{alf} fitting. Note also that below $4960$\AA\, we did not attempt to subtract the sky.

\section{Element abundances and systematic uncertainties}
\label{a:el}

\begin{figure*}
  \includegraphics[angle=0, width=\textwidth]{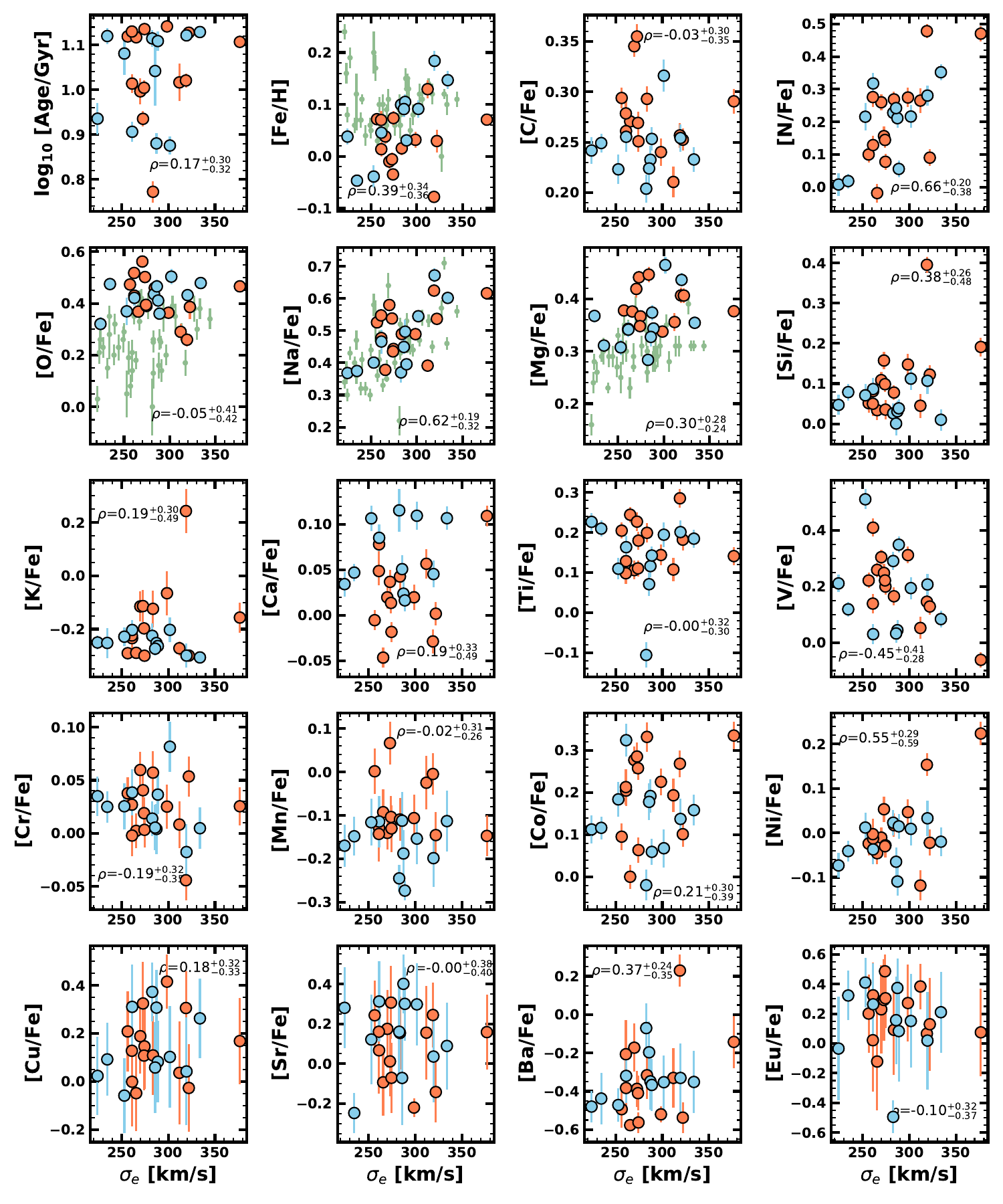}
 \caption{Elemental abundances based on the analysis of MUSE spectra of M3G galaxies with \textsc{alf} as a function of velocity dispersion within the effective radius. BCGs are shown with orange and satellite galaxies with light blue symbols. The top left plot shows the luminosity weighted ($\log_{10}$) age within the same aperture as the elemental abundances. The Pearson correlation coefficient $\rho$ is indicated on each panel. Uncertainties on the velocity dispersion are smaller than the symbols ($\sim2$ km/s). For comparison we include the data from \citet{2022ApJ...932..103G} with small green symbols. Note that [Fe/H], [N/Fe], [Na/Fe], [Mg/Fe], [V/Fe] and [Ni/Fe] are repeated here from Fig.~\ref{fig:X_corr} for completeness. }\label{fig:alf_elem_sig}
\end{figure*}

\begin{figure*}
  \includegraphics[angle=0, width=\textwidth]{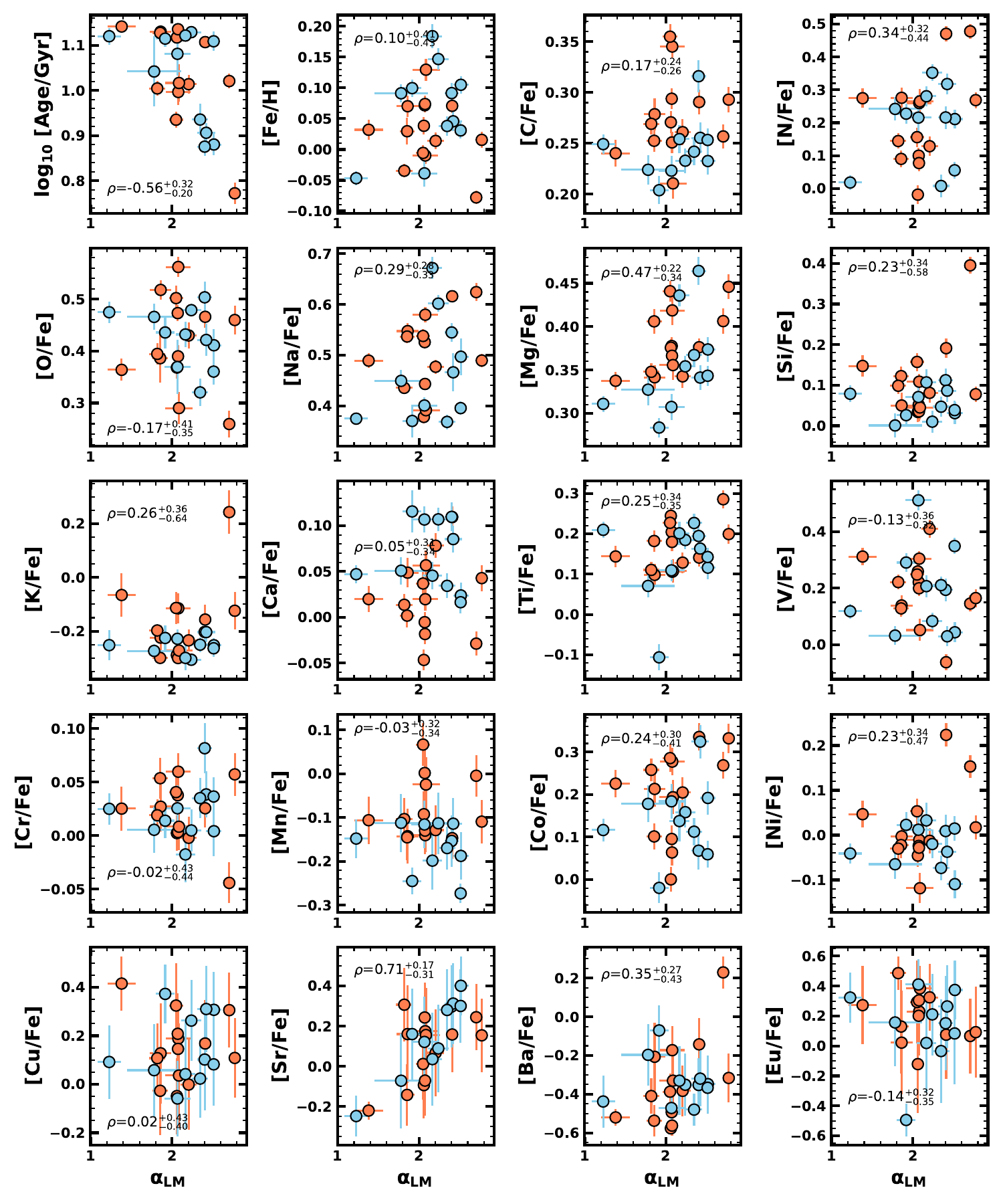}
 \caption{Elemental abundances as a function of the low mass IMF slope for M3G galaxies derived with \textsc{alf}. BCGs are shown with orange and satellite galaxies with light blue symbols. The Pearson correlation coefficient $\rho$ is indicated on each panel. The strongest correlations are found between the IMF slope and [Mg/Fe] and [Sr/Fe]. While there are clear trends for BCGs for some elements (e.g. [Ti/Fe], [Ba/Fe]), there are no statistically secure differences between BCGs and satellite galaxies. Note that [N/Fe], [Mg/Fe], [Ti/Fe], [Co/Fe]. [Sr/Fe] and [Ba/Fe] are repeated here from Fig.~\ref{fig:X_corr} for completeness.}
\label{fig:alf_xfe_imf}
\end{figure*}

We first present here correlations between different parameters resulting from the nominal \textsc{alf} fits as presented in Section~\ref{ss:abund}. In particular, we correlate the elemental abundances (and stellar ages) with the velocity dispersion in Fig.~\ref{fig:alf_elem_sig} and with the low-mass IMF slope in Fig.~\ref{fig:alf_xfe_imf}. On Fig.~\ref{fig:alf_elem_sig} we overplot the values from work by \citet{2022ApJ...932..103G}, also obtained by \textsc{alf}, from Magellan/LDSS3 long-slit spectra of massive galaxies. These spectra have a comparable S/N ($\sim100-200$) to our data, but their wavelength range is somewhat larger (4000-10300$\AA$), and they do not cover the same one-effective radius aperture as we do. The \textsc{alf} set up also differs in several ways.  Notably, they use an IMF model with three slopes, where the slope for the stellar masses higher than 1 M$_\odot$ is fixed to the Salpeter value. Fit is also divided into five spectral ranges, masking several sky lines.

There are some differences in the recovered elemental abundances, notably for [O/Fe] and partially for [Mg/Fe], while the metallicity and [Na/Fe] are in a good correspondence. Placing the observed differences between the two studies into context requires however also a broader comparison. As we show in Fig.~\ref{fig:age_vs_age}, different \textsc{alf} setups deliver different results, with the overall largest difference when the wavelength range is limited to the blue region ($<690$ nm). The most interesting are the four elemental abundances used to constrain the chemical models (see end of Section~\ref{ss:chempy}). These abundances show relatively small changes between various \textsc{alf} runs ($\sim0.05$ dex), largest begin when fitting blue region only ($\sim0.1$ dex). This is consistent with the conclusions of \citet{2020MNRAS.499.2327G}, who state that the results of the full spectral fitting (in their case for age, metallicity and [Mg/Fe]) can be dependant on the spectral range used. A similar results was already seen by \citet{ConGravan14}, when they limited the fitting range to $<580$ nm, except that in their cases [O/Fe] and [Na/Fe] showed larger discrepancies. 

\begin{figure*}
  \includegraphics[angle=0, width=\textwidth]{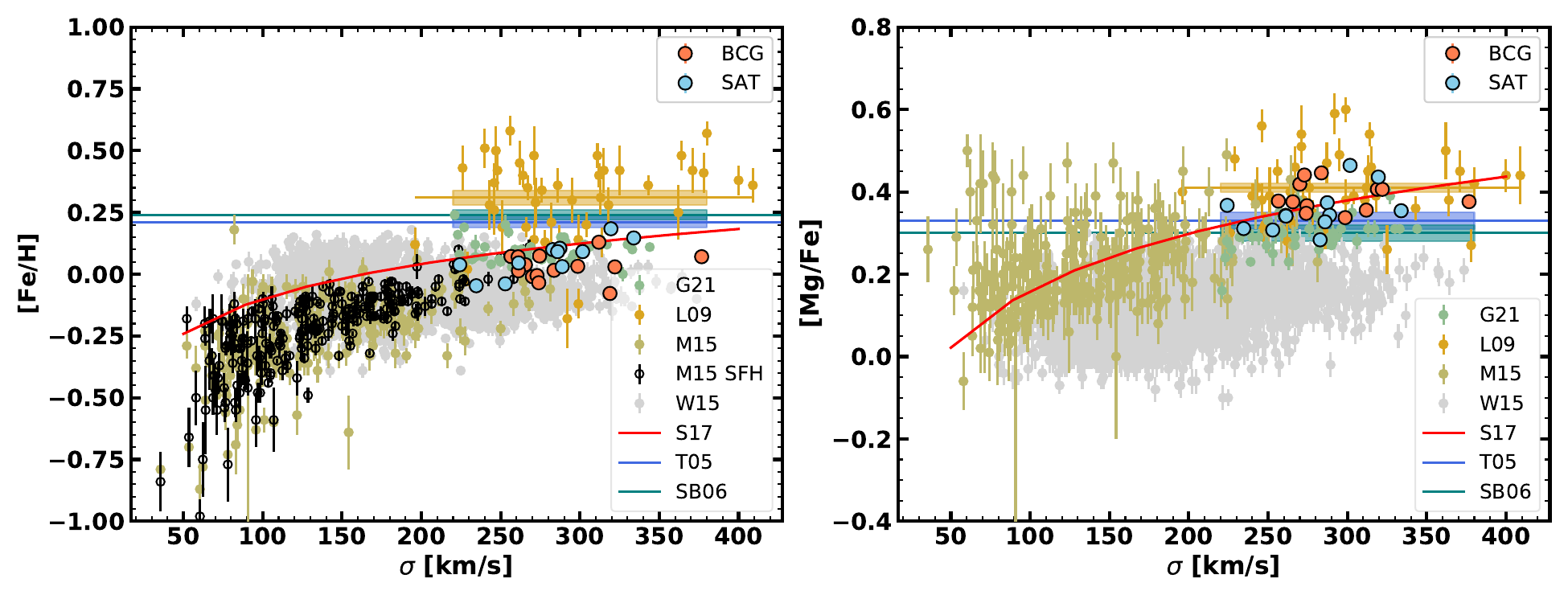}
 \caption{Comparison of the [Fe/H] (left) and [Mg/Fe] (right) between this work and several literature studies, as a function of galaxy velocity dispersion. The following studies were used: \citet{2022ApJ...932..103G} - (G21) green symbols,  \citet{2009MNRAS.398..133L} - (L09) orange symbols, \citet{McDAlaBli15} - (M15) dark khaki symbols, \citet{2015A&A...582A..46W} - (W15) grey symbols, \citet{ThoMarBen05} - (T05) blue line,  \citet{2006A&A...457..809S} - (SB06) green line, and \citet{2017MNRAS.472.2833S} - (S17) red line. T05 and SB06 show average values for metallicity and alpha abundance from these surveys, together with the average uncertainties (shaded regions). S17 shows the functional dependences fitted to their data. We used both full spectral and index fitting methods for comparison, where T05, SB06, L09 and M15 use index fitting methods. The label M15 SFH refers to the metallicities obtained by M15 using the full spectral fitting (the same data as for the index fitting). S17 and W15, and our work, use full spectral fitting. As in the rest of the paper, BCGs are shown with light red and satellites with light blue symbols.}
\label{fig:alf_comp}
\end{figure*}

Given that one can expect differences in recovered parameters depending on various assumptions and the set up of the fit, we compare our results with a compilation of several literature studies which derive the metallicities and $\alpha$-elements abundances in diverse and independent ways. Specifically, we select works that use both the index fitting and the full spectral fitting methods. The results are shown on Fig.~\ref{fig:alf_comp}. There are a few noteworthy points to consider: 

\begin{itemize}
\item The data come from a variety of observations, using both long-slits and integral-field units, and therefore covering different regions of galaxies. \citet{2015A&A...582A..46W} data are based on SDSS 3\arcsec\, fibre spectra, ATLAS$^{\rm3D}$ \citep{McDAlaBli15} and SAMI \citep{2017MNRAS.472.2833S} data are based on one effective elliptical apertures, similar to our case. Other data sets are taken from long-slit observations, and spatially limited. 

\item The uncertainties are often comparable with our values, when the same velocity dispersion range ($\sigma>250$ km/s), as a proxy for galaxy mass, is considered, as already discussed in Section~\ref{ss:unc_imf_low}. 

\item The differences between our and literature results are similar to the differences between various literature results. 

\item A number of literature studies have comparable metallicity ([Fe/H]) or element abundances ([Mg/Fe]), but not both at the same time. Our metalicities are similar to those of \citet{2022ApJ...932..103G} and \citet{2015A&A...582A..46W}, while element abundances are similar to those of \citet{ThoMarBen05, 2006A&A...457..809S} and \citet{2009MNRAS.398..133L}. Our metallicity {\it and} abundances are similar with those of \citet{McDAlaBli15} and \citet{2017MNRAS.472.2833S}, which are also relatively consistent with each other over a larger galaxy mass range. These similarities and differences are not related to the method (full spectral of index fitting): for example the SAMI, our and those of the ATLAS$^{\rm 3D}$ abundances were obtained with two different types of full spectral fitting methods and an index fitting method, respectively, using different single stellar population models, but they still provide consistent results.

\end{itemize}

Overall, we can conclude that our results are consistent with those found in the literature, for the same galaxy mass (velocity dispersion). The comparison is essentially limited to systematic effects, which include the S/N of the data, aperture size, and perhaps most importantly, the assumptions on the stellar populations models. 

In Table~\ref{t:corr} we summarise the values of the Pearson coefficients for relations between element abundances and velocity dispersion, low mass IMF slope, as well as the low and high mass IMF slopes with the velocity dispersion. The correlation coefficients were calculated for three different samples: the full M3G sample, and for BCGs and satellites separately. The errors were estimated at a $2\sigma$ level by taking the lower 2.5\% and upper 97.5\% limit values by bootstrapping. Here we list the most noteworthy relations, where at least one sample shows a significant correlation. An exception is the $\alpha_{\rm LM} - \sigma$ relation which does not show a correlation, but is added for comparison with the $\alpha_{\rm HM} - \sigma$ relation, which is significant for both the full sample and the satellites only. Similar relations are ${\rm [Mg/Fe]} - \sigma$ and ${\rm [Na/Fe]} - \sigma$, while ${\rm [V/Fe]} - \sigma$ relation show high correlation coefficients for the full sample and the BCGs. ${\rm [Fe/H]} - \sigma$ relation is significant for satellites, while ${\rm [Mg/Fe]} - \alpha_{\rm LM}$ and ${\rm [Sr/Fe]} - \alpha_{\rm LM}$ show significant correlations for all three samples.

\begin{table}
   \caption{A summary of relations with a significant Pearson correlation coefficients}
   \label{t:corr}
\begin{tabular}{lccc}
   \hline
    \noalign{\smallskip}
    relation & ALL      &   BCG  & SAT \\
                 & $\rho$ &   $\rho$ & $\rho$ \\
          (1)   & (2)       &   (3)      &  (4)   \\
    \noalign{\smallskip} 
    \hline \hline
    \noalign{\smallskip}
${\rm [Fe/H]} - \sigma$   & $0.39^{+0.35}_{-0.43}$ & $0.09^{+0.35}_{-0.48}$  & $0.82^{+0.10}_{-0.11}$  \\
${\rm [Mg/Fe]} - \sigma$ & $0.30^{+0.26}_{-0.25}$ & $0.08^{+0.28}_{-0.24}$  & $0.44^{+0.24}_{-0.24}$  \\
${\rm [Na/Fe]} - \sigma$ & $0.62^{+0.20}_{-0.33}$ & $0.41^{+0.26}_{-0.49}$  & $0.81^{+0.10}_{-0.13}$  \\
${\rm [V/Fe]} - \sigma$  & $-0.45^{+0.39}_{-0.26}$ & $-0.75^{+0.35}_{-0.14}$  & $-0.18^{+0.33}_{-0.26}$  \\
${\rm [Mg/Fe]} - \alpha_{\rm LM}$  & $0.47^{+0.21}_{-0.30}$ & $0.56^{+0.14}_{-0.46}$  & $0.48^{+0.20}_{-0.32}$  \\
${\rm [Sr/Fe] }- \alpha_{\rm LM}$  & $0.71^{+0.17}_{-0.33}$ & $0.49^{+0.39}_{-0.12}$  & $0.93^{+0.05}_{-0.53}$  \\
$\alpha_{\rm LM} - \sigma$ & $0.27^{+0.30}_{-0.37}$ & $0.24^{+0.29}_{-0.40}$  & $0.32^{+0.29}_{-0.40}$  \\
$\alpha_{\rm HM} - \sigma$ & $-0.55^{+0.31}_{-0.19}$ & $-0.30^{+0.58}_{-0.39}$  & $-0.84^{0.12}_{-0.09}$  \\

    \noalign{\smallskip}
    \hline
\end{tabular}
\\
{Notes: Column (1): Relation type. Columns (2) - (4): Respective Pearson correlation coefficients for the full sample, BCGs and satellites only. The uncertainties mark a $2\sigma$ level.}
\end{table}


\clearpage
\begin{deluxetable}{lcccccccccccccccccccc}
\rotate
\tabletypesize{\scriptsize} 
\tablewidth{0pt}
\tabcolsep=1pt
\tablecaption{Elemental abundances based on the nominal extraction by \textsc{alf}.}
\label{t:alf}

\tablehead{
\colhead{galaxy} &
\colhead{$\log$ Age} &
\colhead{[Fe/H]} &
\colhead{[O/Fe]} &
\colhead{[C/Fe]} &
\colhead{[N/Fe]} &
\colhead{[Na/Fe]} &
\colhead{[Mg/Fe]} &
\colhead{[Si/Fe]} &
\colhead{[K/Fe]} &
\colhead{[Ca/Fe]} &
\colhead{[Ti/Fe]} &
\colhead{[V/Fe]} &
\colhead{[Cr/Fe]} &
\colhead{[Mn/Fe]} &
\colhead{[Co/Fe]} &
\colhead{[Ni/Fe]} &
\colhead{[Cu/Fe]} &
\colhead{[Sr/Fe]} &
\colhead{[Ba/Fe]} &
\colhead{[Eu/Fe]} 
}
\startdata
PGC003342 & 1.00 $\pm$ 0.03 & -0.01 $\pm$ 0.01 & 0.56 $\pm$ 0.02 & 0.35 $\pm$ 0.01 &  0.26 $\pm$ 0.02 & 0.58 $\pm$ 0.01 & 0.42 $\pm$ 0.02 & 0.11 $\pm$ 0.02 & -0.11 $\pm$ 0.06 &  0.02 $\pm$ 0.01 &  0.11 $\pm$ 0.02 &  0.30 $\pm$ 0.03 &  0.06 $\pm$ 0.02 & -0.14 $\pm$ 0.04 &  0.28 $\pm$ 0.03 & -0.01 $\pm$ 0.02 &  0.19 $\pm$ 0.16 &  0.17 $\pm$ 0.19 & -0.17 $\pm$ 0.13 &  0.23 $\pm$ 0.30 \\
PGC004500 & 1.12 $\pm$ 0.01 &  0.07 $\pm$ 0.01 & 0.47 $\pm$ 0.02 & 0.29 $\pm$ 0.01 &  0.10 $\pm$ 0.02 & 0.53 $\pm$ 0.01 & 0.38 $\pm$ 0.01 & 0.05 $\pm$ 0.02 & -0.29 $\pm$ 0.02 & -0.01 $\pm$ 0.01 &  0.20 $\pm$ 0.02 &  0.22 $\pm$ 0.03 &  0.04 $\pm$ 0.02 &  0.00 $\pm$ 0.05 &  0.10 $\pm$ 0.03 & -0.02 $\pm$ 0.03 &  0.21 $\pm$ 0.17 &  0.24 $\pm$ 0.17 & -0.49 $\pm$ 0.10 &  0.20 $\pm$ 0.26 \\
PGC007748 & 1.12 $\pm$ 0.01 &  0.04 $\pm$ 0.01 & 0.37 $\pm$ 0.02 & 0.27 $\pm$ 0.01 & -0.02 $\pm$ 0.03 & 0.38 $\pm$ 0.01 & 0.38 $\pm$ 0.01 & 0.03 $\pm$ 0.02 & -0.29 $\pm$ 0.02 & -0.05 $\pm$ 0.01 &  0.24 $\pm$ 0.02 &  0.26 $\pm$ 0.03 &  0.00 $\pm$ 0.02 & -0.09 $\pm$ 0.05 &  0.00 $\pm$ 0.03 & -0.05 $\pm$ 0.02 & -0.05 $\pm$ 0.16 & -0.09 $\pm$ 0.17 & -0.58 $\pm$ 0.03 & -0.12 $\pm$ 0.33 \\
PGC015524 & 0.94 $\pm$ 0.02 & -0.01 $\pm$ 0.01 & 0.50 $\pm$ 0.02 & 0.35 $\pm$ 0.01 &  0.16 $\pm$ 0.03 & 0.54 $\pm$ 0.01 & 0.44 $\pm$ 0.01 & 0.16 $\pm$ 0.02 & -0.11 $\pm$ 0.06 &  0.04 $\pm$ 0.01 &  0.23 $\pm$ 0.02 &  0.25 $\pm$ 0.03 &  0.04 $\pm$ 0.02 &  0.07 $\pm$ 0.05 &  0.29 $\pm$ 0.03 &  0.05 $\pm$ 0.03 &  0.32 $\pm$ 0.17 &  0.01 $\pm$ 0.18 & -0.39 $\pm$ 0.10 &  0.29 $\pm$ 0.28 \\
PGC018236 & 1.14 $\pm$ 0.01 &  0.07 $\pm$ 0.01 & 0.39 $\pm$ 0.02 & 0.25 $\pm$ 0.01 &  0.08 $\pm$ 0.03 & 0.44 $\pm$ 0.01 & 0.37 $\pm$ 0.01 & 0.04 $\pm$ 0.02 & -0.30 $\pm$ 0.02 & -0.02 $\pm$ 0.01 &  0.18 $\pm$ 0.02 &  0.20 $\pm$ 0.03 &  0.00 $\pm$ 0.02 & -0.13 $\pm$ 0.05 &  0.06 $\pm$ 0.03 & -0.03 $\pm$ 0.03 &  0.15 $\pm$ 0.18 & -0.07 $\pm$ 0.18 & -0.56 $\pm$ 0.05 &  0.30 $\pm$ 0.20 \\
PGC019085 & 1.01 $\pm$ 0.02 &  0.01 $\pm$ 0.01 & 0.43 $\pm$ 0.02 & 0.26 $\pm$ 0.01 &  0.13 $\pm$ 0.03 & 0.48 $\pm$ 0.01 & 0.34 $\pm$ 0.01 & 0.08 $\pm$ 0.02 & -0.23 $\pm$ 0.04 &  0.08 $\pm$ 0.01 &  0.13 $\pm$ 0.03 &  0.41 $\pm$ 0.03 & -0.00 $\pm$ 0.02 & -0.13 $\pm$ 0.06 &  0.20 $\pm$ 0.04 & -0.01 $\pm$ 0.03 & -0.00 $\pm$ 0.19 &  0.07 $\pm$ 0.22 & -0.38 $\pm$ 0.13 &  0.32 $\pm$ 0.22 \\
PGC043900 & 1.11 $\pm$ 0.01 &  0.07 $\pm$ 0.01 & 0.47 $\pm$ 0.02 & 0.29 $\pm$ 0.01 &  0.47 $\pm$ 0.02 & 0.62 $\pm$ 0.01 & 0.38 $\pm$ 0.01 & 0.19 $\pm$ 0.02 & -0.16 $\pm$ 0.06 &  0.11 $\pm$ 0.01 &  0.14 $\pm$ 0.02 & -0.06 $\pm$ 0.03 &  0.03 $\pm$ 0.02 & -0.15 $\pm$ 0.05 &  0.34 $\pm$ 0.03 &  0.22 $\pm$ 0.03 &  0.17 $\pm$ 0.18 &  0.16 $\pm$ 0.19 & -0.14 $\pm$ 0.14 &  0.07 $\pm$ 0.29 \\
PGC046785 & 1.13 $\pm$ 0.01 &  0.15 $\pm$ 0.02 & 0.48 $\pm$ 0.01 & 0.23 $\pm$ 0.01 &  0.35 $\pm$ 0.03 & 0.60 $\pm$ 0.02 & 0.35 $\pm$ 0.01 & 0.01 $\pm$ 0.03 & -0.31 $\pm$ 0.02 &  0.11 $\pm$ 0.01 &  0.18 $\pm$ 0.02 &  0.08 $\pm$ 0.03 &  0.00 $\pm$ 0.02 & -0.11 $\pm$ 0.07 &  0.16 $\pm$ 0.04 & -0.02 $\pm$ 0.03 &  0.26 $\pm$ 0.17 &  0.09 $\pm$ 0.22 & -0.35 $\pm$ 0.16 &  0.21 $\pm$ 0.27 \\
PGC046832 & 1.02 $\pm$ 0.04 &  0.13 $\pm$ 0.02 & 0.29 $\pm$ 0.03 & 0.21 $\pm$ 0.02 &  0.27 $\pm$ 0.04 & 0.39 $\pm$ 0.02 & 0.36 $\pm$ 0.02 & 0.04 $\pm$ 0.03 & -0.27 $\pm$ 0.03 &  0.06 $\pm$ 0.02 &  0.11 $\pm$ 0.03 &  0.05 $\pm$ 0.04 &  0.01 $\pm$ 0.02 & -0.02 $\pm$ 0.06 &  0.19 $\pm$ 0.04 & -0.12 $\pm$ 0.03 &  0.04 $\pm$ 0.21 &  0.16 $\pm$ 0.23 & -0.33 $\pm$ 0.15 &  0.38 $\pm$ 0.15 \\
PGC046860 & 1.11 $\pm$ 0.02 &  0.10 $\pm$ 0.01 & 0.44 $\pm$ 0.03 & 0.20 $\pm$ 0.01 &  0.23 $\pm$ 0.03 & 0.37 $\pm$ 0.03 & 0.28 $\pm$ 0.01 & 0.03 $\pm$ 0.03 & -0.23 $\pm$ 0.05 &  0.12 $\pm$ 0.02 & -0.11 $\pm$ 0.03 &  0.29 $\pm$ 0.03 &  0.01 $\pm$ 0.02 & -0.24 $\pm$ 0.03 & -0.02 $\pm$ 0.04 &  0.02 $\pm$ 0.03 &  0.37 $\pm$ 0.12 &  0.16 $\pm$ 0.23 & -0.07 $\pm$ 0.13 & -0.49 $\pm$ 0.11 \\
PGC047154 & 1.12 $\pm$ 0.01 &  0.18 $\pm$ 0.02 & 0.43 $\pm$ 0.02 & 0.25 $\pm$ 0.01 &  0.28 $\pm$ 0.03 & 0.67 $\pm$ 0.02 & 0.44 $\pm$ 0.01 & 0.11 $\pm$ 0.03 & -0.30 $\pm$ 0.05 &  0.05 $\pm$ 0.01 &  0.20 $\pm$ 0.03 &  0.21 $\pm$ 0.04 & -0.02 $\pm$ 0.03 & -0.20 $\pm$ 0.07 &  0.14 $\pm$ 0.05 &  0.03 $\pm$ 0.04 &  0.04 $\pm$ 0.22 &  0.04 $\pm$ 0.23 & -0.33 $\pm$ 0.18 &  0.02 $\pm$ 0.33 \\
PGC047177 & 0.88 $\pm$ 0.02 &  0.10 $\pm$ 0.01 & 0.41 $\pm$ 0.03 & 0.23 $\pm$ 0.01 &  0.21 $\pm$ 0.03 & 0.50 $\pm$ 0.04 & 0.37 $\pm$ 0.01 & 0.03 $\pm$ 0.03 & -0.25 $\pm$ 0.03 &  0.02 $\pm$ 0.01 &  0.12 $\pm$ 0.03 &  0.04 $\pm$ 0.04 &  0.00 $\pm$ 0.02 & -0.19 $\pm$ 0.05 &  0.19 $\pm$ 0.04 & -0.11 $\pm$ 0.03 &  0.31 $\pm$ 0.16 &  0.40 $\pm$ 0.15 & -0.35 $\pm$ 0.15 &  0.37 $\pm$ 0.20 \\
PGC047197 & 0.88 $\pm$ 0.02 &  0.09 $\pm$ 0.02 & 0.50 $\pm$ 0.03 & 0.32 $\pm$ 0.02 &  0.22 $\pm$ 0.03 & 0.54 $\pm$ 0.02 & 0.46 $\pm$ 0.02 & 0.11 $\pm$ 0.03 & -0.20 $\pm$ 0.05 &  0.11 $\pm$ 0.02 &  0.19 $\pm$ 0.03 &  0.19 $\pm$ 0.04 &  0.08 $\pm$ 0.02 & -0.15 $\pm$ 0.06 &  0.07 $\pm$ 0.05 &  0.01 $\pm$ 0.04 &  0.10 $\pm$ 0.21 &  0.30 $\pm$ 0.21 & -0.35 $\pm$ 0.14 &  0.15 $\pm$ 0.32 \\
PGC047202 & 1.02 $\pm$ 0.02 & -0.08 $\pm$ 0.01 & 0.26 $\pm$ 0.03 & 0.26 $\pm$ 0.01 &  0.48 $\pm$ 0.02 & 0.62 $\pm$ 0.02 & 0.41 $\pm$ 0.01 & 0.40 $\pm$ 0.02 &  0.24 $\pm$ 0.08 & -0.03 $\pm$ 0.01 &  0.29 $\pm$ 0.02 &  0.15 $\pm$ 0.03 & -0.04 $\pm$ 0.02 & -0.00 $\pm$ 0.05 &  0.27 $\pm$ 0.03 &  0.15 $\pm$ 0.03 &  0.31 $\pm$ 0.15 &  0.24 $\pm$ 0.16 &  0.23 $\pm$ 0.08 &  0.07 $\pm$ 0.25 \\
PGC047273 & 0.91 $\pm$ 0.02 &  0.05 $\pm$ 0.02 & 0.42 $\pm$ 0.03 & 0.26 $\pm$ 0.01 &  0.32 $\pm$ 0.03 & 0.47 $\pm$ 0.04 & 0.34 $\pm$ 0.01 & 0.09 $\pm$ 0.03 & -0.20 $\pm$ 0.04 &  0.09 $\pm$ 0.01 &  0.16 $\pm$ 0.03 &  0.03 $\pm$ 0.04 &  0.04 $\pm$ 0.02 & -0.11 $\pm$ 0.06 &  0.32 $\pm$ 0.04 & -0.04 $\pm$ 0.03 &  0.31 $\pm$ 0.18 &  0.31 $\pm$ 0.20 & -0.32 $\pm$ 0.14 &  0.26 $\pm$ 0.28 \\
PGC047355 & 1.08 $\pm$ 0.05 & -0.04 $\pm$ 0.02 & 0.37 $\pm$ 0.05 & 0.22 $\pm$ 0.01 &  0.22 $\pm$ 0.04 & 0.40 $\pm$ 0.02 & 0.31 $\pm$ 0.01 & 0.07 $\pm$ 0.03 & -0.23 $\pm$ 0.04 &  0.11 $\pm$ 0.01 &  0.11 $\pm$ 0.03 &  0.51 $\pm$ 0.04 &  0.03 $\pm$ 0.02 & -0.12 $\pm$ 0.05 &  0.18 $\pm$ 0.04 &  0.01 $\pm$ 0.03 & -0.06 $\pm$ 0.16 &  0.12 $\pm$ 0.23 & -0.47 $\pm$ 0.09 &  0.41 $\pm$ 0.17 \\
PGC047590 & 1.11 $\pm$ 0.02 &  0.03 $\pm$ 0.01 & 0.36 $\pm$ 0.02 & 0.25 $\pm$ 0.01 &  0.06 $\pm$ 0.03 & 0.40 $\pm$ 0.01 & 0.34 $\pm$ 0.01 & 0.04 $\pm$ 0.02 & -0.26 $\pm$ 0.03 &  0.02 $\pm$ 0.01 &  0.14 $\pm$ 0.02 &  0.35 $\pm$ 0.03 &  0.04 $\pm$ 0.02 & -0.27 $\pm$ 0.02 &  0.06 $\pm$ 0.03 &  0.01 $\pm$ 0.03 &  0.08 $\pm$ 0.17 &  0.30 $\pm$ 0.18 & -0.37 $\pm$ 0.14 &  0.08 $\pm$ 0.34 \\
PGC047752 & 1.13 $\pm$ 0.01 &  0.07 $\pm$ 0.02 & 0.52 $\pm$ 0.02 & 0.28 $\pm$ 0.02 &  0.28 $\pm$ 0.03 & 0.55 $\pm$ 0.02 & 0.34 $\pm$ 0.01 & 0.05 $\pm$ 0.03 & -0.22 $\pm$ 0.05 &  0.05 $\pm$ 0.02 &  0.10 $\pm$ 0.03 &  0.14 $\pm$ 0.03 &  0.03 $\pm$ 0.02 & -0.14 $\pm$ 0.06 &  0.21 $\pm$ 0.04 & -0.00 $\pm$ 0.04 &  0.13 $\pm$ 0.21 &  0.16 $\pm$ 0.24 & -0.21 $\pm$ 0.18 &  0.02 $\pm$ 0.35 \\
PGC048896 & 1.13 $\pm$ 0.01 &  0.03 $\pm$ 0.02 & 0.39 $\pm$ 0.05 & 0.25 $\pm$ 0.01 &  0.09 $\pm$ 0.03 & 0.54 $\pm$ 0.01 & 0.41 $\pm$ 0.01 & 0.12 $\pm$ 0.02 & -0.30 $\pm$ 0.02 &  0.00 $\pm$ 0.01 &  0.18 $\pm$ 0.03 &  0.13 $\pm$ 0.03 &  0.05 $\pm$ 0.02 & -0.15 $\pm$ 0.05 &  0.10 $\pm$ 0.03 & -0.02 $\pm$ 0.03 & -0.03 $\pm$ 0.18 & -0.14 $\pm$ 0.15 & -0.54 $\pm$ 0.08 &  0.13 $\pm$ 0.31 \\
PGC049940 & 1.14 $\pm$ 0.01 &  0.03 $\pm$ 0.02 & 0.36 $\pm$ 0.02 & 0.24 $\pm$ 0.01 &  0.27 $\pm$ 0.03 & 0.49 $\pm$ 0.01 & 0.34 $\pm$ 0.01 & 0.15 $\pm$ 0.03 & -0.07 $\pm$ 0.08 &  0.02 $\pm$ 0.01 &  0.14 $\pm$ 0.03 &  0.31 $\pm$ 0.03 &  0.03 $\pm$ 0.02 & -0.11 $\pm$ 0.05 &  0.23 $\pm$ 0.03 &  0.05 $\pm$ 0.03 &  0.42 $\pm$ 0.11 & -0.22 $\pm$ 0.05 & -0.52 $\pm$ 0.04 &  0.27 $\pm$ 0.26 \\
PGC065588 & 1.00 $\pm$ 0.02 & -0.03 $\pm$ 0.01 & 0.39 $\pm$ 0.02 & 0.27 $\pm$ 0.01 &  0.14 $\pm$ 0.02 & 0.44 $\pm$ 0.01 & 0.35 $\pm$ 0.01 & 0.10 $\pm$ 0.02 & -0.20 $\pm$ 0.02 &  0.01 $\pm$ 0.01 &  0.11 $\pm$ 0.02 &  0.22 $\pm$ 0.03 &  0.02 $\pm$ 0.01 & -0.10 $\pm$ 0.05 &  0.26 $\pm$ 0.03 & -0.03 $\pm$ 0.03 &  0.11 $\pm$ 0.14 &  0.31 $\pm$ 0.18 & -0.41 $\pm$ 0.09 &  0.49 $\pm$ 0.11 \\
PGC073000 & 0.77 $\pm$ 0.02 &  0.02 $\pm$ 0.01 & 0.46 $\pm$ 0.03 & 0.29 $\pm$ 0.01 &  0.27 $\pm$ 0.02 & 0.49 $\pm$ 0.01 & 0.45 $\pm$ 0.01 & 0.08 $\pm$ 0.02 & -0.12 $\pm$ 0.07 &  0.04 $\pm$ 0.01 &  0.20 $\pm$ 0.02 &  0.17 $\pm$ 0.03 &  0.06 $\pm$ 0.02 & -0.11 $\pm$ 0.05 &  0.33 $\pm$ 0.03 &  0.02 $\pm$ 0.03 &  0.11 $\pm$ 0.16 &  0.15 $\pm$ 0.18 & -0.32 $\pm$ 0.12 &  0.09 $\pm$ 0.31 \\
PGC097958 & 1.04 $\pm$ 0.08 &  0.09 $\pm$ 0.02 & 0.47 $\pm$ 0.03 & 0.22 $\pm$ 0.01 &  0.24 $\pm$ 0.04 & 0.45 $\pm$ 0.02 & 0.33 $\pm$ 0.02 & 0.00 $\pm$ 0.03 & -0.27 $\pm$ 0.02 &  0.05 $\pm$ 0.02 &  0.07 $\pm$ 0.03 &  0.03 $\pm$ 0.03 &  0.01 $\pm$ 0.02 & -0.11 $\pm$ 0.07 &  0.18 $\pm$ 0.04 & -0.07 $\pm$ 0.03 &  0.06 $\pm$ 0.19 & -0.07 $\pm$ 0.24 & -0.20 $\pm$ 0.16 &  0.16 $\pm$ 0.29 \\
PGC099188 & 0.94 $\pm$ 0.03 &  0.04 $\pm$ 0.02 & 0.32 $\pm$ 0.03 & 0.24 $\pm$ 0.01 &  0.01 $\pm$ 0.03 & 0.37 $\pm$ 0.01 & 0.37 $\pm$ 0.01 & 0.05 $\pm$ 0.02 & -0.25 $\pm$ 0.03 &  0.03 $\pm$ 0.01 &  0.23 $\pm$ 0.02 &  0.21 $\pm$ 0.03 &  0.04 $\pm$ 0.02 & -0.17 $\pm$ 0.05 &  0.11 $\pm$ 0.03 & -0.07 $\pm$ 0.03 &  0.02 $\pm$ 0.16 &  0.28 $\pm$ 0.20 & -0.48 $\pm$ 0.08 & -0.03 $\pm$ 0.35 \\
PGC099522 & 1.12 $\pm$ 0.02 & -0.05 $\pm$ 0.01 & 0.47 $\pm$ 0.02 & 0.25 $\pm$ 0.01 &  0.02 $\pm$ 0.02 & 0.37 $\pm$ 0.01 & 0.31 $\pm$ 0.01 & 0.08 $\pm$ 0.02 & -0.25 $\pm$ 0.06 &  0.05 $\pm$ 0.01 &  0.21 $\pm$ 0.02 &  0.12 $\pm$ 0.03 &  0.03 $\pm$ 0.01 & -0.15 $\pm$ 0.04 &  0.12 $\pm$ 0.03 & -0.04 $\pm$ 0.02 &  0.09 $\pm$ 0.15 & -0.25 $\pm$ 0.10 & -0.44 $\pm$ 0.14 &  0.32 $\pm$ 0.17 \\
\enddata

\end{deluxetable}


\section{\textsc{Chempy} model results}
\label{a:chm}

\begin{figure*}
  \includegraphics[angle=0, width=.99\textwidth]{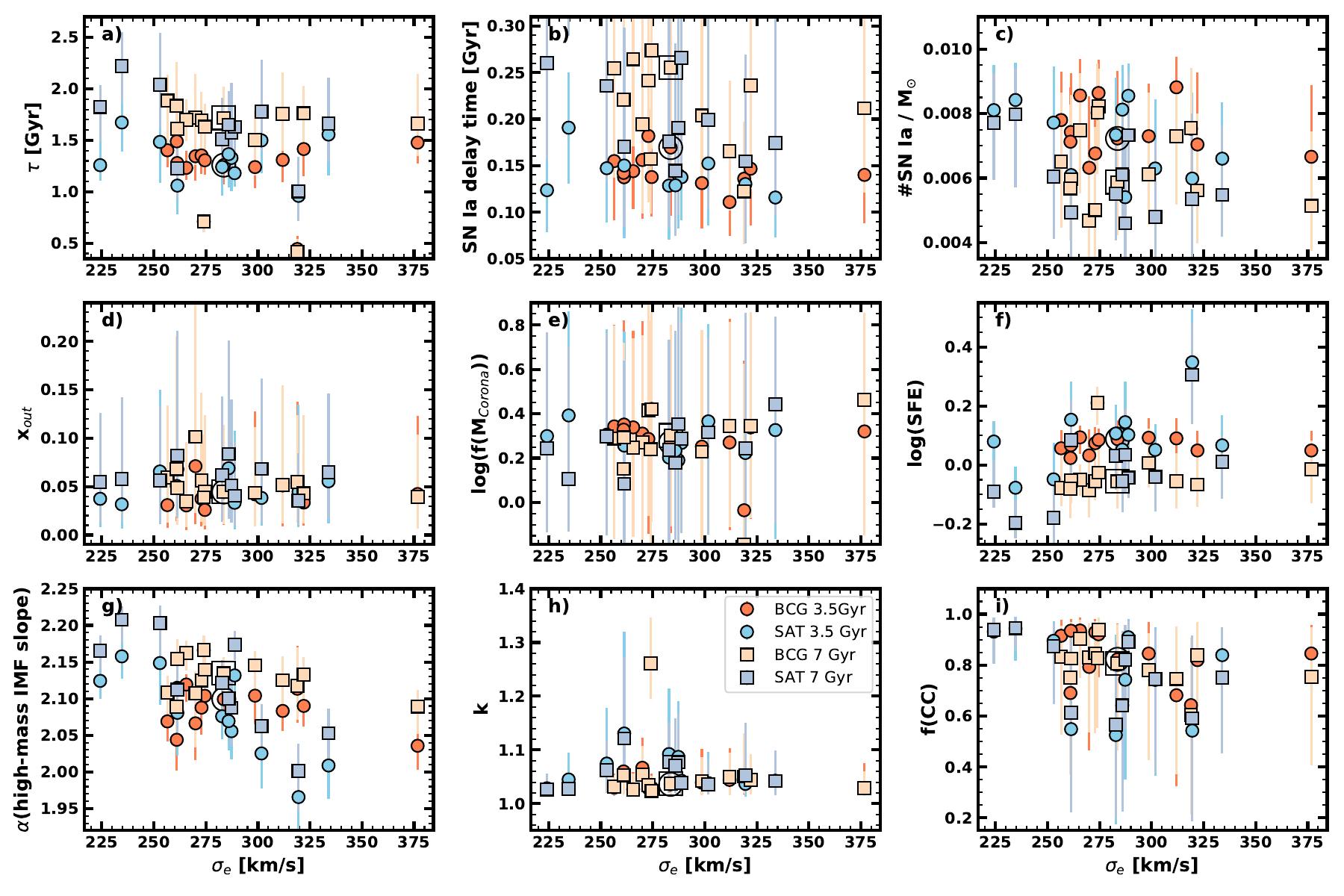}
 \caption{\textsc{Chempy} results as a function of velocity dispersion. We show the results of two chemical models, the nominal models with the chemical evolution lasting for 3.5 Gyr (circles) and a test model with the chemical evolution lasting 7 Gyr (squares). The youngest galaxy, PGC073000 is highlighted with a large circle and a square. From left to right, top to bottom, we show versus the stellar velocity dispersion within R$_{\rm eff}$: a) the star formation time scale $\tau$, b) the best-fit SN Ia delay time, c) the amount of SN Ia exploding over a period of 15 Gyr per solar mass, d) fraction of the outflowing enriched gas deposited in the corona instead of the star forming zone (note that 0.001 is a lower limit imposed through a prior), e) the mass fraction of the corona, f) the star formation efficiency, g) the power law index of the {\it high-mass} IMF, h) the shape parameter $k$ of the Gamma distribution, and i) the fraction of high-mass supernovae exploding as core-collapse supernova instead of as hypernova. BCGs and satellite galaxies are shown with red and blue symbols, respectively, as indicated on the legend of panel h).}\label{fig:chempy_vs_sig}
\end{figure*}

The stellar ages estimated by \textsc{alf} (Section~\ref{ss:abund}) are mostly old, with a few galaxies under 10 Gyr old, and a somewhat outlying PGC073000, which shows evidence for a rejuvenation episode. The question is if a chemical evolution model of a longer duration is more appropriate for these galaxies, and especially for PGC073000. We do not expect major differences because of two main reasons. Firstly, the chemical model is constrained by the elemental abundances and not the age. Specifically, by the abundances of O, Na, Mg and Fe, and our galaxies with younger ages do not stand out as outliers for these abundance. Secondly, our chemical model does not take into account rejuvenation episodes, either via an accretion accretion event or a merger. Nevertheless, we run a \textsc{chempy} model in the same set up as described in Section~\ref{ss:chempy}, including the same method of estimating the uncertainties, but now doubling the time of the simulation to 7 Gyr. The results of both, the 3.5 Gyr and 7 Gyr models, are show in Fig.~\ref{fig:chempy_vs_sig}. 

The overall impression is that of a rather good correspondence between the models, with a few minor difference for the duration of SF, the delay time of SN Ia, the number of SN Ia and the star formation efficiency. The duration of the SF is on average somewhat longer, with a mean age difference of about 0.3 Gyr. Similarly, the SN Ia delay times are also somewhat longer (for about 0.07 Gyr), and this seems to be driven mostly by BCGs, but the uncertainties are large and this result is not robust. The number of exploding SN Ia is somewhat smaller (but the average difference is only 0.001 SNIa/M$_{\odot}$), while the star formation efficient is also somewhat lower ($\sim 0.1$ dex). The observed changes suggest a compensation effect between the duration of SF and the star formation efficiency, as the shorter formation scales require higher SF efficiency. Similarly, the SN Ia delay time and the number of SN Ia explosions seem to be inversely correlated. Nevertheless, all these differences are within the estimated uncertainties. Specifically, as already stated in Section~\ref{ss:chem}, the trends in the high mass IMF slope also do not change, and the correlations between $\alpha_{\rm HM}$ and $\sigma_e$ remain the same, even though there are small differences in slope values for each galaxy, resulting in an offset of $\sim0.04$ towards steeper slopes, but still well below the Salpeter value. This difference is fully within the estimated uncertainty. For consistency, we list here the correlation coefficients for the $\alpha_{\rm HM} - \sigma$ anti-correlations for the 7 Gyr \textsc{chempy} run, which can be compared with the nominal results listed in Table~\ref{t:corr}. The Pearson correlation coefficient for the full sample is $\rho = -0.57^{0.25}_{-0.18}$, for BCGs $\rho = -0.34^{0.54}_{-0.38}$ and for SAT $\rho = -0.82^{0.12}_{-0.09}$, supporting the conclusion that the satellite galaxies drive the anti-correlation. 

PGC07300 was highlighted as the most significant outlier in age, which most likely experienced a rejuvenation episode. We highlight its positions in the panels of Fig.~\ref{fig:chempy_vs_sig} by a large circle and a large square. The galaxy does not show any atypical differences in the parameters from the two chemical models with respect to other galaxies in the sample. Overall, this allows us to conclude that the longer chemical model, while certainly being somewhat different, does not change any of the conclusion reached by our nominal (shorter) model. 

As a further test of the robustness of the recovered high-mass IMF slope, we also run two \textsc{chempy} models in which the outflow feedback fractions (x$_{out}$) were set to zero and to a value of 0.2. In practice, setting x$_{out}=0$ means that there is no return of the enriched gas to corona (the regions surrounding the galaxy) and the newly supplied gas (from corona) is always only the primordial gas, inflowing at the rate of the SFR. This results in two effects: all the enriched gas remains in ISM available for star formation, and newly inflowing gas has zero metallicity \citep{RybJusRix17}. Similarly, setting  setting x$_{out}=0.2$ determines the fraction of enriched gas that is removed from the ISM, and therefore lost for direct star formation. While this gas will return to the ISM from corona, its chemical composition will be first diluted with the primordial gas. The results of these chemical models (not shown) are actually similar to previous tests: there are certain changes to some parameters (i.e. the mass fraction of corona), but they are no more pronounced than the changes induced by longer duration of the chemical models. Crucially, the high-mass end IMF slope $\alpha_{\rm HM}$ changes very little. For the case of no outflow feedback fraction there is a minor overall steepening (for an average of 0.02), while the correlations between $\alpha_{\rm HM}$ and $\sigma_e$ remain the identical. The new Pearson correlations coefficients are: the full sample is $\rho = -0.57^{+0.31}_{-0.18}$, for BCGs $\rho = -0.28^{+0.69}_{-0.36}$ and for SAT $\rho = -0.86^{+0.11}_{-0.07}$. In the case of the fixed outflow feedback fraction, the slope actually decreases (for an average of 0.05), but the correlations remain the same. The new Pearson correlations coefficients are: the full sample is $\rho = -0.49^{+0.41}_{-0.26}$, for BCGs $\rho = -0.12^{+0.66}_{-0.41}$ and for SAT $\rho = -0.85^{+0.11}_{-0.08}$. We conclude that the main chemical modelling results pertaining to the high-mass end IMF slopes are robust.

\begin{table*}
   \caption{Parameters of the nominal \textsc{Chempy} model.}
\label{t:chempy}

\begin{tabular}{lccccccccc}
   \hline
    \noalign{\smallskip}

galaxy & tau & t$_{\rm SN Ia}$  &SN Ia &outflow &f$_{\rm Mcor}$  &$\log$ SFE &k &f(CC)\\

 & Gyr & Gyr &1/M$_odot$ & & & &&\\

(1)& (2)& (3)  & (4) &(5) &(6)& (7) & (8)& (9)\\
    \noalign{\smallskip} 
    \hline \hline
    \noalign{\smallskip}

PGC003342 & $1.34^{0.40}_{-0.24}$ & $0.156^{0.126}_{-0.063}$ & $0.0063^{0.0022}_{-0.0018}$ & $0.07^{0.10}_{-0.06}$ & $0.31^{0.51}_{-0.43}$ & $0.03^{0.09}_{-0.10}$ & $1.07^{0.06}_{-0.05}$ & $0.79^{0.16}_{-0.33}$ \\
PGC004500 & $1.40^{0.18}_{-0.17}$ & $0.155^{0.067}_{-0.064}$ & $0.0078^{0.0015}_{-0.0016}$ & $0.03^{0.06}_{-0.02}$ & $0.34^{0.46}_{-0.48}$ & $0.06^{0.06}_{-0.04}$ & $1.03^{0.03}_{-0.02}$ & $0.92^{0.06}_{-0.16}$ \\
PGC007748 & $1.23^{0.16}_{-0.12}$ & $0.144^{0.044}_{-0.041}$ & $0.0086^{0.0011}_{-0.0014}$ & $0.03^{0.06}_{-0.03}$ & $0.34^{0.43}_{-0.41}$ & $0.09^{0.04}_{-0.06}$ & $1.03^{0.03}_{-0.01}$ & $0.94^{0.05}_{-0.12}$ \\
PGC015524 & $1.35^{0.28}_{-0.18}$ & $0.182^{0.108}_{-0.071}$ & $0.0068^{0.0018}_{-0.0018}$ & $0.04^{0.06}_{-0.03}$ & $0.29^{0.52}_{-0.43}$ & $0.07^{0.06}_{-0.09}$ & $1.03^{0.04}_{-0.02}$ & $0.93^{0.05}_{-0.24}$ \\
PGC018236 & $1.31^{0.24}_{-0.14}$ & $0.138^{0.046}_{-0.043}$ & $0.0086^{0.0010}_{-0.0018}$ & $0.03^{0.05}_{-0.02}$ & $0.42^{0.40}_{-0.48}$ & $0.08^{0.04}_{-0.07}$ & $1.03^{0.03}_{-0.02}$ & $0.92^{0.06}_{-0.17}$ \\
PGC019085 & $1.28^{0.36}_{-0.26}$ & $0.137^{0.063}_{-0.061}$ & $0.0074^{0.0018}_{-0.0019}$ & $0.05^{0.09}_{-0.04}$ & $0.35^{0.44}_{-0.42}$ & $0.07^{0.11}_{-0.12}$ & $1.06^{0.05}_{-0.03}$ & $0.94^{0.05}_{-0.11}$ \\
PGC043900 & $1.48^{0.35}_{-0.20}$ & $0.140^{0.116}_{-0.052}$ & $0.0067^{0.0022}_{-0.0014}$ & $0.04^{0.08}_{-0.03}$ & $0.32^{0.47}_{-0.44}$ & $0.05^{0.07}_{-0.09}$ & $1.03^{0.05}_{-0.02}$ & $0.85^{0.11}_{-0.25}$ \\
PGC046785 & $1.56^{0.43}_{-0.40}$ & $0.116^{0.083}_{-0.043}$ & $0.0066^{0.0018}_{-0.0013}$ & $0.06^{0.09}_{-0.04}$ & $0.33^{0.51}_{-0.49}$ & $0.07^{0.10}_{-0.10}$ & $1.04^{0.05}_{-0.03}$ & $0.84^{0.11}_{-0.32}$ \\
PGC046832 & $1.31^{0.32}_{-0.21}$ & $0.111^{0.049}_{-0.036}$ & $0.0088^{0.0009}_{-0.0015}$ & $0.05^{0.07}_{-0.04}$ & $0.27^{0.56}_{-0.39}$ & $0.09^{0.07}_{-0.09}$ & $1.04^{0.06}_{-0.03}$ & $0.68^{0.25}_{-0.36}$ \\
PGC046860 & $1.24^{0.34}_{-0.28}$ & $0.129^{0.073}_{-0.058}$ & $0.0073^{0.0018}_{-0.0019}$ & $0.06^{0.07}_{-0.05}$ & $0.20^{0.53}_{-0.41}$ & $0.11^{0.10}_{-0.10}$ & $1.09^{0.12}_{-0.06}$ & $0.53^{0.39}_{-0.35}$ \\
PGC047154 & $0.96^{0.29}_{-0.21}$ & $0.130^{0.076}_{-0.063}$ & $0.0060^{0.0027}_{-0.0018}$ & $0.04^{0.09}_{-0.04}$ & $0.22^{0.55}_{-0.50}$ & $0.35^{0.18}_{-0.17}$ & $1.04^{0.05}_{-0.02}$ & $0.54^{0.32}_{-0.32}$ \\
PGC047177 & $1.33^{0.46}_{-0.32}$ & $0.140^{0.120}_{-0.058}$ & $0.0054^{0.0020}_{-0.0015}$ & $0.05^{0.07}_{-0.03}$ & $0.19^{0.63}_{-0.51}$ & $0.15^{0.14}_{-0.11}$ & $1.09^{0.10}_{-0.06}$ & $0.74^{0.21}_{-0.39}$ \\
PGC047197 & $1.50^{0.30}_{-0.32}$ & $0.152^{0.092}_{-0.067}$ & $0.0063^{0.0022}_{-0.0012}$ & $0.04^{0.07}_{-0.03}$ & $0.37^{0.44}_{-0.46}$ & $0.05^{0.09}_{-0.08}$ & $1.04^{0.06}_{-0.02}$ & $0.74^{0.21}_{-0.38}$ \\
PGC047202 & $0.44^{0.13}_{-0.06}$ & $0.136^{0.052}_{-0.056}$ & $0.0075^{0.0019}_{-0.0021}$ & $0.05^{0.08}_{-0.04}$ &-$0.04^{0.68}_{-0.55}$ & $0.85^{0.12}_{-0.20}$ & $1.05^{0.06}_{-0.03}$ & $0.64^{0.25}_{-0.38}$ \\
PGC047273 & $1.06^{0.32}_{-0.28}$ & $0.150^{0.061}_{-0.079}$ & $0.0061^{0.0020}_{-0.0016}$ & $0.07^{0.13}_{-0.05}$ & $0.26^{0.51}_{-0.52}$ & $0.15^{0.13}_{-0.11}$ & $1.13^{0.19}_{-0.09}$ & $0.55^{0.31}_{-0.33}$ \\
PGC047355 & $1.48^{0.46}_{-0.40}$ & $0.147^{0.073}_{-0.058}$ & $0.0077^{0.0017}_{-0.0022}$ & $0.07^{0.08}_{-0.05}$ & $0.31^{0.47}_{-0.44}$ &-$0.05^{0.12}_{-0.10}$ & $1.07^{0.10}_{-0.05}$ & $0.90^{0.08}_{-0.15}$ \\
PGC047590 & $1.18^{0.18}_{-0.13}$ & $0.137^{0.052}_{-0.046}$ & $0.0085^{0.0010}_{-0.0016}$ & $0.03^{0.07}_{-0.03}$ & $0.27^{0.61}_{-0.41}$ & $0.10^{0.05}_{-0.06}$ & $1.04^{0.05}_{-0.02}$ & $0.91^{0.07}_{-0.14}$ \\
PGC047752 & $1.49^{0.40}_{-0.31}$ & $0.142^{0.066}_{-0.058}$ & $0.0071^{0.0017}_{-0.0015}$ & $0.05^{0.08}_{-0.04}$ & $0.33^{0.49}_{-0.44}$ & $0.02^{0.09}_{-0.10}$ & $1.06^{0.07}_{-0.04}$ & $0.69^{0.26}_{-0.29}$ \\
PGC048896 & $1.42^{0.33}_{-0.26}$ & $0.147^{0.095}_{-0.061}$ & $0.0070^{0.0022}_{-0.0019}$ & $0.03^{0.07}_{-0.02}$ & $0.34^{0.46}_{-0.41}$ & $0.05^{0.07}_{-0.10}$ & $1.04^{0.05}_{-0.03}$ & $0.82^{0.14}_{-0.24}$ \\
PGC049940 & $1.24^{0.22}_{-0.20}$ & $0.131^{0.061}_{-0.049}$ & $0.0073^{0.0016}_{-0.0017}$ & $0.04^{0.09}_{-0.03}$ & $0.25^{0.52}_{-0.40}$ & $0.09^{0.07}_{-0.08}$ & $1.04^{0.06}_{-0.02}$ & $0.85^{0.12}_{-0.34}$ \\
PGC065588 & $0.71^{0.08}_{-0.11}$ & $0.157^{0.053}_{-0.059}$ & $0.0080^{0.0015}_{-0.0019}$ & $0.04^{0.06}_{-0.03}$ & $0.24^{0.58}_{-0.44}$ & $0.21^{0.05}_{-0.07}$ & $1.26^{0.09}_{-0.07}$ & $0.83^{0.13}_{-0.27}$ \\
PGC073000 & $1.26^{0.18}_{-0.19}$ & $0.169^{0.071}_{-0.073}$ & $0.0072^{0.0021}_{-0.0015}$ & $0.04^{0.07}_{-0.03}$ & $0.27^{0.50}_{-0.41}$ & $0.09^{0.06}_{-0.06}$ & $1.04^{0.04}_{-0.02}$ & $0.82^{0.14}_{-0.32}$ \\
PGC097958 & $1.37^{0.36}_{-0.32}$ & $0.129^{0.054}_{-0.050}$ & $0.0081^{0.0014}_{-0.0017}$ & $0.07^{0.09}_{-0.05}$ & $0.23^{0.49}_{-0.40}$ & $0.05^{0.11}_{-0.09}$ & $1.07^{0.07}_{-0.05}$ & $0.64^{0.28}_{-0.36}$ \\
PGC099188 & $1.26^{0.31}_{-0.15}$ & $0.124^{0.049}_{-0.045}$ & $0.0081^{0.0014}_{-0.0017}$ & $0.04^{0.07}_{-0.03}$ & $0.30^{0.46}_{-0.40}$ & $0.08^{0.07}_{-0.09}$ & $1.03^{0.03}_{-0.01}$ & $0.93^{0.05}_{-0.10}$ \\
PGC099522 & $1.67^{0.31}_{-0.28}$ & $0.191^{0.060}_{-0.060}$ & $0.0084^{0.0012}_{-0.0014}$ & $0.03^{0.04}_{-0.03}$ & $0.39^{0.47}_{-0.47}$ &-$0.08^{0.07}_{-0.07}$ & $1.05^{0.05}_{-0.02}$ & $0.94^{0.04}_{-0.13}$ \\
    \noalign{\smallskip}
    \hline
\end{tabular}

{Notes: Column (1): name of the galaxy; Column (2): star formation time scale. Column (3): best fit SN Ia delay time. Column (4): amount of SN Ia exploding over a period of 15 Gyr per solar mass.  Column (5): fraction of the outflowing enriched gas deposited in the corona instead of the star forming zone (0.001 is a lower limit imposed through a prior). Column (6):  the mass fraction of the corona. Column (7): the star formation efficiency. Column (8): the shape parameter k of the Gamma distribution. Column (9): the fraction of high-mass supernovae exploding as core-collapse supernova instead of as hypernova.}
\end{table*}


\section{\textsc{alf} spectral fits}

\begin{figure*}
  \includegraphics[angle=0, width=0.22\textwidth]{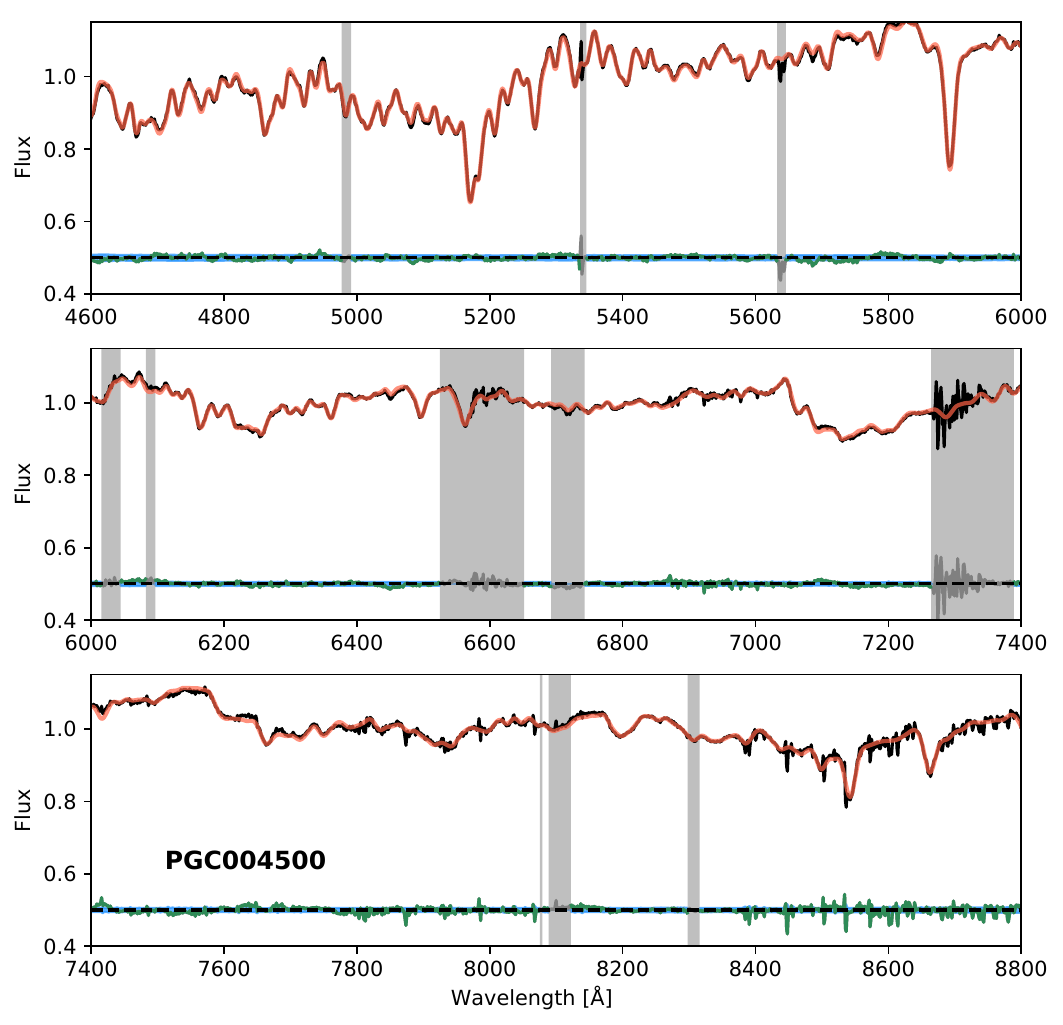}
  \includegraphics[angle=0, width=0.22\textwidth]{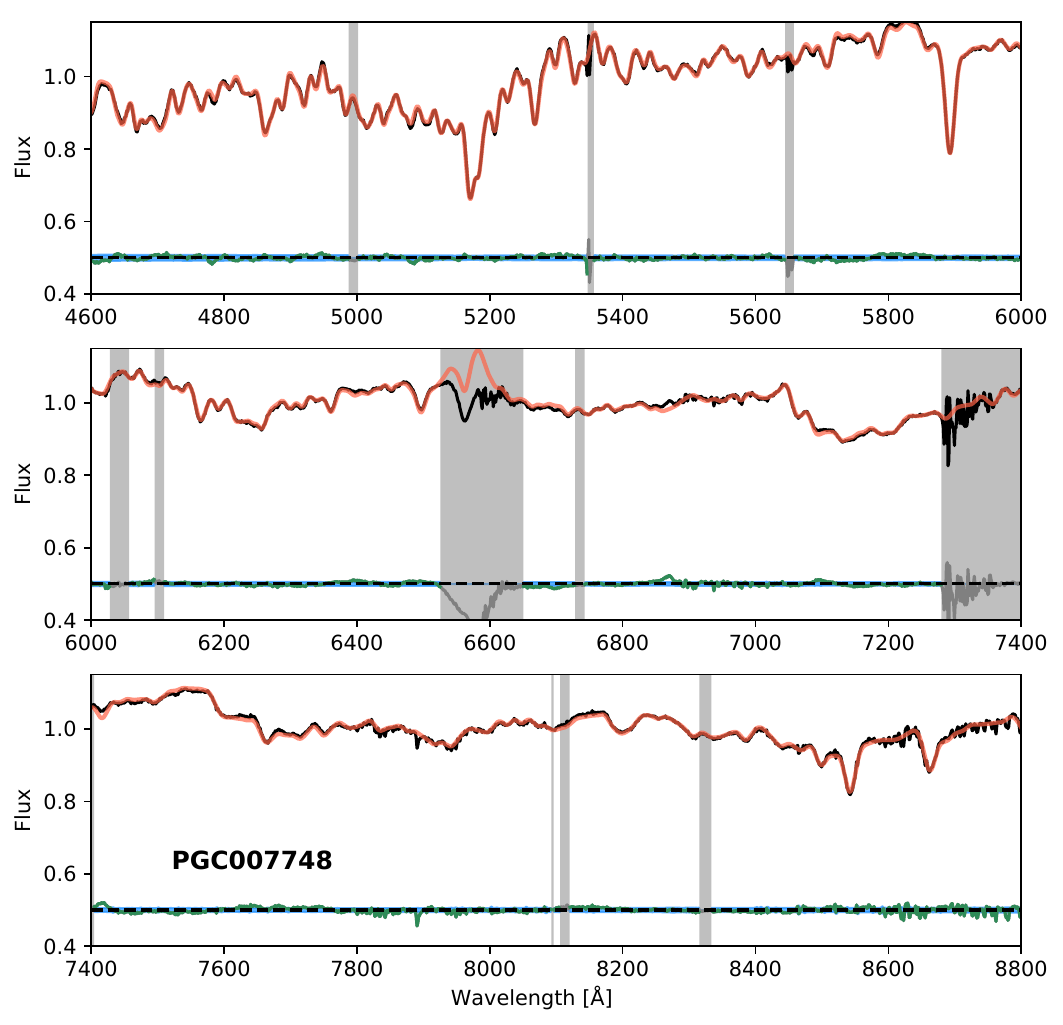}
  \includegraphics[angle=0, width=0.22\textwidth]{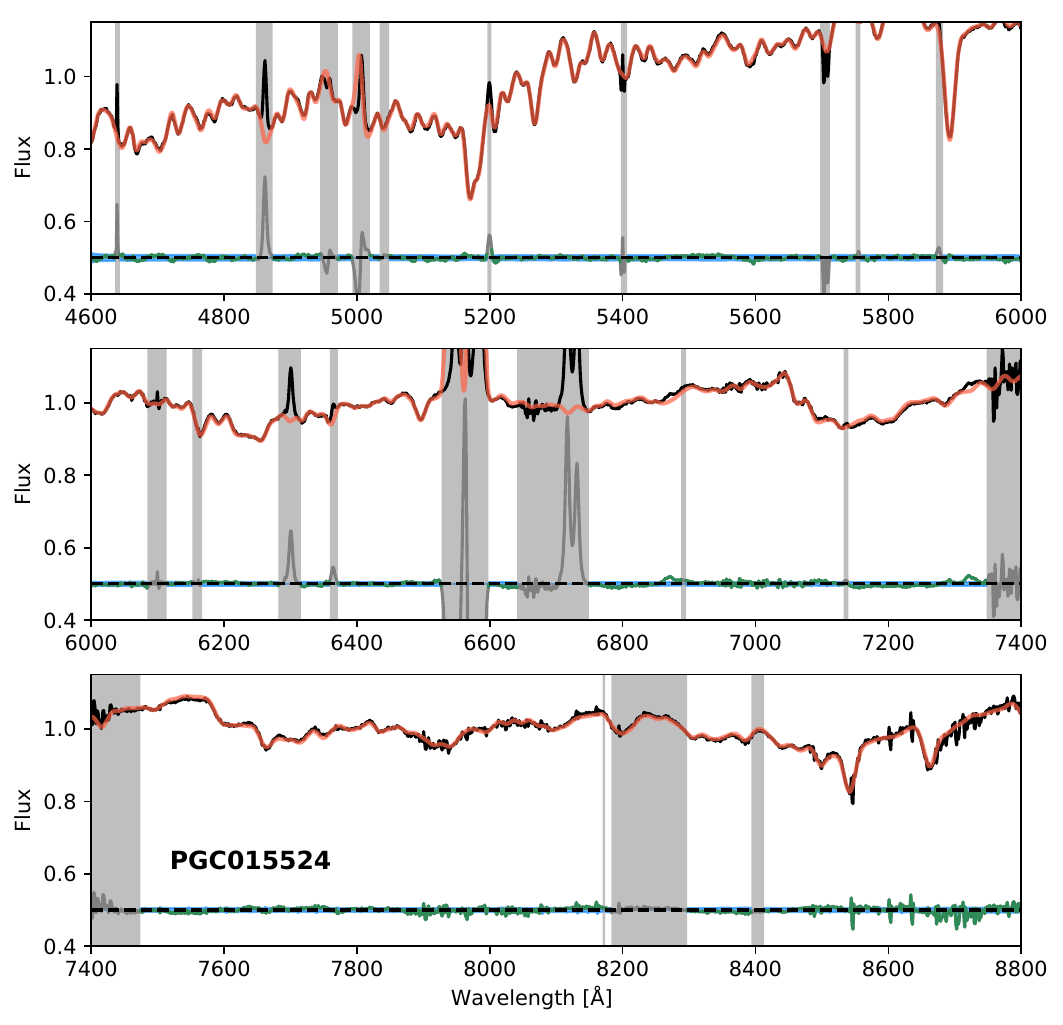}
  \includegraphics[angle=0, width=0.22\textwidth]{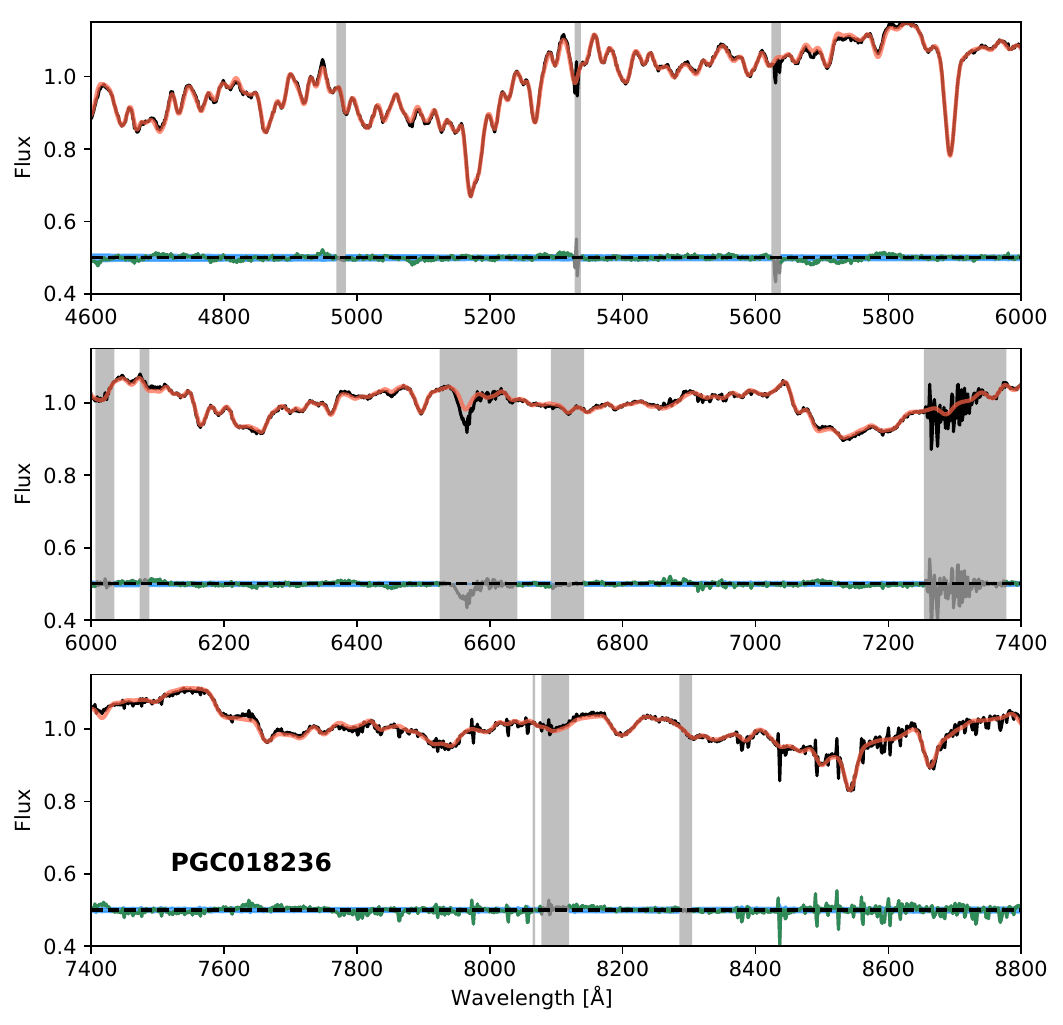}
  \includegraphics[angle=0, width=0.22\textwidth]{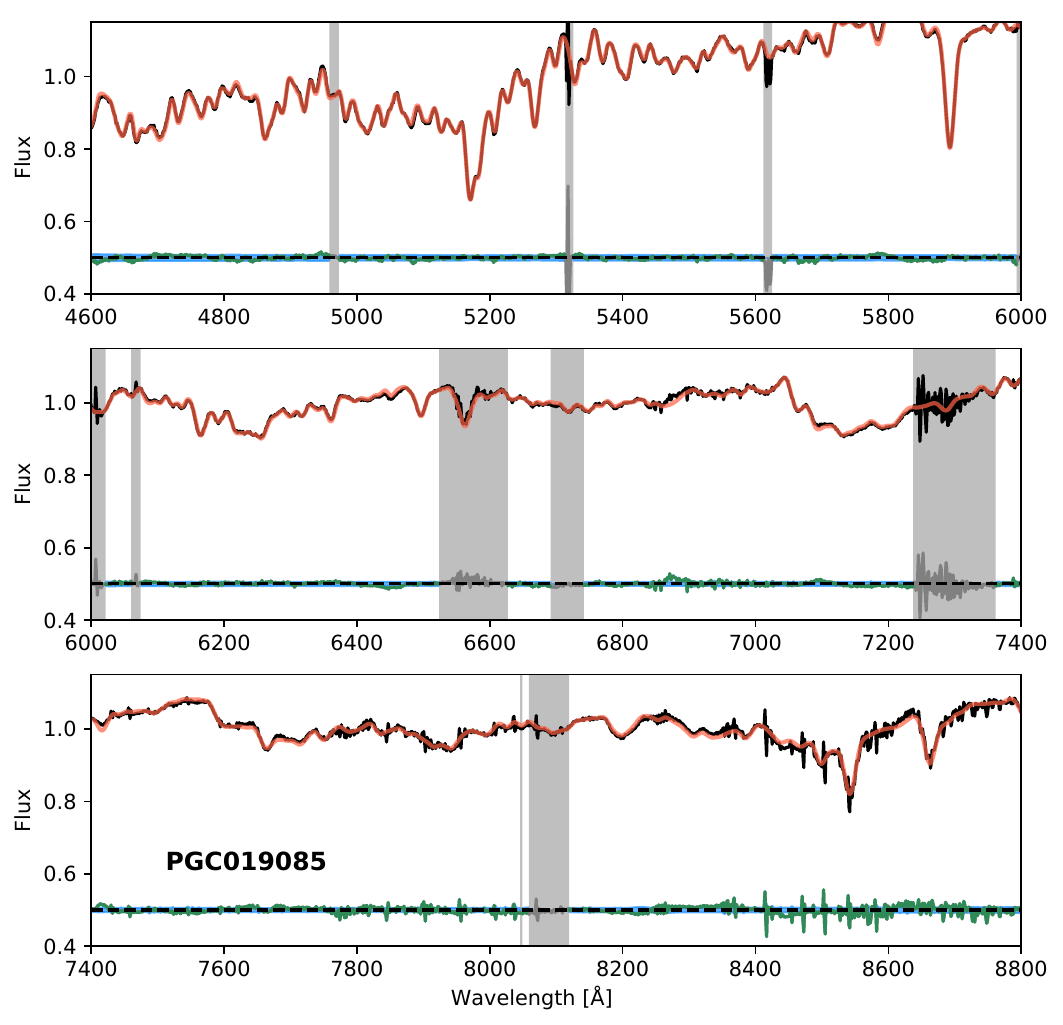}
  \includegraphics[angle=0, width=0.22\textwidth]{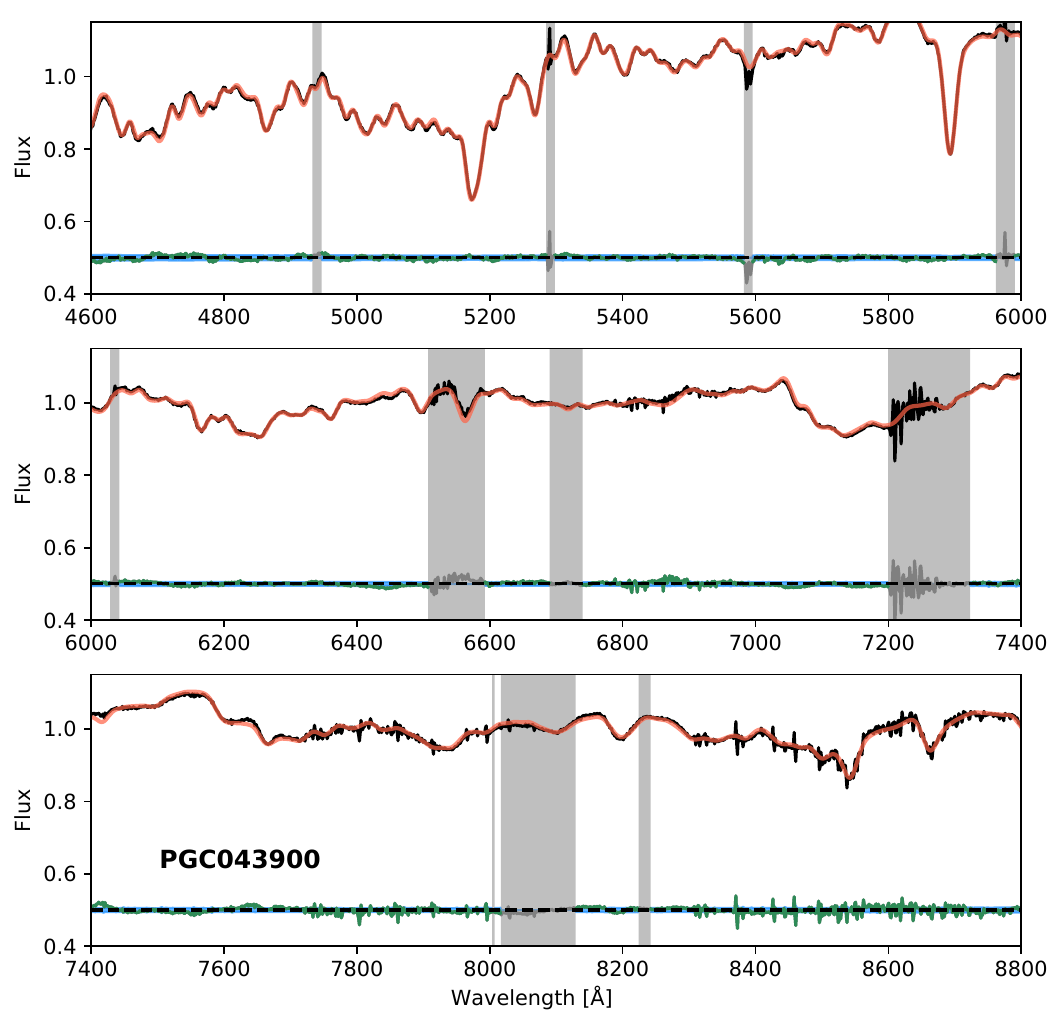}
  \includegraphics[angle=0, width=0.22\textwidth]{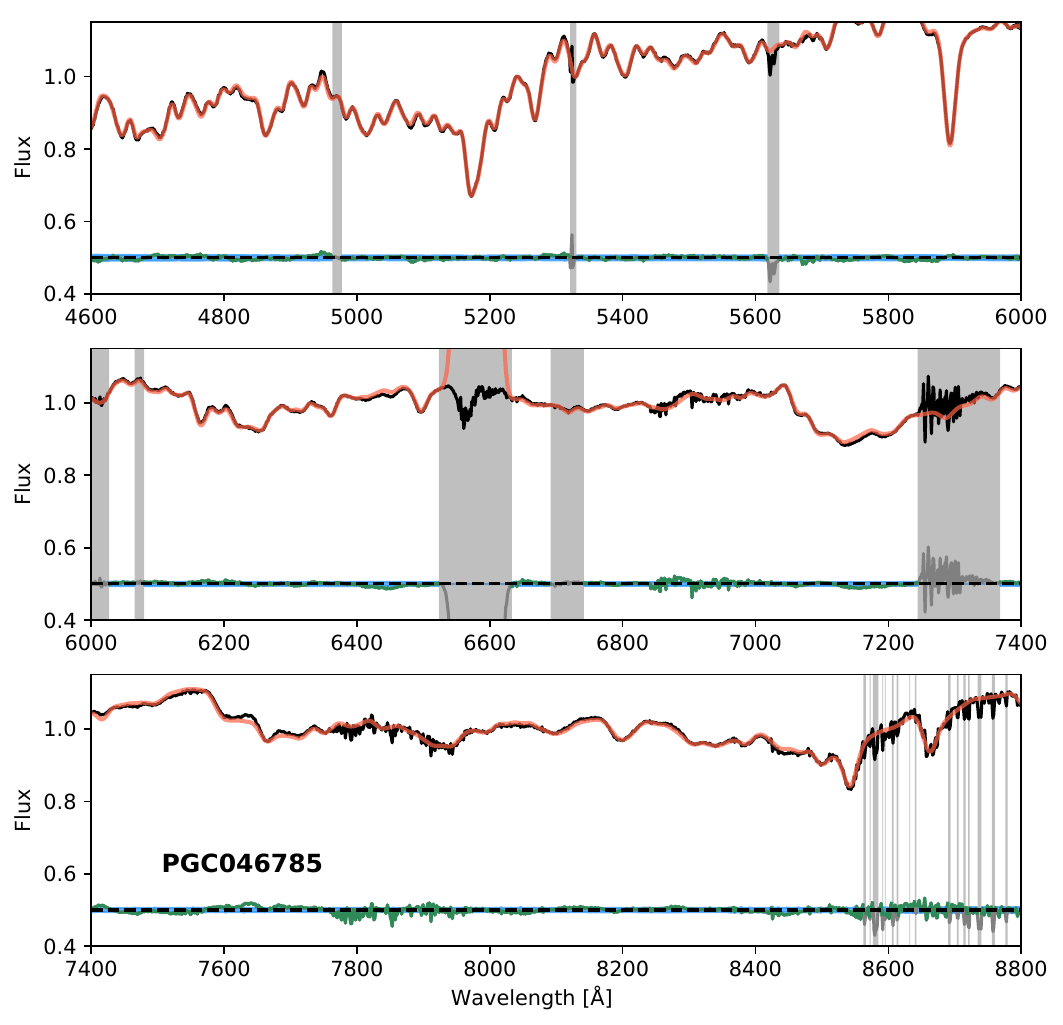}
  \includegraphics[angle=0, width=0.22\textwidth]{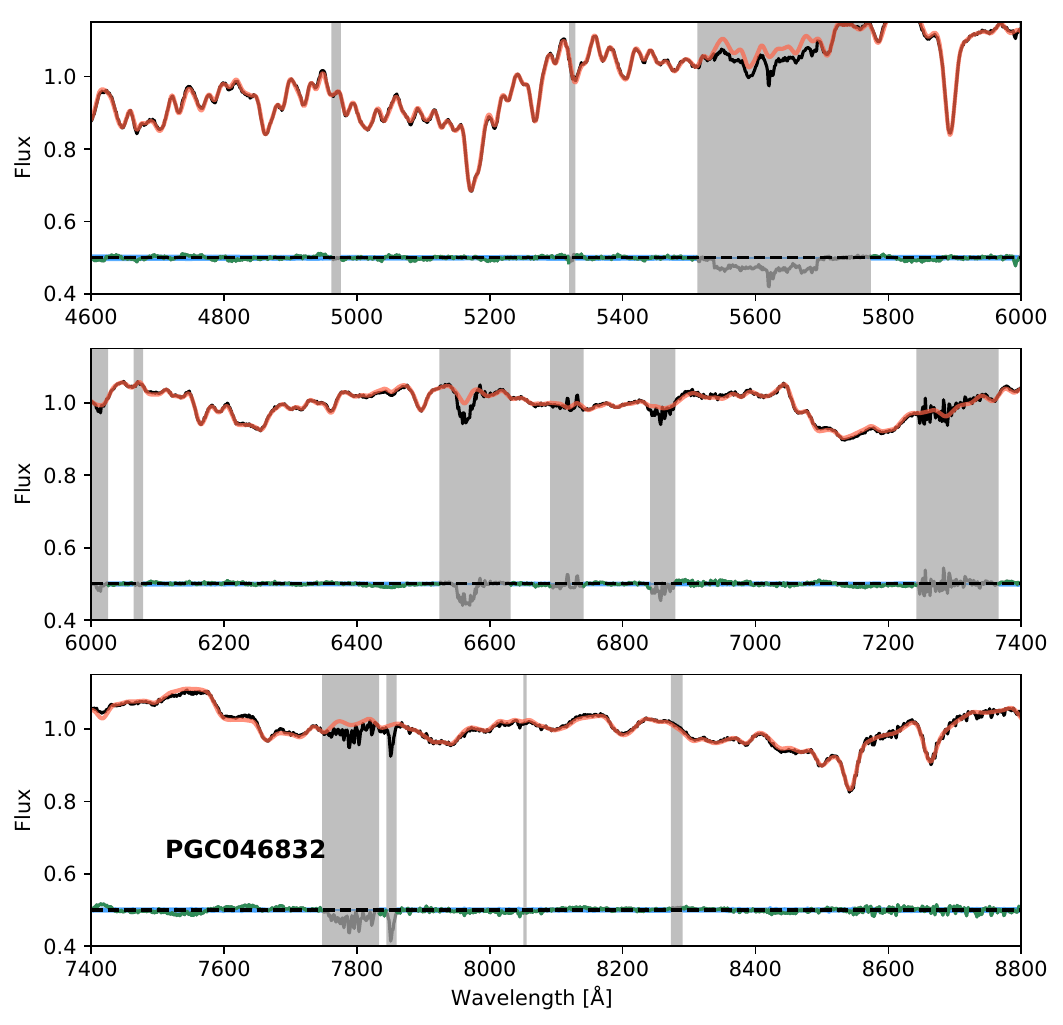}
  \includegraphics[angle=0, width=0.22\textwidth]{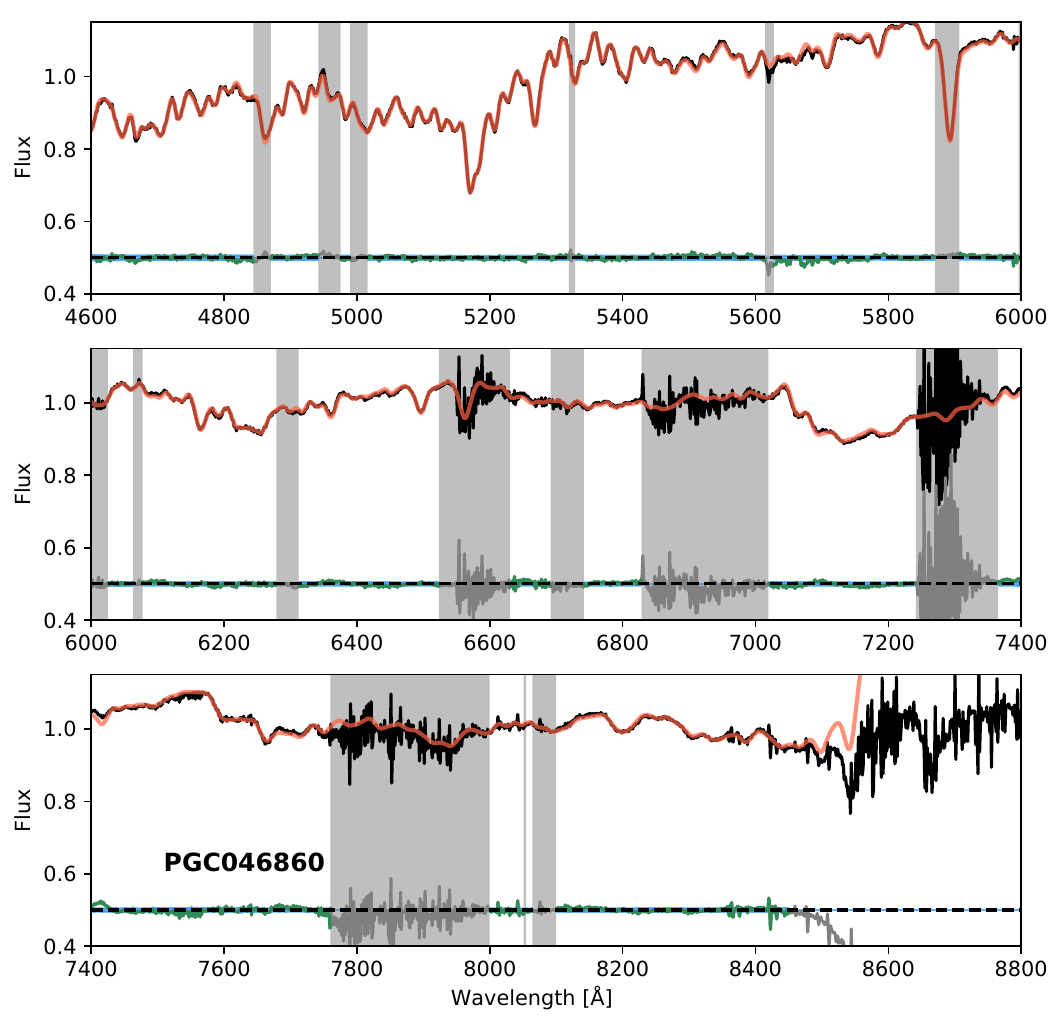}
  \includegraphics[angle=0, width=0.22\textwidth]{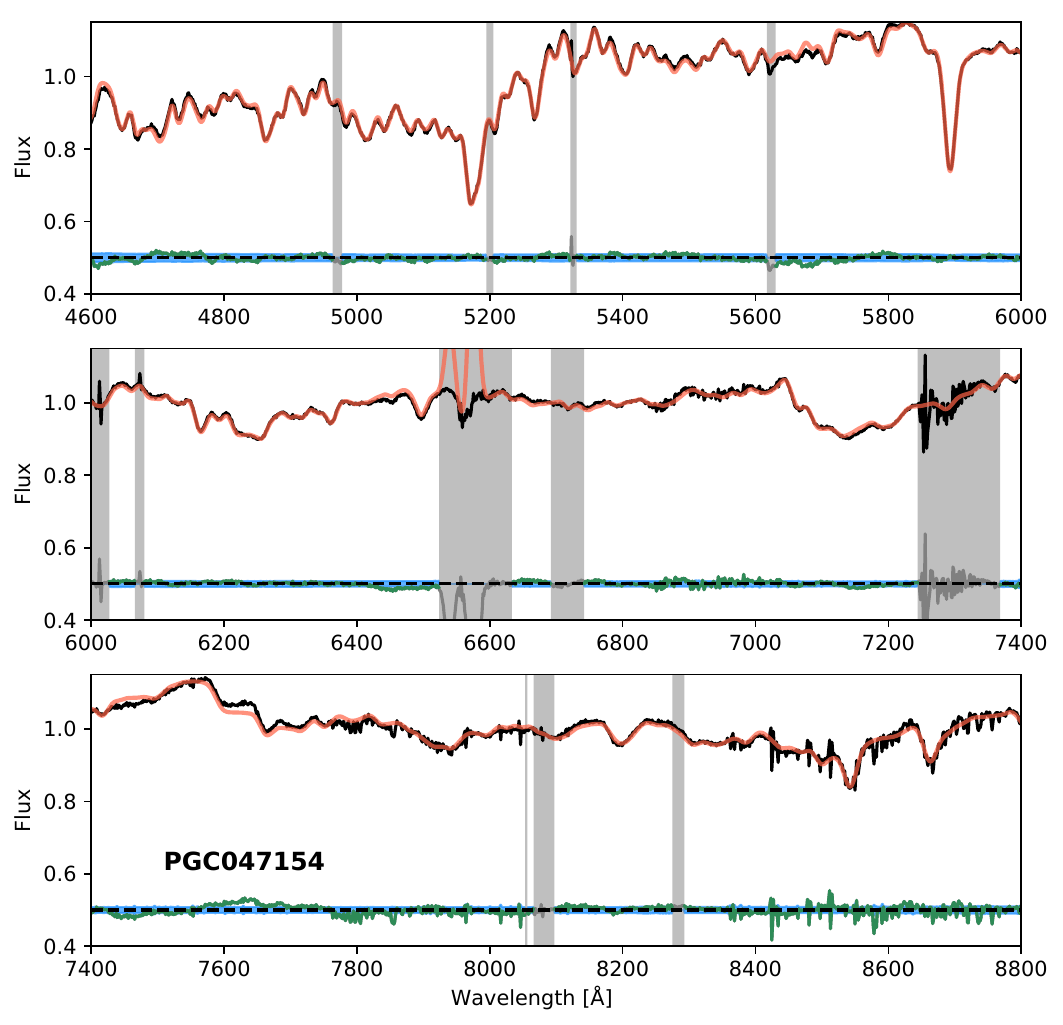}
  \includegraphics[angle=0, width=0.22\textwidth]{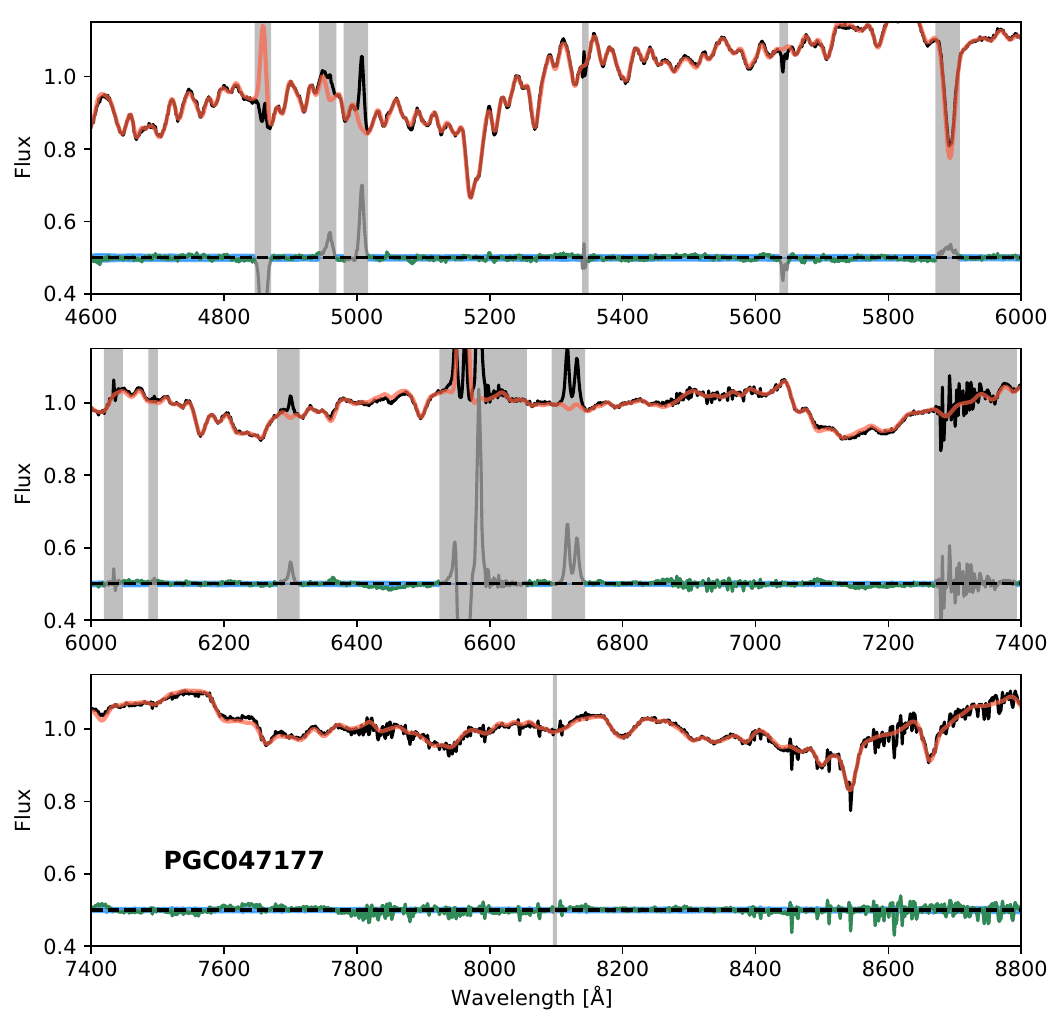}
  \includegraphics[angle=0, width=0.22\textwidth]{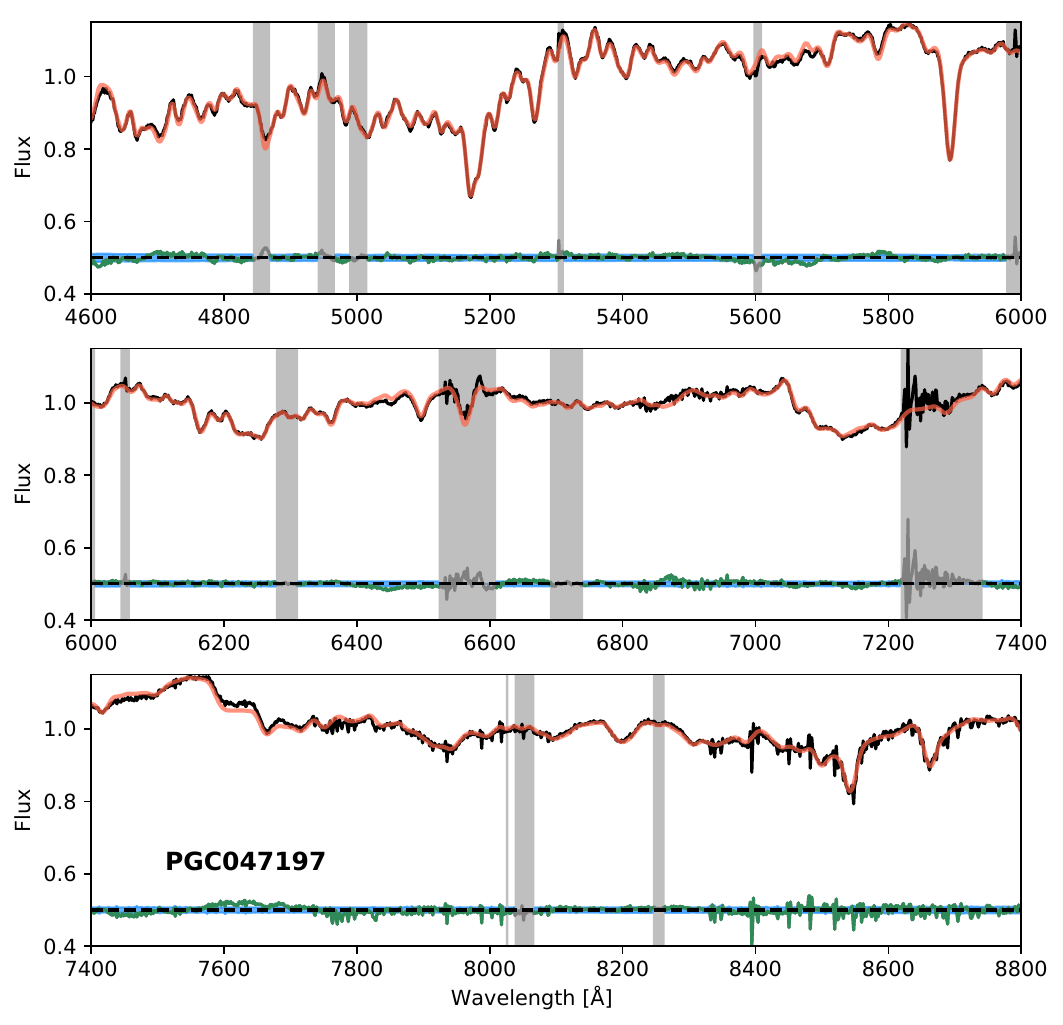}
  \includegraphics[angle=0, width=0.22\textwidth]{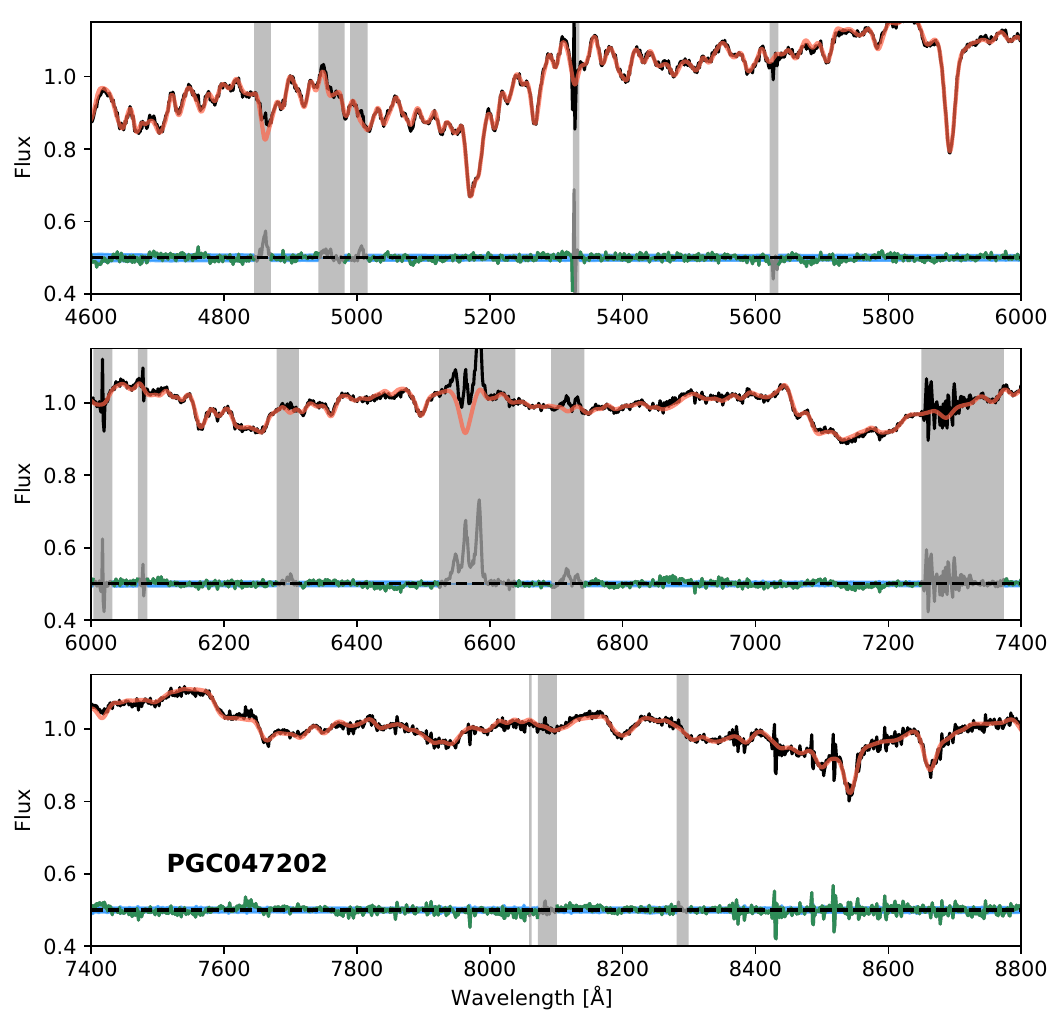}
  \includegraphics[angle=0, width=0.22\textwidth]{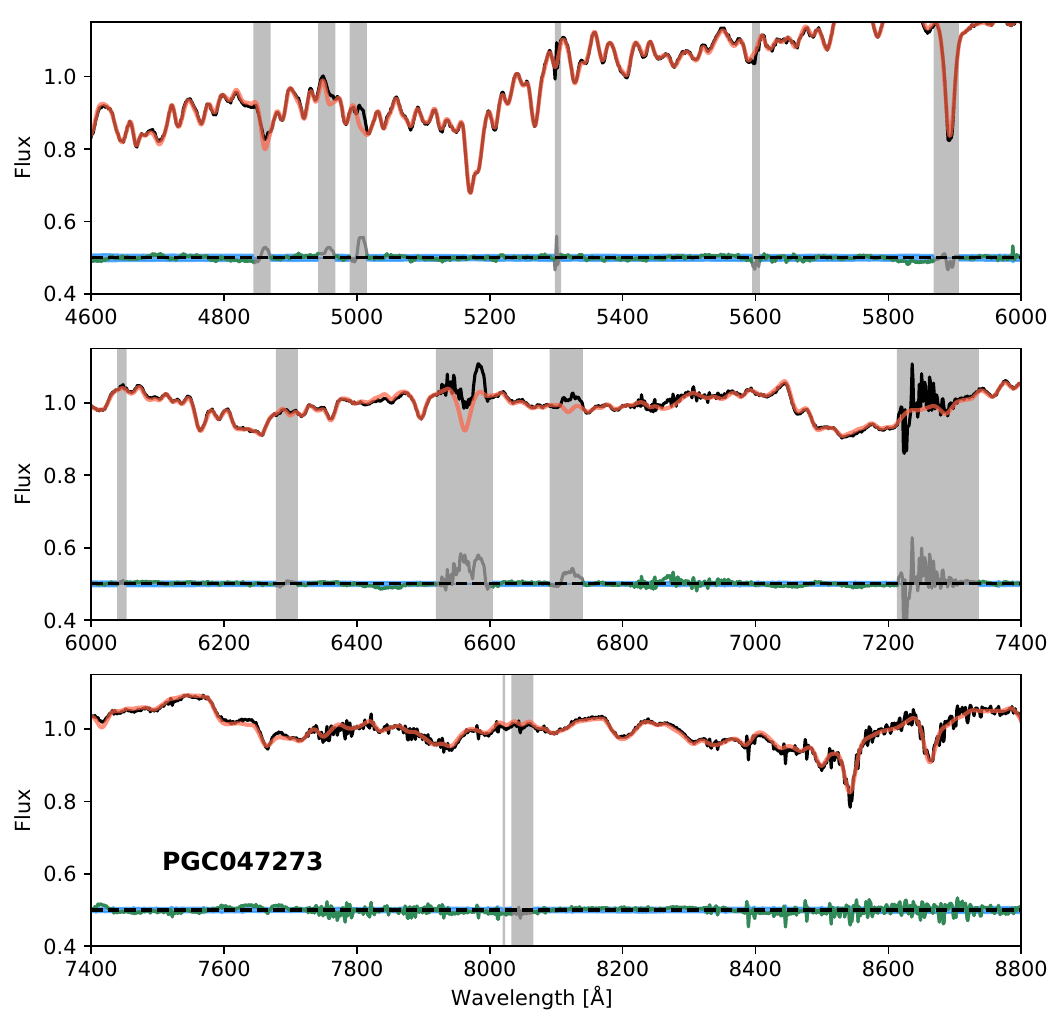}
  \includegraphics[angle=0, width=0.22\textwidth]{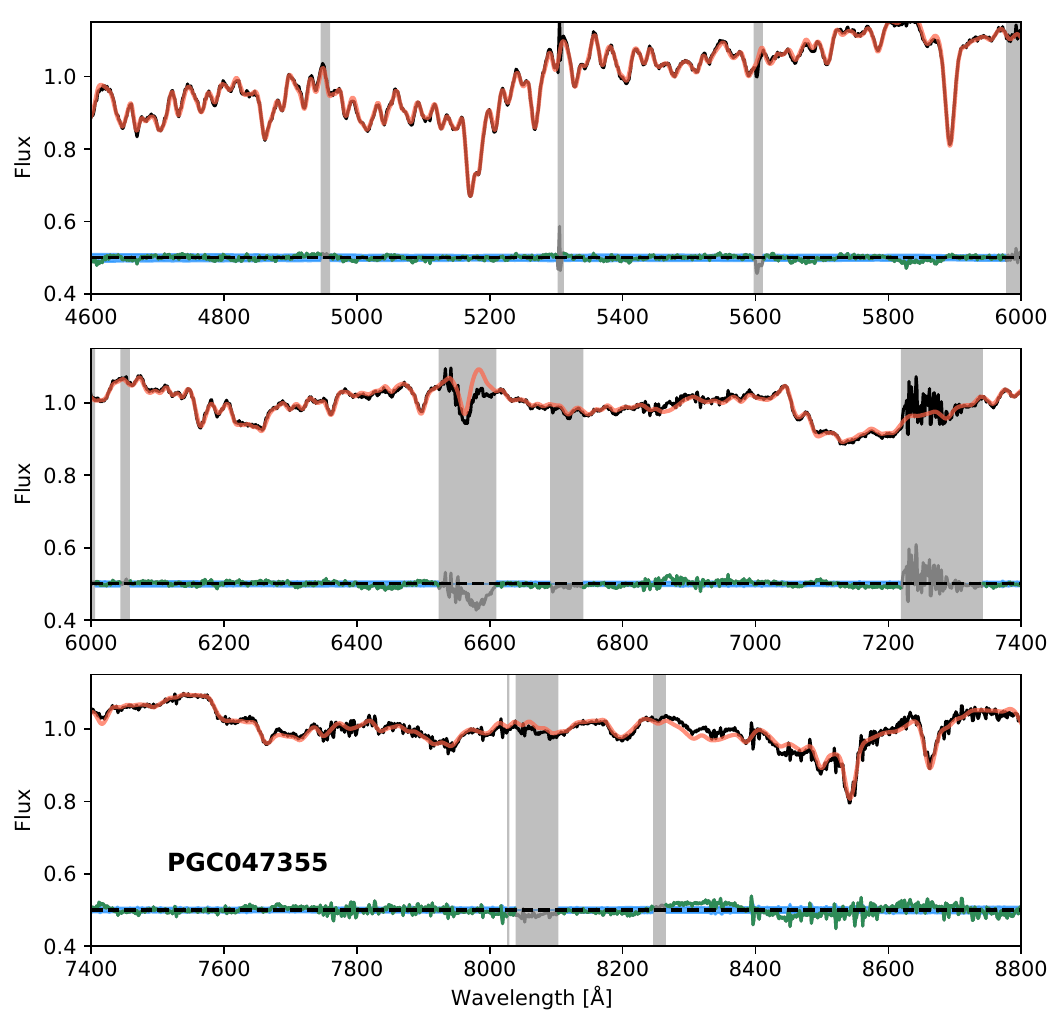}
  \includegraphics[angle=0, width=0.22\textwidth]{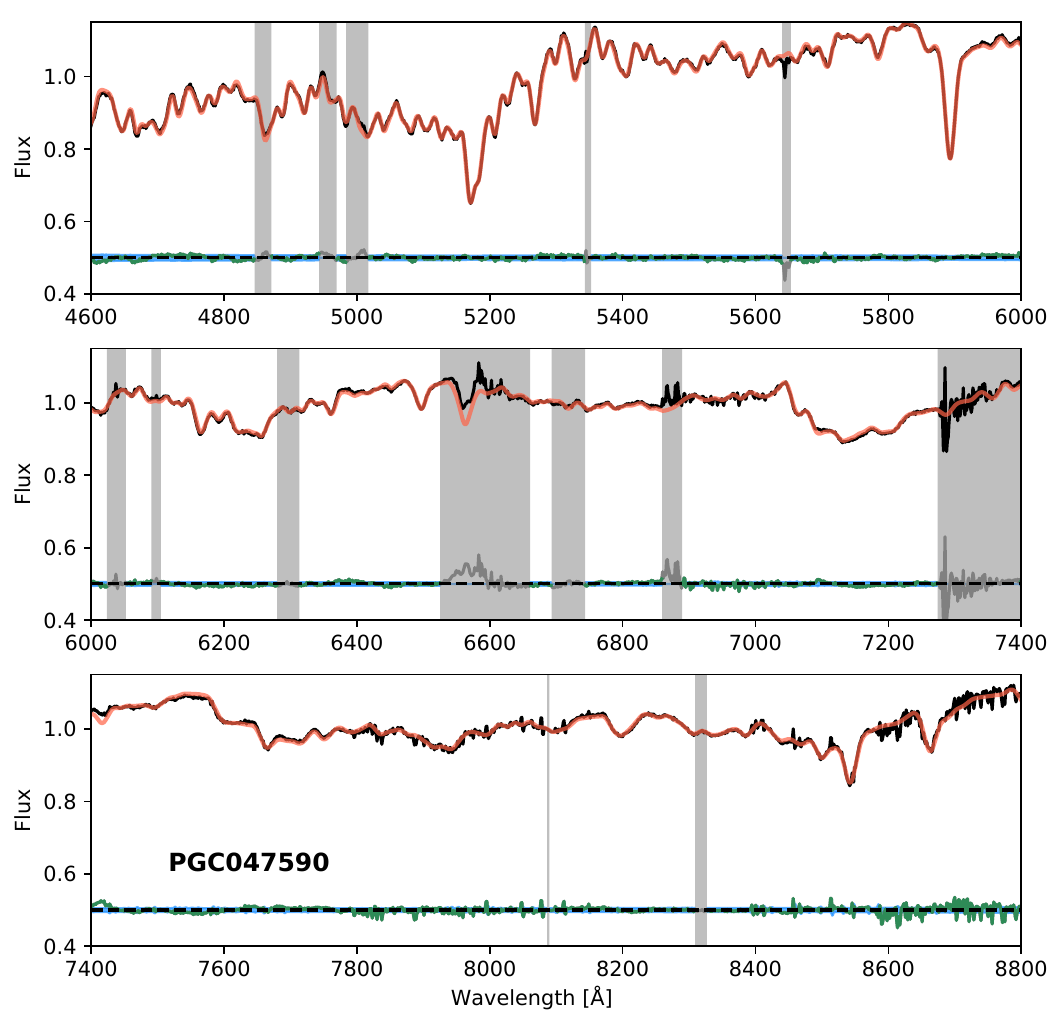}
  \includegraphics[angle=0, width=0.22\textwidth]{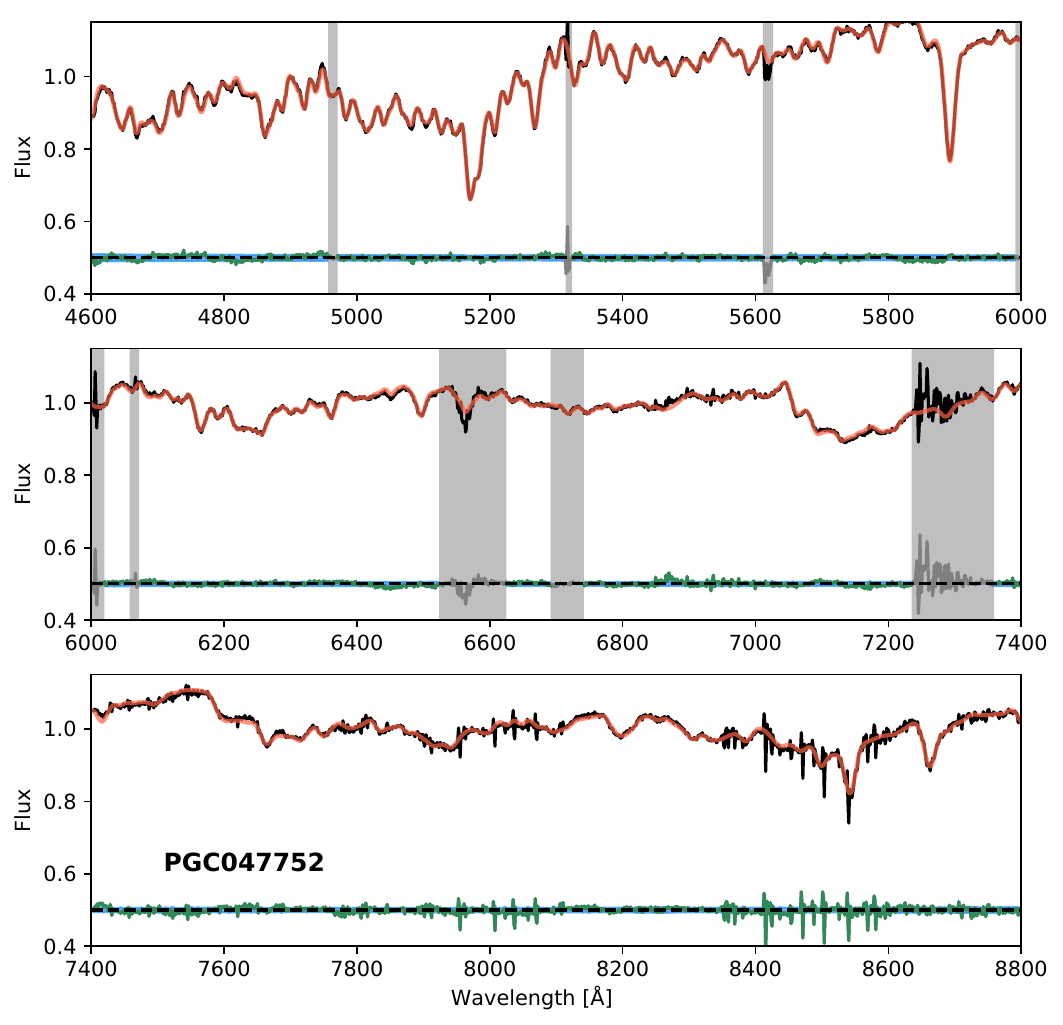}
  \includegraphics[angle=0, width=0.22\textwidth]{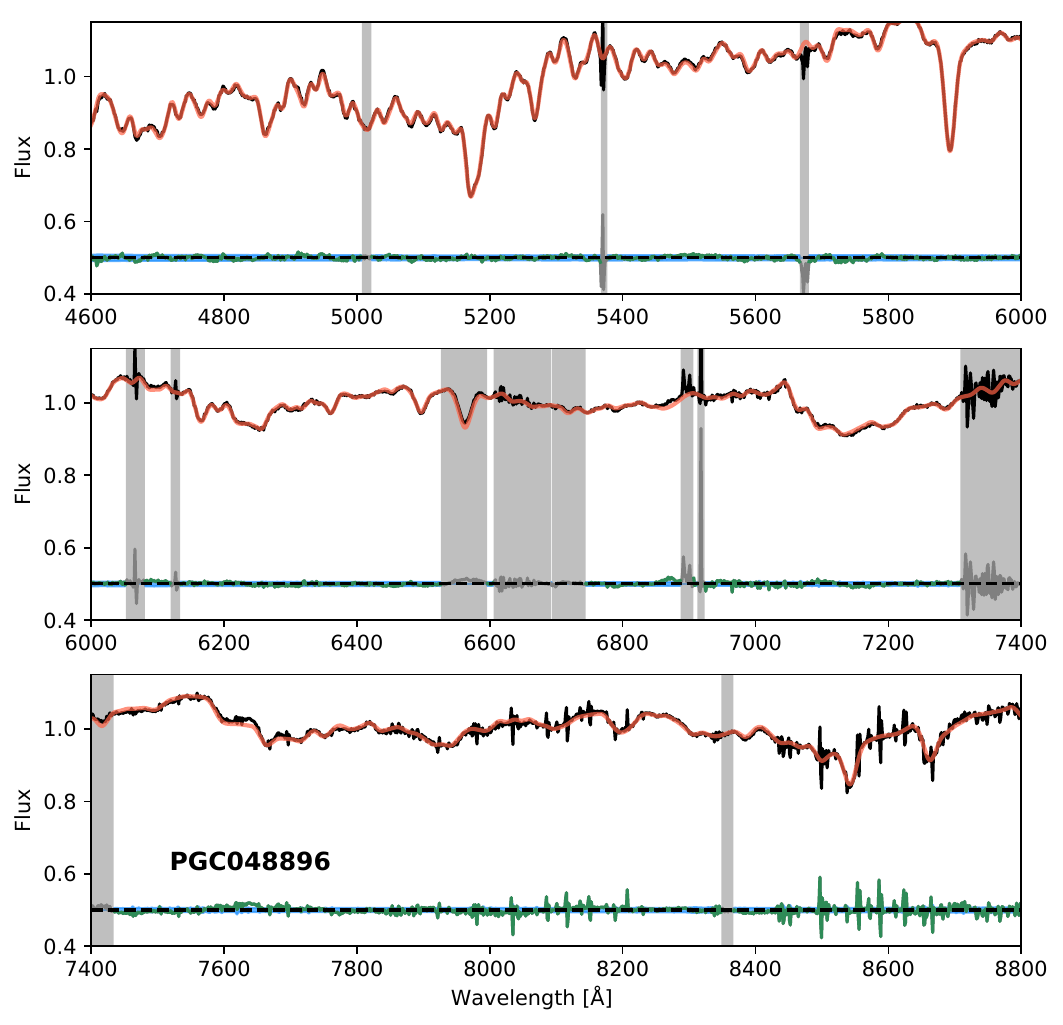}
  \includegraphics[angle=0, width=0.22\textwidth]{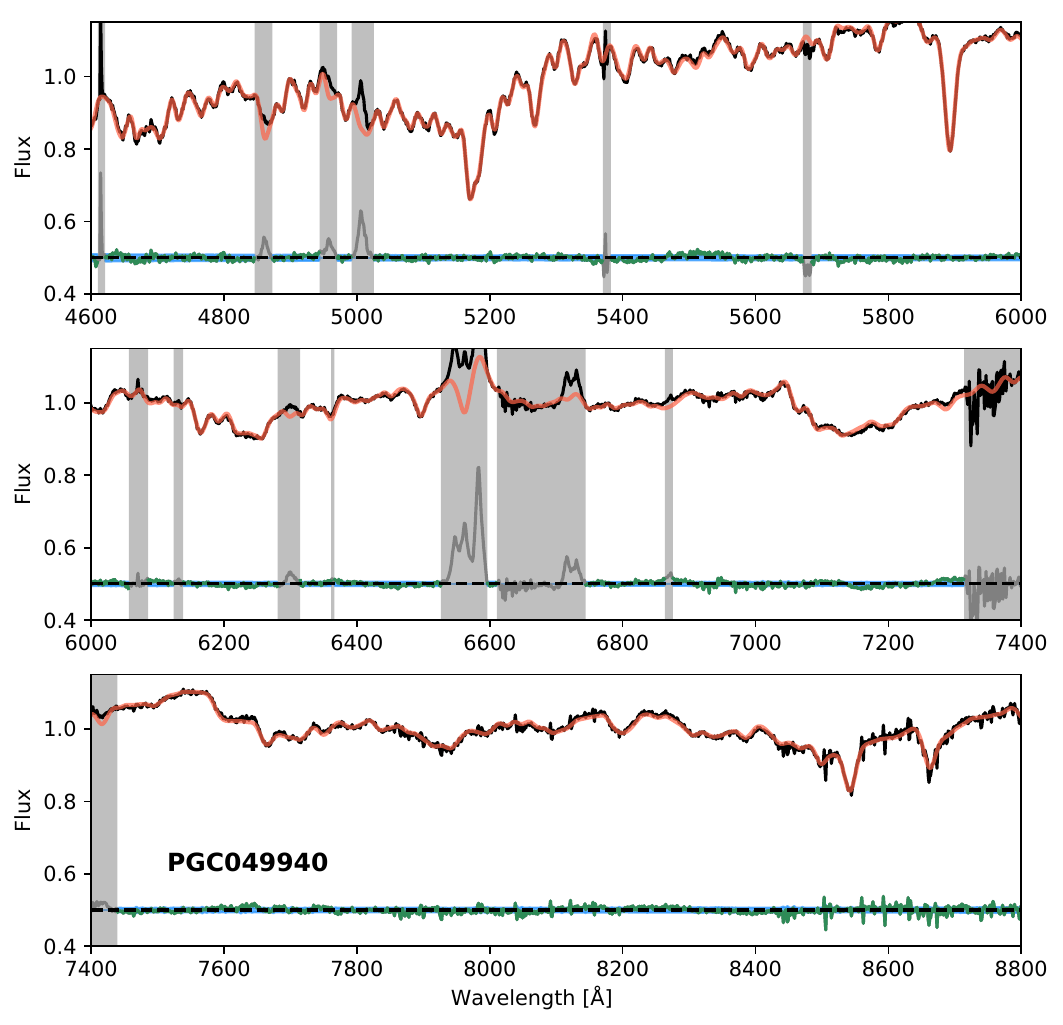}
  \includegraphics[angle=0, width=0.22\textwidth]{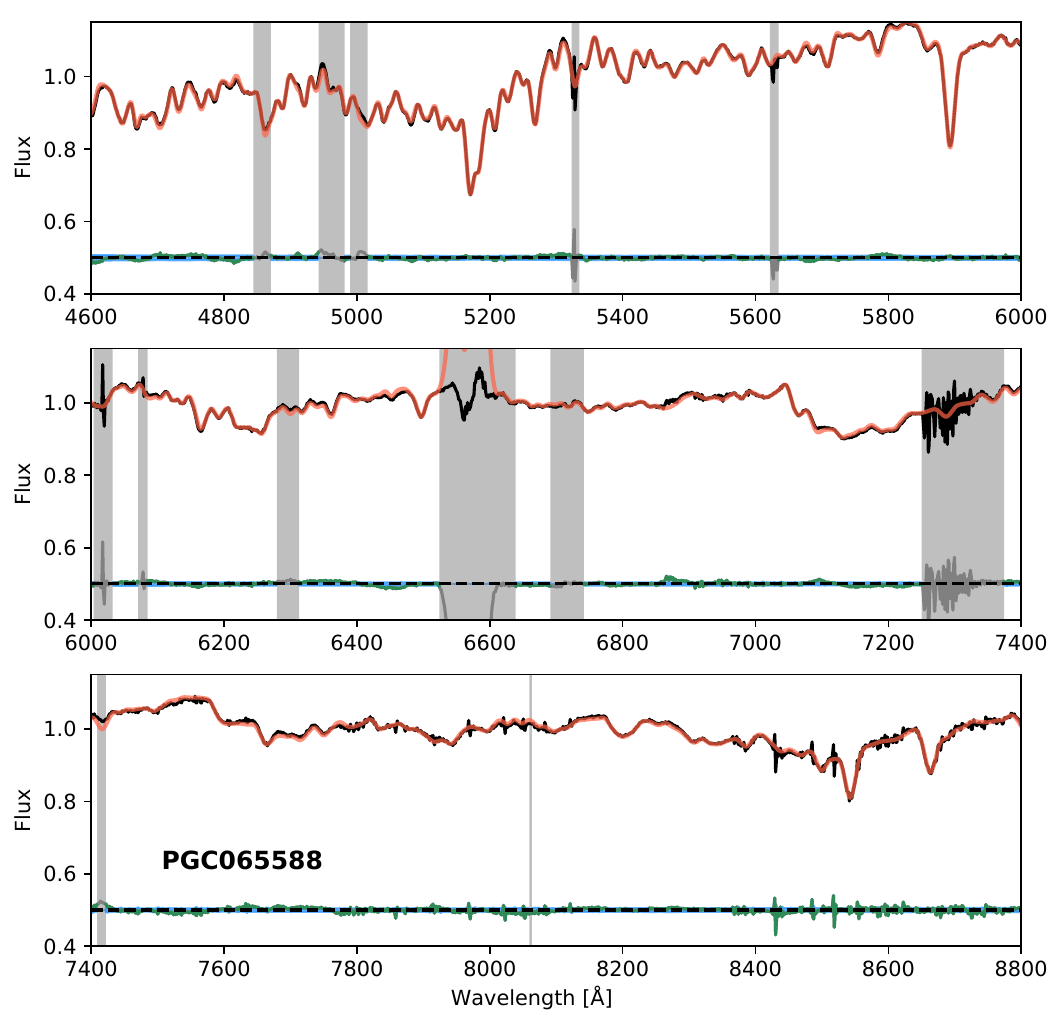}
  \includegraphics[angle=0, width=0.22\textwidth]{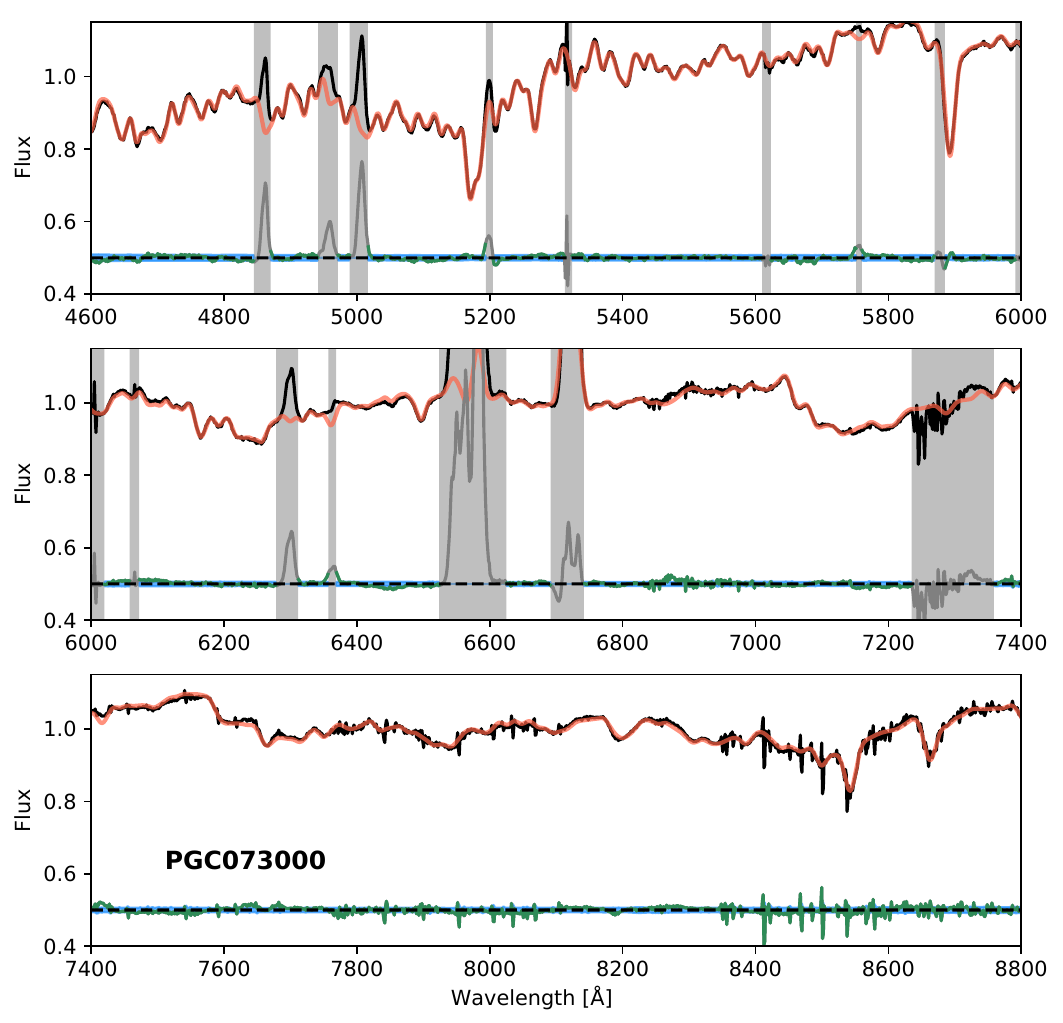}
  \includegraphics[angle=0, width=0.22\textwidth]{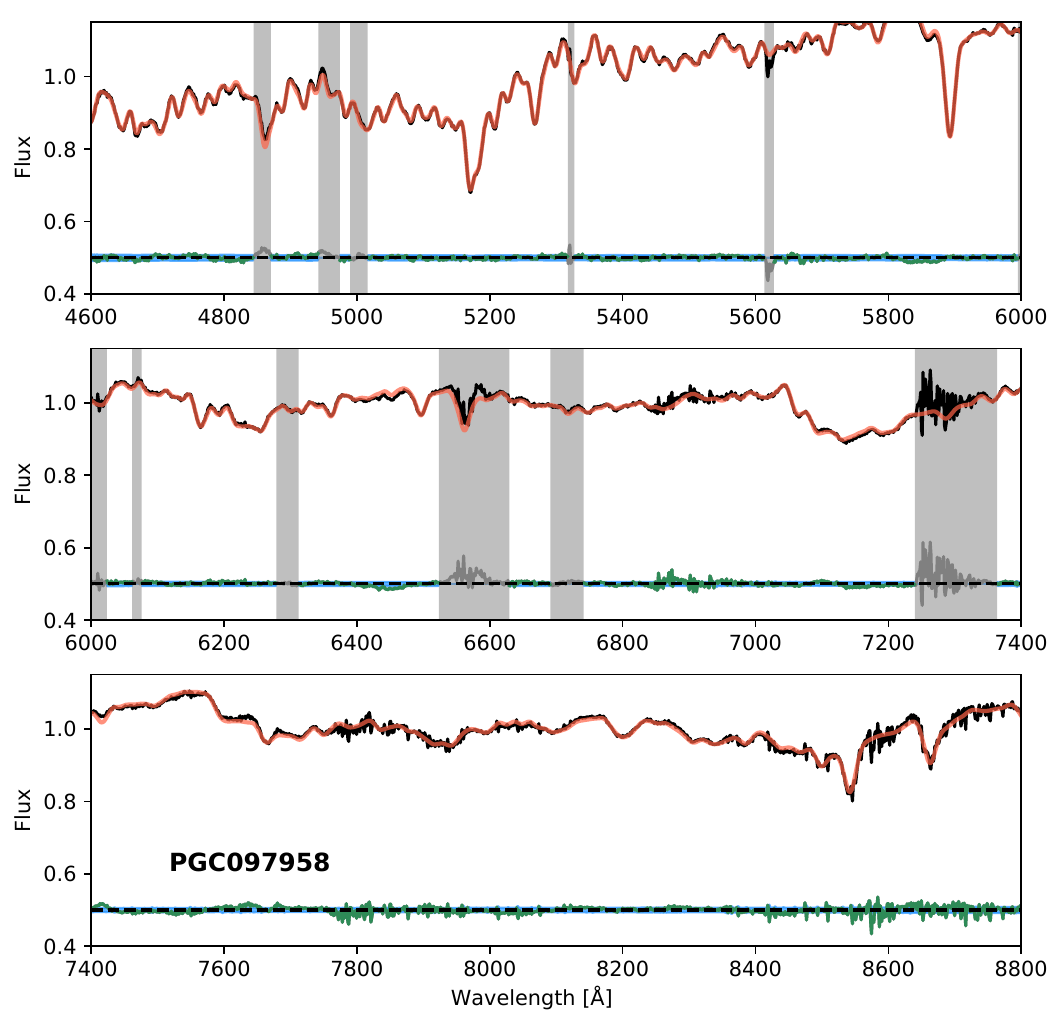}
  \includegraphics[angle=0, width=0.22\textwidth]{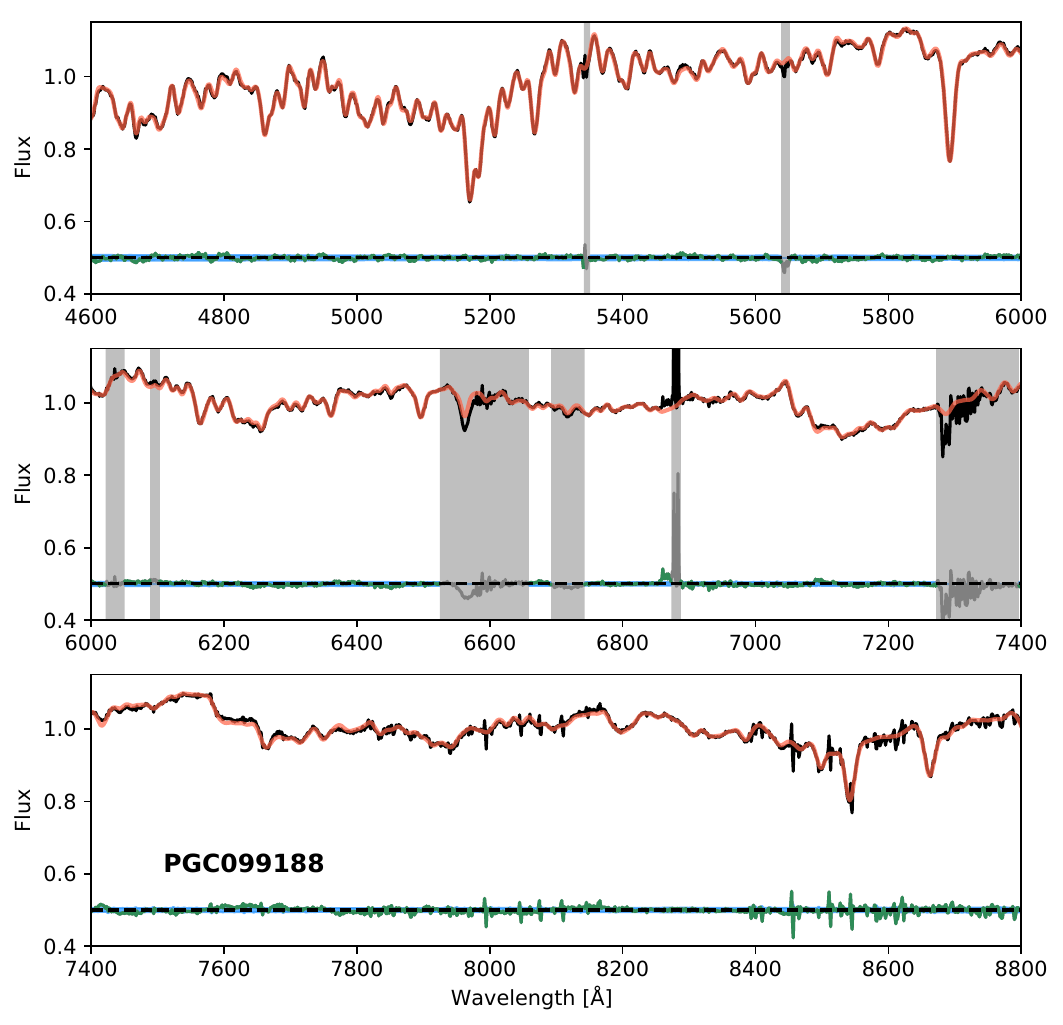}
  \includegraphics[angle=0, width=0.22\textwidth]{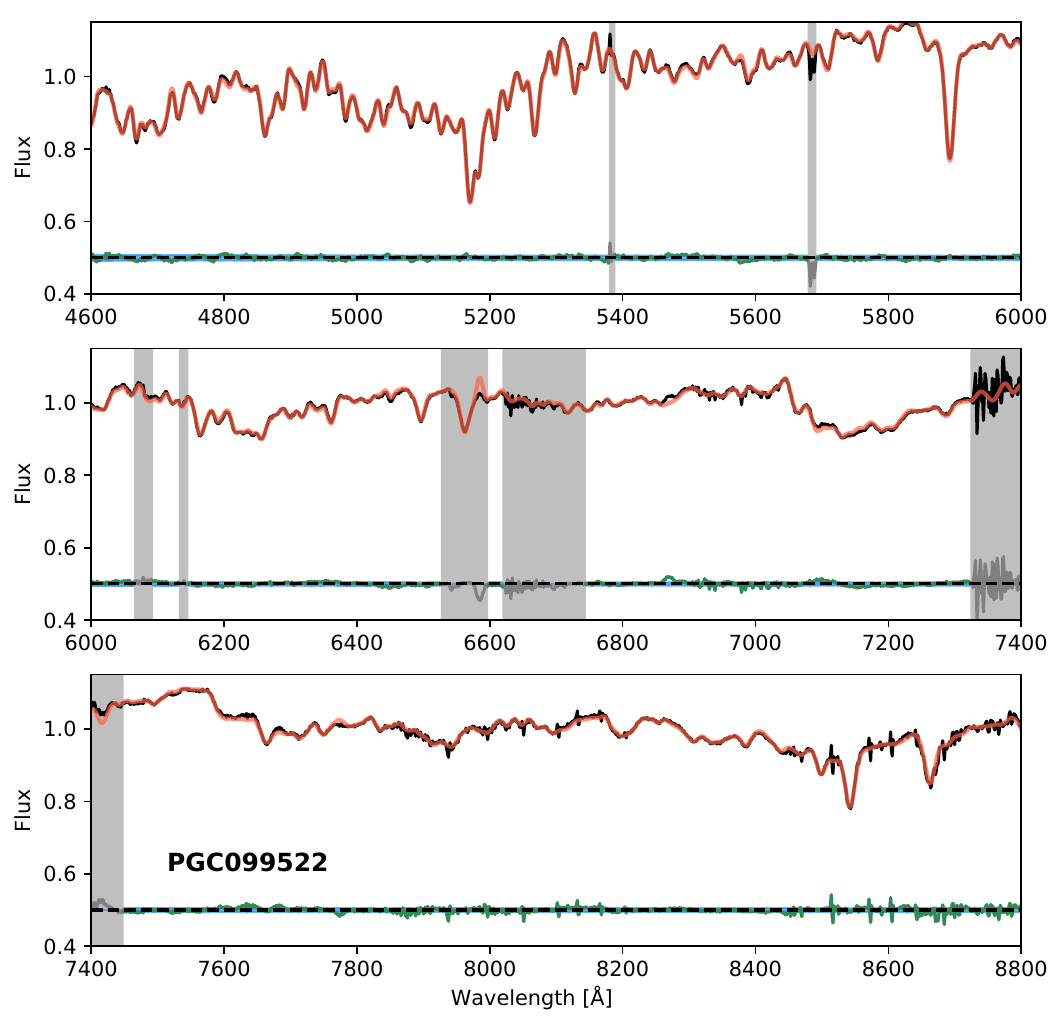}
 \caption{The best-fit \textsc{alf} models fitted to the M3G spectra extracted with in the elliptical half-right regions, each consisting of three panels, as in Fig.~\ref{fig:example_spec}. The panel with the galaxy name is the lowest of the three. Grey patches show spectral regions that were masked because of telluric absorption, sky emission or line emission. The data-model residual is shown in green (grey) in unmasked (masked) regions, and shifted to 0.5 the average flux value. Where there are no green residual points, the fit was not performed (e.g. beyond 8450\AA\, for PGC46860). The blue region denotes the 1-$\sigma$ uncertainty on the data.}
\label{fig:alf_fits}
\end{figure*}

\bsp	
\label{lastpage}
\end{document}